\documentclass[a4paper,oneside,11pt]{amsart}

\usepackage{amsmath,amssymb,epsf}
\usepackage{amsfonts,amsbsy,amscd,stmaryrd}
\usepackage{latexsym,amsopn}
\usepackage{amsthm}
\usepackage{esint}
\usepackage{mathrsfs}
\allowdisplaybreaks
\usepackage{geometry}
\usepackage{setspace}
\usepackage{graphicx}
\usepackage{color}
\definecolor{Blue}{rgb}{0.,0.,1.}
\definecolor{Red}{rgb}{1.,0.,0.}

\newcounter{smallarabics}
\newenvironment{arabicenumerate}
{\begin{list}{{\normalfont\textrm{(\arabic{smallarabics})}}}
  {\usecounter{smallarabics}\setlength{\itemindent}{0cm}
   \setlength{\leftmargin}{5ex}\setlength{\labelwidth}{4ex}
   \setlength{\topsep}{0.75\parsep}\setlength{\partopsep}{0ex}
   \setlength{\itemsep}{0ex}}}
{\end{list}}

\newcounter{smallroman}

\newcommand{\ben}{\begin{arabicenumerate}}  
\newcommand{\een}{\end{arabicenumerate}}

\def\init{\setcounter{equation}{0}}

\newtheorem{theorem}{Theorem}[section]
\newtheorem{assumption}{Hypothesis}[section]
\newtheorem{proposition}[theorem]{Proposition}
\newtheorem{lemma}[theorem]{Lemma}
\newtheorem{definition}[theorem]{Definition}

\newtheorem{remark}[theorem]{Remark}
\newtheorem{example}[theorem]{Example}
\newcommand{\beq}{\begin{equation}}
\newcommand{\eeq}{\end{equation}}

\newcommand{\bea}{\begin{aligned}}
\newcommand{\eea}{\end{aligned}}

\newcommand{\bex}{\begin{example}}
\newcommand{\eex}{\end{example}}
\def\bel{\begin{lemma}}
\def\eel{\end{lemma}}
\def\bet{\begin{theoreme}}
\def\eet{\end{theoreme}}
\def\bed{\begin{definition}}
\def\eed{\end{definition}}
\def\ber{\begin{remark}}
\def\eer{\end{remark}}

\def\rr{{\mathbb R}}
\def\zz{{\mathbb Z}}
\def\cc{{\mathbb C}}
\def\nn{{\mathbb N}}

\def\ss{{\mathbb S}}

\def\part{{\rm par}}

\def\H{{\rm H}}

\def\bar{\overline}

\def\c0inf{C_0^\infty}

\def\proof{
\noindent{\bf Proof.}\ \ }

\usepackage{calrsfs}
\DeclareMathAlphabet{\pazocal}{OMS}{zplm}{m}{n}

\def\cY{{\pazocal Y}}
\def\cZ{{\pazocal Z}}

\def\cS{{\pazocal S}}
\def\kS{{\mathcal S}}
\def\kC{{\mathcal C}}
\def\kV{{\mathcal V}}

\def\cR{{\pazocal R}}

\def\cI{{\pazocal I}}
\def\cM{{\pazocal M}}
\def\kM{{\mathfrak M}}

\def\cC{{\pazocal C}}
\def\cW{{\pazocal W}}

\def\wf{{\rm WF}}
\def\mod{{\rm \ \ mod \ }}

\def\i{{\rm i}}

\def\sgn{{\rm sgn}}

\let\Im\relax
\let\Re\relax
\DeclareMathOperator{\Ker}{Ker}
\DeclareMathOperator{\Im}{Im}
\DeclareMathOperator{\Re}{Re}

\newcommand{\qeds}{\qed\medskip}

\def \p{ \partial}

\def\12{\frac{1}{2}}
\def\14{\frac{1}{4}}

\def\supp{{\rm supp}}
\def\e{{\rm e}}

\DeclareMathOperator{\Ran}{Ran}

\def\bbbone{{\mathchoice {\rm 1\mskip-4mu l} {\rm 1\mskip-4mu l}
{\rm 1\mskip-4.5mu l} {\rm 1\mskip-5mu l}}}
\newcommand{\one}{\boldsymbol{1}}

\def\sgn{{\rm sgn}}

\def\cP{{\pazocal U}}
\def\tcP{\tilde{\pazocal U}}
\def\cJ{{\pazocal J}}
\def\c{{\rm c}}

\def\cC{{\pazocal C}}
\def\cF{{\pazocal F}}

\def\cX{{\pazocal X}}
\def\cK{{\pazocal K}}

\def\12{\frac{1}{2}}

\def\supp{{\rm supp}}

\def\e{{\rm e}}

\def\Ran{{\rm Ran}}

\def\Diff{{\rm Diff}}

\def\bz{{\rm z}}

\def\bep{\begin{proposition}}
\def\eep{\end{proposition}}
\def\b{{\rm b}}

\newcommand{\mat}[4]{\left(\begin{array}{cc}#1 &#2  \\ #3 &#4 \end{array}\right)}

\newcommand{\bra}{\langle} 
\newcommand{\ket}{\rangle}

\def\init{\setcounter{equation}{0}}

\DeclareSymbolFont{boldoperators}{OT1}{cmr}{bx}{n}
\SetSymbolFont{boldoperators}{bold}{OT1}{cmr}{bx}{n}

\makeatletter
\newcommand*{\defeq}{\mathrel{\rlap{%
                     \raisebox{0.3ex}{$\m@th\cdot$}}%
                     \raisebox{-0.3ex}{$\m@th\cdot$}}%
                     =}
                     
\makeatother
\makeatletter
\newcommand*{\eqdef}{=\mathrel{\rlap{%
                     \raisebox{0.3ex}{$\m@th\cdot$}}%
                     \raisebox{-0.3ex}{$\m@th\cdot$}}%
                     }
\makeatother

\def\sc{{\rm sc}}

\def\SolI{{\rm Sol}_{I}}
\def\Sol{{\rm Sol}}

\def\WF{{\rm WF}}

\newcommand{\traa}[1]{\mskip-6mu\upharpoonright_{#1}}

\newcommand{\metrics}{{\rm Sym}^2 T^* M}
\newcommand{\scmetrics}{{\rm Sym}^2 \,{}^{\sc} T^* M}
\def\cf{\pazocal{C}^\infty}
\def\cfd{\dot{\pazocal{C}}^\infty}
\def\cmf{{\pazocal{C}}^{-\infty}}
\def\pM{{\p M}}

\def\sources{{}^{\b}S N^{*{\scriptscriptstyle{-}}} S}
\def\sinks{{}^{\b}S N^{*{\scriptscriptstyle{+}}} S}
\def\sos{{}^{\b}S N^{*{\scriptscriptstyle{\pm}}} S}
\def\bconormal{{}^{\b}S N^* S}
\def\zero{\!{\rm\textit{o}}}
\def\be{{}^{\rm b}}
\def\c{{\rm c}}
\def\wfb{{\rm WF}_{\rm b}}

\def\ff{{\rm ff}}

\def\inti{{\circ}}
\def\Hb{H_{\b}}
\def\diag{{\rm diag}}

\newcommand{\pap}{+}
\newcommand{\pam}{-}
\newcommand{\papm}{{\pm}}

\newcommand{\ssb}{{\scriptscriptstyle{\rm b}}}

\let\origmaketitle\maketitle
\def\maketitle{
  \begingroup
  \def\uppercasenonmath##1{} % this disables uppercasing title
  \let\MakeUppercase\relax % this disables uppercasing authors
	\origmaketitle
  \endgroup
}

\newcommand\Const{2\pi}
\newcommand\FourConst{8\pi}

\begin{document}

\title[Quantum fields from global propagators on asymptotically Minkowski and de Sitter spacetimes]{\large Quantum fields from global propagators on asymptotically Minkowski \\ and extended de Sitter spacetimes}

\author{}
\address{Stanford University, CA 94305-2115, USA}
\email{andras@math.stanford.edu}
\author{\normalsize Andr\'as \textsc{Vasy} \& Micha{\l} \textsc{Wrochna}}
\address{Universit\'e Grenoble Alpes, Institut Fourier, UMR 5582 CNRS, CS 40700, 38058 Grenoble \textsc{Cedex} 09, France}
\email{michal.wrochna@univ-grenoble-alpes.fr}
%\subjclass[2010]{81T13, 81T20, 35S05, 35S35}
\begin{abstract}We consider the wave equation on asymptotically Minkowski spacetimes and the Klein-Gordon equation on even asymptotically de Sitter spaces. In both cases we show that the extreme difference of propagators (i.e.~retarded propagator minus advanced, or Feynman minus anti-Feynman), defined as Fredholm inverses, induces a symplectic form on the space of solutions with wave front set confined to the radial sets. Furthermore, we construct isomorphisms between the solution spaces and symplectic spaces of asymptotic data. As an application of this result we obtain distinguished Hadamard two-point functions from asymptotic data. Ultimately, we prove that non-interacting Quantum Field Theory on asymptotically de Sitter spacetimes extends across the future and past conformal boundary, i.e.~to a region represented by two even asymptotically hyperbolic spaces. Specifically, we show this to be true both at the level of symplectic spaces of solutions and at the level of Hadamard two-point functions.
\end{abstract}

%\begin{nouppercase}
\maketitle
%\end{nouppercase}

\section{Introduction and summary of results}

\subsection{Introduction}
As understood nowadays, the rigorous construction of a non-intera\-cting Quantum Field Theory associated to a hyperbolic differential operator $P$ on a given spacetime ($M^\inti,g$) is crucially based on two ingredients. The first one is the existence of advanced  and retarded (also called backward and forward) propagators $P_{\pm}^{-1}$, i.e.~inverses of $P$ that solve the inhomogeneous problem $Pu=f$ for $f$ vanishing at respectively future or past infinity\footnote{The convention for the signs in $P_\pm^{-1}$ is taken to be different from the one used typically in the QFT literature, for the sake of consistency with e.g.~\cite{positive}.}. The relevant properties of the propagators that one seeks to prove crucially rely on decay estimates (or support properties) of $P^{-1}_\pm f$ given decay (or compact support) of $f$. Specifically, one needs for instance to show that the formal adjoint of $P_+^{-1}$ is $P_-^{-1}$, so that $P_+^{-1}-P_-^{-1}$ is anti-hermitian, and thus defines a symplectic form using the volume density. Then by acting with $P_+^{-1}-P_-^{-1}$ on say, test functions, one gets a space of solutions equipped with the induced symplectic form. One obtains this way a \emph{symplectic space of solutions} of $P$ that physically represents the classical field theory.

The second ingredient one needs is a way to specify a quantum state. Without going into details (cf. Appendix \ref{secapp1}), this can be conveniently reformulated as the problem of constructing \emph{two-point functions} (here more specifically \emph{bosonic} ones), which in the present setup will be pairs of operators $\Lambda^\pm$ acting, say, on test functions, such that
\beq\label{eq:d2pf}
P\Lambda^\pm=\Lambda^\pm P=0, \ \ \Lambda^+-\Lambda^-=\i(P_+^{-1}-P_-^{-1}), \ \ \Lambda^\pm \geq 0,
\eeq
where positivity refers to the canonical sesquilinear pairing obtained from the volume form. The physical interpretation is then that $\Lambda^+ + \Lambda^-$ defines the one-particle Hilbert space of the quantum theory, with $\Lambda^+$ and $\Lambda^-$ representing its particle, respectively, anti-particle content. In the case of globally hyperbolic spacetimes (cf. recent reviews \cite{HW,KM}), the present consensus is that physically reasonable two-point functions should in addition satisfy the \emph{Hadamard condition}
\beq\label{eq:hadamard1}
\wf'(\Lambda^\pm)=\textstyle\bigcup_{t\in\rr}\Phi_t(\diag_{T^*M^\inti})\cap\pi^{-1}\Sigma^\pm,
\eeq
where $\bigcup_{t\in\rr}\Phi_t(\diag_{T^*M^\inti})$ is the flowout of
the diagonal in $(T^*M^\inti\times T^*M^\inti)\setminus\zero$ by the
bicharacteristic flow of the wave operator $\Box_g$ ($\Phi_t$ acts on
the left component), $\Sigma^\pm$ are the two connected components of its characteristic set and $\pi$ projects to the left component. The basic example are the \emph{vacuum two-point functions} for the Klein-Gordon operator $-\p_{z_0}^2-\Delta_\bz-{\rm \it{m}}^2$ on $1+d$-dimensional Minkowski space $\rr_{z_0}\times\rr^d_{\bz}$, i.e.:
\[
({\Lambda}^\pm f)(z_0) = \int_{\rr} \frac{\e^{\pm\i(z_0-z_0')\sqrt{\Delta_\bz+{\rm \it{m}}^2}}}{2\sqrt{\Delta_\bz+{\rm \it{m}}^2}} f(z_0') dz_0'.
\]
 More generally, pairs of operators satisfying \eqref{eq:d2pf} and \eqref{eq:hadamard1}  are known to exist in the case of the Klein-Gordon and wave equation on globally hyperbolic spacetimes \cite{FNW,GW} and are unique modulo smooth terms (i.e.~modulo operators with smooth kernel) \cite{radzikowski}. This key result is fundamentally based on Duistermaat and H\"ormander's real principal type propagation of singularities theorem \cite{DH}. Since one is however interested in setting up QFTs on more general manifolds \cite{Kay,wald}, potentially with boundary \cite{KL,rehren,zahn}, and understanding how \eqref{eq:hadamard1} can be controlled in terms of asymptotic data, one is naturally led to revisit propagation of singularities theorems and their connections to inverses of $P$. 

Incidentally, all these ingredients are reassembled in a recent approach to propagation estimates that uses microlocal analysis in a \emph{global} setup \cite{kerrds,semilinear,HV,GHV}. The main technical feature are propagation of singularities theorems that (in contrast to H\"ormander's work) are also valid near \emph{radial sets}, where the bicharacteristic flow degenerates. These are expressed as estimates microlocalized along the bicharacteristic flow, which then can be combined to yield a global estimate, at least if one can get around potential issues induced by trapping. Ultimately, if this is the case, the estimate in question translates to the Fredholm property of $P$ acting between several choices of Hilbert spaces $\cX_I$, $\cY_I$, whose precise definition depends on the details of the setup and refers in particular to the bicharacteristic flow. One obtains this way generalized inverses $P_I^{-1}$, whose wave front set can be deduced from their mapping properties. Apart from generalized inverses $P_\pm^{-1}$ that generalize the advanced and retarded propagators, one gets globally defined \emph{Feynman} and \emph{anti-Feynman} propagators \cite{GHV,positive}, whose mathematical properties and physical interpretation are an interesting subject of study in its own right \cite{GHV,positive}, cf.~\cite{BS,BS2,GW3,DS} for related works.

Before introducing any details of the setup, let us point out the main difficulty in adapting this strategy to the construction of two-point functions. Although one could fairly easily define a pair of operators $\Lambda^\pm$ satisfying the Hadamard condition \eqref{eq:hadamard1} by taking the difference of two adequately chosen inverses of $P$, one would not expect the positivity condition $\Lambda^\pm\geq 0$ to hold apart from exceptional cases (even though under quite general assumptions it is actually possible to get this way $\Lambda^+ + \Lambda^-\geq0$, see \cite{positive}). One possible alternative is to define $\Lambda^\pm$ by specifying its asymptotic data, in terms of which positivity can be hoped to be realized explicitly. In fact, this strategy has already been successfully applied indeed in the case of the conformal wave equation on a class of asymptotically flat spacetimes \cite{Mo1,Mo2,characteristic} (see also \cite{BJ,DMP1,DMP2} for other classes of spacetimes), where one can consider as data at future null infinity the characteristic Cauchy data for a conformally rescaled metric. Recent advances also show that one can define Hadamard states for asymptotically static spacetimes using tools from scattering theory \cite{GW4}. An additional important motivation for this point of view is that in QFT one is interested in constructing two-point functions with specific global or asymptotic properties (including symmetries): this has been a very active field of study recently \cite{BJ,DD,DMP1,Mo1,sanders2} and is still the subject of many conjectures \cite{KW}. 

In the present paper we consider the (rescaled, see below) wave operator $P$ on asymptotically Min\-ko\-wski spacetimes and the Klein-Gordon operator $\hat P_X$ on a class of asymptotically de Sitter spacetimes. Asymptotic data of solutions will be realized by regarding solutions as conormal distributions of a certain type, and then global inverses of $P$ and $\hat P_X$ (also called propagators) will serve us to construct the associated \emph{Poisson operators}, i.e.~the maps that assign to given asymptotic data the corresponding solution.

\subsection*{QFT on asymptotically Minkowski spacetimes} As an illustration of our setup, we start with the special case of the radial compactification of Minkowski space. 

Namely, if $M^\inti=\rr^{1+d}$ is Minkowski space with its metric $g=dz_0^2 -(d z_1^2 +\dots + d z_d^2)$, we replace it by a compact
manifold with boundary $M$ by making the change of coordinates
$z_i=\rho^{-1}\vartheta_i$ (with $\vartheta_i$ coordinates on the sphere
$\ss^{d}$) away from the origin, and then gluing a sphere at infinity,
i.e.~the boundary of $M$ is $\p M=\{\rho=0\}$ with $\rho=(z_0^2+z_1^2
+ \dots + z_d^2)^{-1/2}$. In the setup of Melrose's $\b$-analysis
\cite{melrose}, which lies at the heart of our approach, regularity
and decay are measured relatively to weighted $\b$-Sobolev spaces
$H_\b^{m,l}(M)=\rho^l H_\b^{m}(M)$, where (away from the origin, and
in a particular spherical coordinate chart $U_i$, say $\vartheta_j$,
$j=0,\ldots,n$, $j\neq i$) the
$\b$-Sobolev space $H_\b^m(M)$ is the Sobolev space
$H^m(\rr^{1+d})$ in coordinates $(-\log \rho,
\{\vartheta_j:\ j\neq i\})\in\rr\times U\subset\rr\times\rr^d$ (see e.g.~\cite[3.3]{hintz} for the detailed definition and Subsect. \ref{ss:breg} for an equivalent one). The space of smooth functions vanishing
to arbitrary order at the boundary can be conveniently characterized
as $\cfd(M)=\bigcap_{m,l\in\rr}H_\b^{m,l}(M)$ and its dual 
provides a
useful space of distributions denoted by $\cC^{-\infty}(M)$.

The definition of $H_\b^{m,l}(M)$ can be modified to allow for orders
$m$ that vary on $M$ and in the dual variables
\cite{propagation}. Specifically, we will need here $m$ to be monotone
along the (suitably reinterpreted, cf. Subsect. \ref{ss:wb})
bicharacteristic flow and for each of the two connected components
$\Sigma^\pm$, $m$ needs to be larger than the \emph{threshold value}
$\12-l$ near one of the ends and smaller than $\12-l$ near the
other. This gives in total four distinct choices that we label by a
subset $I\subset\{+,-\}$ that indicates the components of
$\Sigma^+\cup\Sigma^-$ along which $m$ is taken to be
increasing. For any such $(m,l)$, the choice of $m$ is actually immaterial in terms of the
Fredholm/invertibility properties discussed below, as long as the
properties described above, including the ends at which the particular
inequalities hold, are kept unchanged. The main outcome of the recent work of Gell-Redman, Haber and Vasy \cite{GHV} that we use here is that the rescaled wave operator
\[
P\defeq \rho^{-(d-1)/2} \rho^{-2} \Box_g \rho^{(d-1)/2} :\cX_I \to \cY_I
\]
is Fredholm as an operator acting on the Hilbert spaces
\[
\cX_I\defeq \left\{ u \in  H_\b^{m,l}(M) : \ Pu\in H_\b^{m-1,l}(M) \right\}, \quad \cY_I\defeq H_\b^{m-1,l}(M),
\]
for \emph{any} $m,l$ consistent with the choice of $I\subset\{ + ,
-\}$, apart from a discrete set of values of $l$; $P$ is actually invertible for
$|l|$ small; and the same holds true if $M$ is a small perturbation of (radially compactified) Minkowski spacetime. With the (non-standard, see Footnote 1) conventions used in the present paper, the operators $P^{-1}_{\{\pm\}}$, denoted also $P^{-1}_{\pm}$, are precisely the advanced/retarded propagators. On the other hand, the remaining two, $P_\emptyset^{-1}$ and $P_{\{+,-\}}^{-1}$, are named Feynman and anti-Feynman propagator \cite{GHV} and we show that they have indeed the same wave front set as the Feynman/anti-Feynman parametrices of Duistermaat and H\"ormander \cite{DH}.

Our first result directly relevant for QFT on perturbations of
Minkowski space is that, for $l$ not in the discrete set mentioned
above, the extreme propagator difference defines a bijection
\beq\label{eq:fbij}
P_I^{-1}-P_{I^{\rm c}}^{-1} : \ \frac{H_{\b}^{\infty,l}(M)}{P H_{\b}^{\infty,l}(M)}\longrightarrow\Sol(P),
\eeq  
where $H_\b^{\infty,l}(M)=\bigcap_{m\in\rr}H_\b^{m,l}(M)$ and $\Sol(P)$ consists of solutions of $P$ that are smooth in the interior $M^\inti$ of $M$ (more precisely, with $\b$-wave front set only at the radials sets). Furthermore, $P_I^{-1}-P_{I^\c}^{-1}$ is formally anti self-adjoint \cite{positive}, therefore by \eqref{eq:fbij}, for $l=0$ this induces a symplectic form on $\Sol(P)$. In the advanced/retarded case $I=\{\pm\}$ the resulting symplectic space of solutions represents the classical (bosonic) field theory (in fact, in our setup it plays the same role as the space of smooth space-compact solutions in standard formulations, cf. \cite{BGP}). On the other hand, the validity of \eqref{eq:fbij} in the Feynman/anti-Feynman case  ($I=\emptyset/\{+,-\}$) is far more puzzling as it seems to have no direct analogue in well-known QFT constructions, it serves us however as the first ingredient in the proof of several auxiliary results on the Feynman propagator. 

%Let us point out, however, that by a recent result \cite{positive}, $\i^{-1} (P_I^{-1}-P_{I^\c}^{-1})$ is positive for $I=\emptyset$ (when appropriately identified with a sesquilinear form, cf. Subsect. \ref{ss:ssaf}), therefore in that case the spaces \eqref{eq:fbij} meet all the formal requirements for a \emph{fermionic} classical field theory (indeed from the mathematical point of view one needs a pre-Hilbert space, as opposed to bosonic field theory where a symplectic space suffices, cf. for instance \cite{derger}). While we restrain ourselves from interpreting this observation too literally, the use of fermionic terminology will turn out to be helpful in our discussion of two-point functions constructed from asymptotic data.

Before discussing the construction of Hadamard two-point functions, let us point out that
after suitable modifications our result \eqref{eq:fbij} also applies
to the class of asymptotically Minkowski spacetimes considered by
Baskin, Vasy and Wunsch \cite{BVW} and Hintz and Vasy \cite{HV}, which
includes (globally hyperbolic) small perturbations of Minkowski space,
but is also believed to include some non globally hyperbolic
examples, cf.~Section \ref{sec:asmin} for the precise assumptions.  In this
greater generality, the work of Gell-Redman, Haber and Vasy gives the
Fredholm  property of $P_I\defeq P:\cX_I\to\cY_I$ rather than its
invertibility (unless for instance $I=\{\pm\}$ and $M^\inti$ is
globally hyperbolic) for all $l\in\rr$ except for a discrete subset
corresponding to \emph{resonances}. Consequently $P_I^{-1}$ makes sense merely as a generalized inverse, mapping from the range of $P_I$ to a predefined complement of the kernel of $P_I$. Nevertheless, the spaces in \eqref{eq:fbij} can be modified by removing some finite dimensional subspaces in such way that one still gets an isomorphism of symplectic spaces and thus a reasonable field theory. %Furthermore, we show that the generalized inverses $P_I^{-1}$ are distinguished parametrices in the sense of Duistermaat and H\"ormander \cite{DH} (i.e.~have the correct wave front set) 

An important r\^ole is played by the assumption that the kernel consists of smooth elements, specifically
\beq
\label{eq:as2p}
\Ker P_I\subset H_\b^{\infty,l}(M), 
\eeq
where strictly speaking $\Ker P_I$ is the intersection of the kernel of $P_I$ over all choices of the orders $m$ compatible with $I$. Although this assumption still needs to be better understood in the advanced/retarded case (unless $M^\inti$ is globally hyperbolic, in which case (\ref{eq:as2p}) is trivial), we prove that \eqref{eq:as2p} is actually automatically satisfied in the Feynman/anti-Feynman case at least for $l=0$.

\subsection*{Four types of asymptotic data}

Our construction of distinguished Hadamard two-point functions (as
well as the proof of \eqref{eq:as2p} in the (anti-)Feynman case) is
based on making explicit an isomorphism between the space of solutions $\Sol(P)$ and the symplectic space of their asymptotic data, to a large extent basing on the work of Baskin, Vasy and Wunsch on asymptotics of the radiation field \cite{BVW}. If $M^\inti$ is actual Minkowski space, we thus introduce the coordinate $v=\rho^2(z_0^2-(z_1^2+\dots+z_d^2))$ and then the submanifold $\{\rho=0,v=0\}$ is the union of two connected components denoted $S_\pm$ and representing the \emph{lightcone at future/past null infinity} (Figure 1). More generally, on asymptotically Minkowski spacetimes there is a coordinate $v$ with similar features, with two components of $\{\rho=0,v=0\}$ also denoted $S_\pm$.

\def\svgwidth{5cm}
%\sskip
\begin{figure}[h]\label{fig1}
\begingroup%
  \makeatletter%
  \providecommand\color[2][]{%
    \errmessage{(Inkscape) Color is used for the text in Inkscape, but the package 'color.sty' is not loaded}%
    \renewcommand\color[2][]{}%
  }%
  \providecommand\transparent[1]{%
    \errmessage{(Inkscape) Transparency is used (non-zero) for the text in Inkscape, but the package 'transparent.sty' is not loaded}%
    \renewcommand\transparent[1]{}%
  }%
  \providecommand\rotatebox[2]{#2}%
  \ifx\svgwidth\undefined%
    \setlength{\unitlength}{336.55242521bp}%
    \ifx\svgscale\undefined%
      \relax%
    \else%
      \setlength{\unitlength}{\unitlength * \real{\svgscale}}%
    \fi%
  \else%
    \setlength{\unitlength}{\svgwidth}%
  \fi%
  \global\let\svgwidth\undefined%
  \global\let\svgscale\undefined%
  \makeatother%
  \begin{picture}(1,0.98787576)%
    \put(0,0){\includegraphics[width=\unitlength]{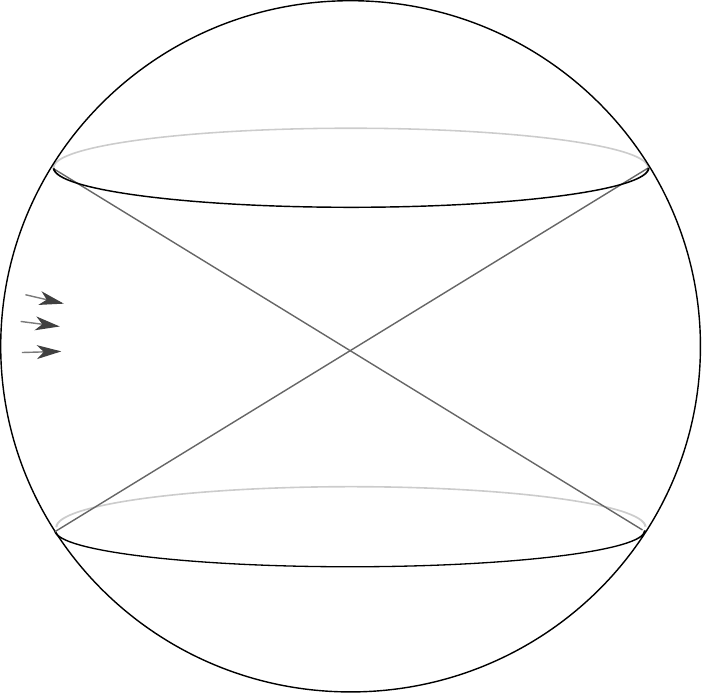}}%
    \put(0.41304778,0.70725926){\color[rgb]{0,0,0}\makebox(0,0)[lb]{\smash{$S_+$}}}%
    \put(0.37294303,0.19752967){\color[rgb]{0,0,0}\makebox(0,0)[lb]{\smash{$S_-$}}}%
    \put(0.09906525,0.50117284){\color[rgb]{0,0,0}\makebox(0,0)[lb]{\smash{$\rho\p_\rho$}}}%
    \put(0.45476981,0.8713377){\color[rgb]{0.2,0.2,0.2}\makebox(0,0)[lb]{\smash{$v>0$}}}%
    \put(0.4494318,0.07651267){\color[rgb]{0.2,0.2,0.2}\makebox(0,0)[lb]{\smash{$v>0$}}}%
    \put(0.6009072,0.46336202){\color[rgb]{0.2,0.2,0.2}\makebox(0,0)[lb]{\smash{$v<0$}}}%
  \end{picture}%
\endgroup 
\caption{Radially compactified Minkowski space $M$.}
\end{figure}
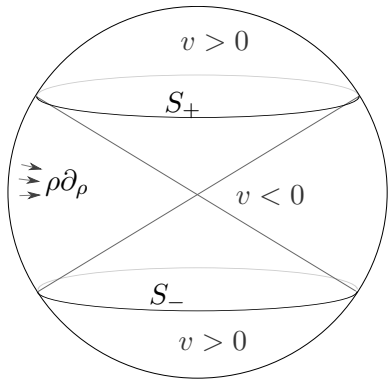

Completing the coordinates $\rho$,$v$ with some $y$ and denoting $\gamma$ the dual variable of $v$, one has as a direct consequence of \cite{BVW} that near $S_+$ (and similarly near $S_-$), any solution $u\in\Sol(P)$ can be written as the sum of two integrals of the form
\[
\int \rho^{\i\sigma} \e^{\i v \gamma} |\gamma|^{\i\sigma-1} a^{\pm}_+(\sigma,y)\chi^{\pm}(\gamma) d\gamma d\sigma
\]
modulo terms with above-threshold regularity (i.e.~in $H^{m,l}(M)$ for
some $m>\12-l$), with $\chi^\pm$ smooth and supported in
$\pm[0,\infty)$. Here $a_+^\pm(\sigma,y)$ are holomorphic functions of
$\sigma$ in a half plane with values in $\cf(S_+)$, rapidly decaying
in $\Re\sigma$, and they define a pair of asymptotic data of $u$ that
we denote $\varrho_+ u$. Similarly one can define data at past null
infinity $\varrho_-u=(a_-^+,a_-^-)$, or consider one piece of data at
future infinity and the other at past infinity: we call this
\emph{Feynman} $\varrho_\emptyset u \defeq(a_+^+,a_-^+)$ and
\emph{anti-Feynman data} $\varrho_{\{+,-\}}u\defeq
(a_+^-,a_-^-)$. Note that in all cases $\gamma>0$ corresponds to
sinks, $\gamma<0$ to sources, of the bicharacteristic flow, so in the Feynman
case the data are at the sinks, while in the anti-Feynman case at the
sources. The corresponding propagators $P_I^{-1}$ are then used to
construct Poisson operators $\cP_I$, i.e.~inverses of
$\varrho_I$. Most importantly, for any choice of $I$, if any  of the
two pieces of $\varrho_I$-data of a solution $u\in\Sol(P)$ vanishes
then $u$ has wave front set only in one of the two connected
components $\Sigma^\pm$ of the characteristic set of $P$ (in the sense
of the usual wave front set in the interior $M^\inti$). (This
is related to $(a_+^+,a_-^-)$ not being appropriate data: they are at the
sink and source in the {\em same} component of $\Sigma$.) As a consequence, denoting  $\pi^\pm$ the projections to the respective piece of data, by letting
\beq\label{eq:defla}
\Lambda_I^\pm\defeq (P_I^{-1}-P_{I^\c}^{-1})^*\varrho_I^* \pi^\pm \varrho_I (P_I^{-1}-P_{I^\c}^{-1})
\eeq
(see Subsect. \ref{ssec:asymptoticdata}--\ref{ss:hada} for details of the construction), we eventually obtain pairs of operators that satisfy $\Lambda^\pm_I\geq 0$, $P\Lambda_I^\pm=\Lambda_I^\pm P=0$ and the Hadamard condition \eqref{eq:hadamard1}. Moreover, by means of a pairing formula we show that they satisfy the relation
\beq
\Lambda_I^+-\Lambda^-_I=\i(P_+^{-1}-P_{-}^{-1})
\eeq
exactly if $I=\{\pm\}$, and modulo possible terms smooth in $M^\inti$ if $I=\emptyset$ or $I=\{+,-\}$, and thus we conclude: 

\begin{theorem}The operators $\Lambda_I^\pm$ with $I=\{+\}$ and $I=\{-\}$ are Hadamard two-point functions, i.e.~they satisfy \eqref{eq:d2pf} and \eqref{eq:hadamard1}.\end{theorem}

 These can be interpreted as the analogues of two-point functions
 constructed in \cite{Mo1,Mo2,characteristic} from data at future or
 past infinity in the case of the conformal wave equation, and in
 \cite{GW4} from scattering data in the case of the massive
 Klein-Gordon equation, even though the methods are  very
 different. On the other hand, the operators $\Lambda_I^\pm$
 in the Feynman/anti-Feynman case are a side product of our analysis and are primarily of mathematical interest (though they coincide with the vacuum two-point functions in the case of exact Minkowski space): we show indeed the identity $\Lambda_I^+ + \Lambda_I^- = \i^{-1} (P_I^{-1} - P_{I^\c}^{-1})$, which provides a refinement of the positivity result from \cite{positive}.

\subsection*{QFT on extended asymptotically de Sitter spacetimes} Our results for asymptotically de Sitter spacetimes are to some extent analogous to the case of asymptotically Minkowski ones, thanks to the duality between the Klein-Gordon equation on the former and the wave equation on the latter, made explicit in \cite{poisson} by means of a Mellin transform in $\rho$. Considering for simplicity the case of exact (radially compactified) Minkowski space $M$ of dimension $d+1$, recall that the $d$-dimensional de Sitter spacetime $(X_0,g_{X_0})$ is by definition the hyperboloid $z_0^2-(z_1^2+\dots + z_d^2)=-1$ in $M$ equipped with the induced metric. In the compactified picture it can be conveniently identified with the subregion $\{\rho=0,v<0\}$ of the sphere at infinity (i.e.~of the boundary $\p M=\{\rho=0\}=\ss^{d}$). In a similar vein, the hyperboloids $z_{0}^2-(z_1^2+\dots + z_d^2) =1$ with either $z_0>0$ or $z_0<0$ are two copies of hyperbolic space $(X_\pm,g_{X_\pm})$ (also called `Euclidean AdS' in the physics literature) and are identified with the two connected components of the region $\{\rho=0, v>0\}$. Here we consider $(X_0,g_{X_0})$, resp. $(X_\pm,g_{X_\pm})$ as compact manifolds with boundary, i.e.~$\p X_0=S_+\cup S_-$ and $\p X_\pm = S_\pm$. The boundary of de Sitter, $\p X_0$, is called traditionally the \emph{conformal infinity} (or \emph{conformal boundary}), thus the whole boundary of $M$,
\beq
\label{eq:components}
\p M = X_ + \cup X_0 \cup X_-,
\eeq      
represents de Sitter spacetime extended across conformal infinity (which we simply call  extended de Sitter spacetime), see Figure 2. 

\def\svgwidth{13cm}
%\sskip
\begin{figure}[h]\label{fig2}
\begingroup%
  \makeatletter%
  \providecommand\color[2][]{%
    \errmessage{(Inkscape) Color is used for the text in Inkscape, but the package 'color.sty' is not loaded}%
    \renewcommand\color[2][]{}%
  }%
  \providecommand\transparent[1]{%
    \errmessage{(Inkscape) Transparency is used (non-zero) for the text in Inkscape, but the package 'transparent.sty' is not loaded}%
    \renewcommand\transparent[1]{}%
  }%
  \providecommand\rotatebox[2]{#2}%
  \ifx\svgwidth\undefined%
    \setlength{\unitlength}{751.36991694bp}%
    \ifx\svgscale\undefined%
      \relax%
    \else%
      \setlength{\unitlength}{\unitlength * \real{\svgscale}}%
    \fi%
  \else%
    \setlength{\unitlength}{\svgwidth}%
  \fi%
  \global\let\svgwidth\undefined%
  \global\let\svgscale\undefined%
  \makeatother%
  \begin{picture}(1,0.44248775)%
    \put(0,0){\includegraphics[width=\unitlength]{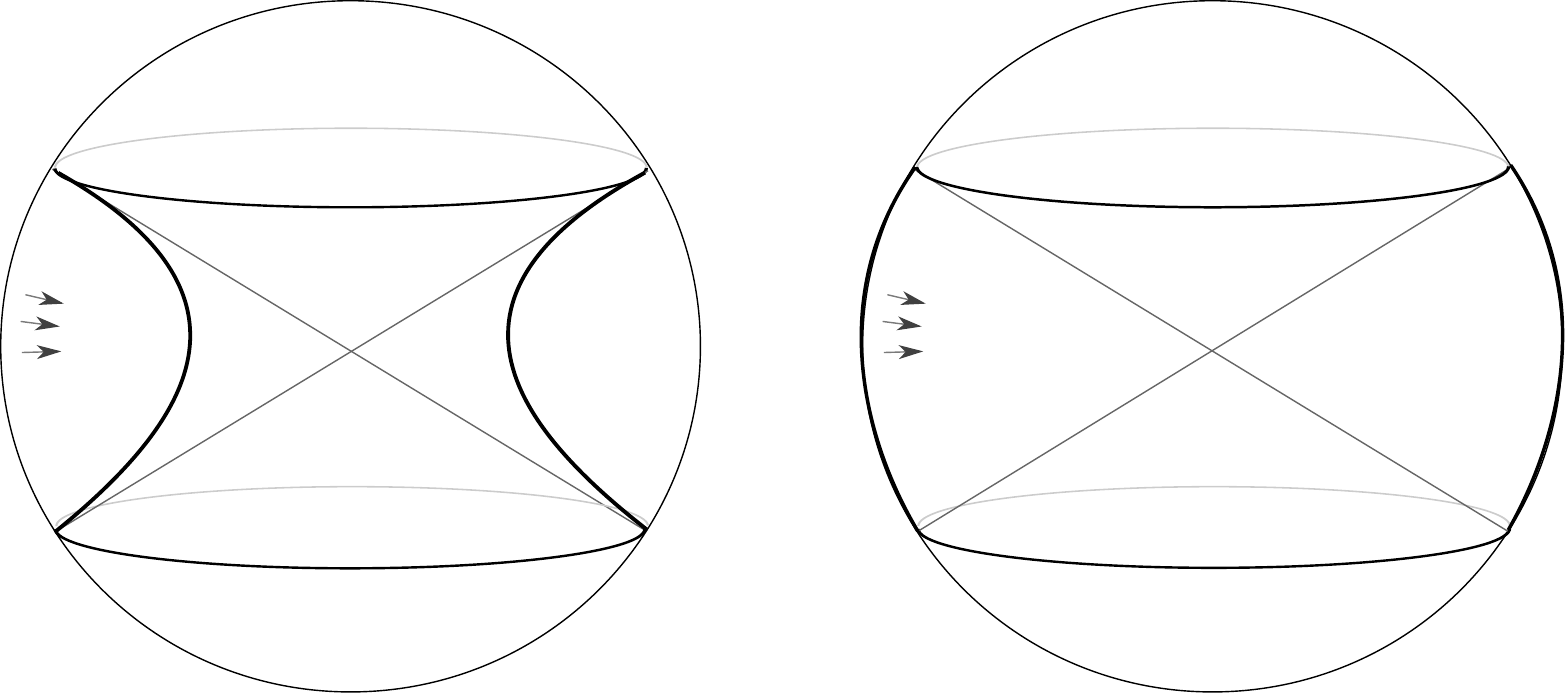}}%
    \put(0.18501171,0.31679445){\color[rgb]{0,0,0}\makebox(0,0)[lb]{\smash{$S_+$}}}%
    \put(0.16704805,0.08847718){\color[rgb]{0,0,0}\makebox(0,0)[lb]{\smash{$S_-$}}}%
    \put(0.04437315,0.22448454){\color[rgb]{0.2,0.2,0.2}\makebox(0,0)[lb]{\smash{$\rho \p_\rho$}}}%
    \put(0.20369978,0.39028821){\color[rgb]{0.275,0.275,0.275}\makebox(0,0)[lb]{\smash{$v>0$}}}%
    \put(0.2073929,0.03275038){\color[rgb]{0.275,0.275,0.275}\makebox(0,0)[lb]{\smash{$v>0$}}}%
    \put(0.22436556,0.16275654){\color[rgb]{0.275,0.275,0.275}\makebox(0,0)[lb]{\smash{$v<0$}}}%
    \put(0.71738971,0.3213576){\color[rgb]{0,0,0}\makebox(0,0)[lb]{\smash{$S_+$}}}%
    \put(0.71463641,0.09151928){\color[rgb]{0,0,0}\makebox(0,0)[lb]{\smash{$S_-$}}}%
    \put(0.59500356,0.22448454){\color[rgb]{0.2,0.2,0.2}\makebox(0,0)[lb]{\smash{$\rho\p_\rho$}}}%
    \put(0.74041425,0.39485132){\color[rgb]{0.275,0.275,0.275}\makebox(0,0)[lb]{\smash{$v>0$}}}%
    \put(0.74410744,0.0312294){\color[rgb]{0.275,0.275,0.275}\makebox(0,0)[lb]{\smash{$v>0$}}}%
    \put(0.7543488,0.16731962){\color[rgb]{0.275,0.275,0.275}\makebox(0,0)[lb]{\smash{$v<0$}}}%
    \put(0.13329194,0.21751388){\color[rgb]{0,0,0}\makebox(0,0)[lb]{\smash{\scalebox{1.2}{${X_0}$}}}}%
    \put(0.64540707,0.24485976){\color[rgb]{0,0,0}\makebox(0,0)[lb]{\smash{\scalebox{1.2}{${X_0}$}}}}%
    \put(0.84209247,0.03194807){\color[rgb]{0,0,0}\makebox(0,0)[lb]{\smash{\scalebox{1.1}{${X_-}$}}}}%
    \put(0.86208129,0.38734683){\color[rgb]{0,0,0}\makebox(0,0)[lb]{\smash{\scalebox{1.1}{${X_+}$}}}}%
  \end{picture}%
\endgroup 
\caption{The de Sitter hyperboloid $X_0$ before and after identification with the `equatorial belt' region of the boundary $\{\rho=0\}$ of radially compactified Minkowski space. The two other regions are two copies $X_\pm$ of hyperbolic space.}
\end{figure}
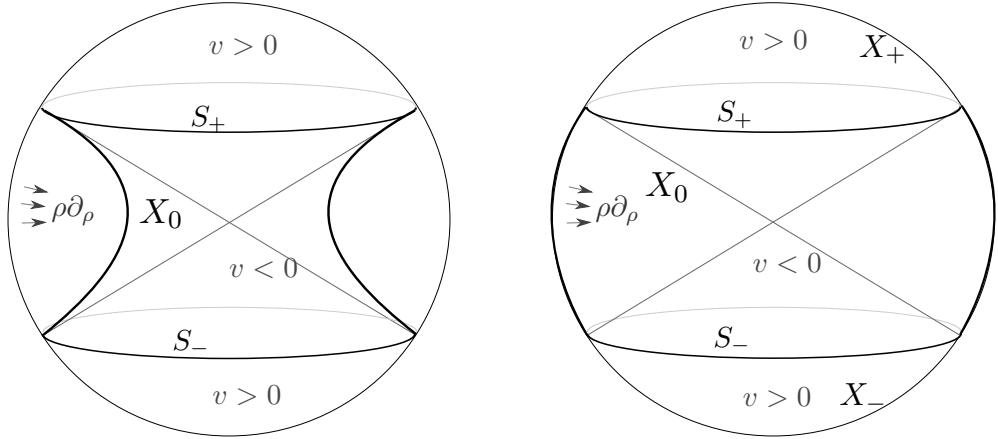
Following \cite{poisson}, we consider the differential operator on $X\defeq \pM=X_ + \cup X_0 \cup X_-$
\[
\hat P_X(\sigma)\defeq \cM_\rho P \cM_\rho^{-1}=\cM_\rho\rho^{-(d-1)/2} \rho^{-2} \Box_g \rho^{(d-1)/2}\cM_\rho^{-1},
\]
obtained from $P$ by conjugating it with the Mellin transform\footnote{Recall that the Mellin transform of $u\in\cC_\c^\infty((0,\infty))$ is defined by $(\cM_\rho u)(\sigma)\defeq \int_0^\infty \rho^{-\i\sigma-1}u(\rho) d\rho.$} $\cM_\rho$ in $\rho$ and thus depending on a complex variable $\sigma$. The crucial ingredient in our analysis are the two identities
\beq\label{eq:twodualities}
\bea
\hat P_{X}\traa{X_0}&=x^{-\i\sigma-(d-1)/2-2}_{X_0}(\Box_{X_0} - \sigma^2 -(d-1)^2/4) x_{X_0}^{\i\sigma+(d-1)/2},\\
\hat P_{X}\traa{X_\pm}&=x^{-\i\sigma-(d-1)/2-2}_{X_\pm}(-\Delta_{X_\pm} + \sigma^2 +(d-1)^2/4) x_{X_\pm}^{\i\sigma+(d-1)/2},
\eea
\eeq
to the very best of our knowledge made explicit the first time in \cite{poisson},
where
\[
x_{X_0}=\left(\frac{z_1^2+\dots + z_d^2 - z_0^2}{z_1^2+\dots + z_d^2 + z_0^2}\right)^{\12}, \quad x_{X_\pm}=\left(\frac{z_0^2 -(z_1^2+\dots + z_d^2) }{z_1^2+\dots + z_d^2 + z_0^2}\right)^{\12}.
\]
As the first identity in \eqref{eq:twodualities} connects $P$ with the Klein-Gordon operator on $X_0$, this suggests a sort of duality\footnote{Let us mention that the possibility of a duality between quantum fields on Minkowski and de Sitter spacetimes has attracted widespread interest in the physics literature, see \cite{BGMS,boer} for proposals somehow close in spirit to our approach, though technically different (cf. the work of Strominger \cite{strominger} for an entirely different proposal that relates the QFT on $X_0$ to a conformal field theory on the conformal boundary). It is also interesting to note that a work of Moschella and Schaeffer \cite{MS} discusses the Laplace operator  on hyperbolic space in connection to QFT on de Sitter space, although it provides no construction of a QFT on hyperbolic space.}  between QFT on $M$ and QFT on de Sitter space $X_0$ and one can wonder if that would mean that there is also a duality between QFT on $M$ and a hypothetical QFT on hyperbolic space $X_+$ (or $X_-$). Instead of addressing the question directly, in the present paper we set a QFT on the whole extended de Sitter space $X$ and shows that it extends the QFT on the de Sitter region $X_0$. Beside the case of exact de Sitter space, our results do also apply to even asymptotically de Sitter spacetimes (Definition \ref{def:as}), introduced in \cite{poisson} (extended by two even asymptotically hyperbolic spaces, cf. the work of Guillarmou \cite{guillarmou}), where a direct analogue of \eqref{eq:components} and \eqref{eq:twodualities} is available in terms of some asymptotically Minkowski spacetime $M$.

The relevant feature of the operator $\hat P_X$ on extended asymptotically de Sitter spacetimes is that it fits into the framework of \cite{kerrds,semilinear} and thus possesses various inverses in a similar way as $P$ does (here as meromorphic functions of $\sigma$), the main difference being that one only needs to consider regularity in the sense of Sobolev spaces $H^s(X)$ (note that $X$ is a closed manifold), where $s$ varies in phase space. This allows us to obtain in a very analogous way an isomorphism
\beq\label{eq:isoi1}
\hat P_{X,I}^{-1}-\hat P_{X,I^c}^{-1}: \ \frac{\cf(X)}{\hat P_{X} \cf(X)}\longrightarrow{\Sol}(\hat P_{X})
\eeq 
with $\Sol(\hat P_X)$ the space of solutions of $\hat P_X u=0$ such that $\wf(u)\subset N^*(S_+\cup S_-)$. Moreover, the definition of Hadamard two-point functions transports directly to this case, thus once their existence is proved one gets a perfectly reasonable QFT on $X$ (at least if $\sigma\in\rr$ so that $\hat P_{X,I^c}^{-1}$ is the formal adjoint of $\hat P_{X,I}^{-1}$), despite it being governed by a differential operator $\hat P_X$ that is hyperbolic only in the asymptotically de Sitter region $\{ v < 0\}$. In order to understand the relation of this new QFT with the well-established theory on $X_0$, let us recall that the latter is based on the isomorphism
\[
\hat P_{X_{0},+}^{-1} - \hat P_{X_{0},-}^{-1} : \ \frac{\cf_\c(X_0^\inti)}{\hat P_{X_0} \cf_\c(X_0^\inti)}\longrightarrow{\Sol}(\hat P_{X_0})
\]
where $\Sol(\hat P_{X_0})$ is the space of solutions of $\hat P_{X_0}$ that are smooth in the interior $X_0^\inti$. On the other hand, we prove that the map
\beq\label{eq:supiso}
\traa{X_0}\circ \, x_{X_0}^{\i\sigma+(d-1)/2}: \ \Sol(\hat P_{X})\to \Sol(\hat P_{X_0})
\eeq
is an isomorphism (i.e.~symplectomorphism), which allows to conclude that QFT on $X_0$ \emph{extends across the boundary}. Even more specifically, we show:

\begin{theorem} Any pair of Hadamard two-point functions $\Lambda_{X_0}^\pm$ on an even asymptotically de Sitter spacetime $(X_0,g_{X_0})$ extends canonically to Hadamard two-point functions $\Lambda_X^\pm$ on $X$ via the isomorphism \eqref{eq:supiso}.\end{theorem}

In our terminology, the statement that the two-point functions $\Lambda_X^\pm$ on $X$ are \emph{Hada\-mard} means that they satisfy a wave front set condition which is formulated using the bicharacteristic flow of $\hat P_X$ in an analogous way to how the usual Hadamard condition on $X_0$ is expressed using the bicharacteristic flow of $\hat P_{X_0}$ (see Definition \ref{def:onX}). In particular, the restriction of $\Lambda_X^\pm$ to the two regions with Euclidean signature is an operator with smooth Schwartz kernel.

 Furthermore, we construct Hadamard two-point functions $\Lambda_{X_0,I}^\pm$ on $X_0$  from asymptotic data in a similar fashion as in the Minkowski case: these then extend to Hadamard two-point functions on $X$ and we give a direct formula for the latter in terms of the $X_0$ asymptotic data.

\subsection*{QFT on asymptotically hyperbolic space}

Since the two-point functions on asymptotically de Sitter space $X_0$ give rise to two-point functions on the extended space $X$, one can wonder whether on the two copies $X_+$, $X_-$ of asymptotically hyperbolic space there is a structure that resembles the symplectic space on $X_0$. We show that in fact there is an isomorphism
\beq\label{eq:isoi3}
\hat P_{X_{\pm},+}^{-1} - \hat P_{X_{\pm},-}^{-1} : \ \frac{\cfd(X_\pm)}{\hat P_{X_\pm}\cfd(X_\pm)}\longrightarrow{\Sol}(\hat P_{X_\pm})
\eeq   
where $\hat P_{X_{\pm},+}^{-1}$, $\hat P_{X_{\pm},-}^{-1}$ are defined by analytic continuation of the resolvent of $\Delta_{X_\pm}$ starting from positive, resp. negative large values of the imaginary part of complex parameter $\sigma$, and $\Sol(\hat P_{X_\pm})$ is a space of solutions (defined more precisely in Subsect. \ref{ssec:asymptoticdata}) of $\hat P_{X_\pm}$ that are smooth in the interior $X_\pm^\inti$. We show that by taking \emph{two} copies of either ${\Sol}(\hat P_{X_+})$ or ${\Sol}(\hat P_{X_-})$ one obtains a symplectic space that is isomorphic to the one in the de Sitter region, $\Sol(\hat P_{X_0})$. Thus, on the level of non-interacting quantum fields, one field on $X_0$ corresponds to a pair of fields on $X_+$ or $X_-$.

Let us stress that the QFT obtained this way, although of course defined with fundamentally Euclidean objects, is crucially different from Euclidean QFTs often considered in the physics literature and obtained by a Wick rotation (i.e.~complex scaling) of the time variable in a relativistic QFT, cf. \cite{JJM,JR1,JR2} for the case of curved spacetimes and other recent developments. For instance, our two-point functions on $X_\pm$ are subject to a positivity condition reminiscent of relativistic QFT, as opposed to the reflection positivity in Euclidean QFT. 

\subsection*{Outlook} Since the two-point functions $\Lambda_{X_+^2,+}^\pm$ that we consider on two copies of an asymptotically hyperbolic space (see Subsect. \ref{lastss}) have smooth Schwartz kernel, we expect that this could serve as a basis to construct a very regular interacting (i.e.~non-linear) QFT. We plan to follow on this idea in a future work.

One can also wonder whether the strategy adopted in the present paper
extends to other classes of spacetimes, possibly with trapping; it is
plausible that this question could be addressed using the recent
advances in \cite{semilinear,kerrds,BW,dyatlov}.

A further aspect to look into is the relation of the Feynman and
anti-Feynman asymptotic data that we consider here with the generalized
Atiyah-Patodi-Singer and anti-Atiyah-Patodi-Singer boundary data
adapted recently to the Lorentzian case by B\"ar and Strohmaier
\cite{BS,BS2} in the context of the Dirac equation on globally hyperbolic manifolds with a compact Cauchy surface, where the boundary conditions are considered at finite times. Although the setup is clearly different, there are many
striking analogies to be explored \cite{GW3}, which suggest a direct link of Feynman asymptotic data with particle creation, in particular it would
be thus beneficial to have a Dirac version of our results. (Cf.\ the
differential forms setup of \cite{forms}.)

\subsection{Summary of results}  Our main results can be summarized as follows.

In the case of the wave equation on an asymptotically Minkowski
spacetime $M$, we assume that $l=0$ is not a resonance (i.e., of the
Mellin transformed normal operator family of the relevant function
space setup corresponding to $I$, $I^\c$, see Subsect. \ref{ss:invop}), and we assume `smoothness of kernels' \eqref{eq:as2p}.

 \begin{enumerate}\itemsep1.5mm 
\item[\textbf{1)}]\label{res1} In Proposition \ref{prop:exact} we prove that the propagator difference $P_I^{-1}-P_{I^\c}^{-1}$ induces an isomorphism that generalizes \eqref{eq:fbij}. 
\item[\textbf{2)}]\label{res2} In Proposition \ref{prop:bijec} we show bijectivity of the maps $\varrho_I$ that assign to a solution its asymptotic data (strictly speaking, in order to have a bijection we consider a space of solutions $\Sol_I(P)$ with elements of $\Ker P_I$, $\Ker P_{I^\c}$ removed) and then Theorem \ref{prop:iso} provides an explicit formula for the induced symplectic form on asymptotic data.
\item[\textbf{3)}]\label{res3} In Theorem \ref{prop:Had}  we show that the operators $\Lambda^\pm_I$ constructed from asymptotic data \eqref{eq:defla} satisfy a condition which for $(M^\inti,g)$ globally hyperbolic reduces to the Hadamard condition \eqref{eq:hadamard1}. In particular we get then two pairs of Hadamard two-point functions $\Lambda_-^\pm$, $\Lambda_+^\pm$ from data at past and future null infinity. 
\end{enumerate}

Next, for any even asymptotically de Sitter spacetime $X_0$, we consider the Klein-Gordon operator $\hat P_{X_0}=\Box_{X_0} - \sigma^2 -(d-1)^2/4$ and the associated operators on the extended space $X$ and on the asymptotically hyperbolic spaces $X_\pm$. We assume that $\sigma\in \rr\setminus\{0\}$  is not a pole of $\hat P_{X,I}^{-1}(\sigma)$.

 \begin{enumerate}\itemsep1.5mm 
\item[\textbf{4)}]\label{res4} In Propositions \ref{prop:isoxxo} and \ref{prop:usio3} we prove the isomorphisms \eqref{eq:isoi1}, \eqref{eq:isoi3} induced by respective propagator differences, and the isomorphism \eqref{eq:supiso} between solution spaces on $X$ and on $X_0$. The construction of Hadamard two-point functions is summarized in Theorem \ref{thm:newhada}.
\item[\textbf{5)}]\label{res5} In Propositions \ref{newsdj} and \ref{newprop2} we relate symplectic spaces and two-point functions on $X_0$ to analogous objects on two copies of the asymptotically hyperbolic space $X_+$.
\end{enumerate}

In particular,  the results summarized in \textbf{4)} mean that non-interacting scalar fields on even asymptotically de Sitter spacetime canonically extend across the conformal boundary.

\section{Asymptotically Minkowski spacetimes and propagation of singularities}\init\label{sec:asmin}

\subsection{Notation} If $M$ is a smooth manifold with boundary $\p M$, we denote by $M^\inti$ its interior. We denote by $\cf(M)$ the space of smooth functions on $M$ (in the sense of extendability across the boundary). The space of smooth functions vanishing with all derivatives  at the boundary $\p M$ are denoted $\dot\cC^{\infty}(M)$ and their dual $\cC^{-\infty}(M)$. The signature of Lorentzian metrics is taken to be $(+,-,\dots,-)$. We adopt the convention that sesquilinear forms $\bra\cdot,\cdot\ket$ are linear in the second argument.

\subsection{Geometrical setup}

The spacetime of interest is modelled by an $n$-dimensional smooth manifold $M$ with boundary $\pM$ ($n\geq 2$), equipped with a \emph{Lorentzian scattering metric} $g$. 

To define this class of metrics, let $\rho$ be a boundary-defining function of $\p M$, meaning that $\p M=\{\rho=0\}$ and $d\rho\neq 0$ on $\p M$, and let $w=(w_1,\dots,w_{n-1})$ be coordinates on $\p M$. Then ${}^{\rm sc}T^*M$ is the bundle whose sections are locally given by the $\cf(M)$-span of the differential forms $\rho^{-2}d\rho, \rho^{-1}d w=(\rho^{-1}dw_1,\ldots,\rho^{-1}dw_{n-1})$. Lorentzian scattering metrics are by definition non-degenerate sections of ${\rm Sym}^2 {}^{\rm sc}T^*M$ of Lorentzian signature \cite{melrose2}, and they define an open subset of $\cf(M;\scmetrics)$ (equipped with the $\cf$ topology).

A more refined structure near the boundary $\p M$ can be specified as follows \cite{BVW,HV,GHV}.

\begin{definition}\label{def:sc}One says that $(M,g)$ is a \emph{Lorentzian scattering space} if there exists $v\in\cf(M)$ s.t. $v\traa{\p M}$ has non-degenerate differential at $S\defeq\{\rho=0,v=0 \}$ and moreover:
\begin{itemize}
\item on $\p M$, $g(\rho^2\p_\rho,\rho^2\p_\rho)$ has the same sign as $v$;
\item $g$ has the form
\beq\label{eq:formscmetrics}
g= v \frac{d\rho^2}{\rho^4} - \left(\frac{d\rho}{\rho^2}\otimes\frac{\alpha}{\rho}+\frac{\alpha}{\rho}\otimes\frac{d\rho}{\rho^2}\right)-\frac{\tilde{g}}{\rho^2},
\eeq
where $\tilde g\in \cf(M;\metrics)$ with
$\tilde{g}\traa{(d\rho,dv)^{\rm ann}}$ positive definite\footnote{Here
  $\tilde{g}\traa{(d\rho,dv)^{\rm ann}}$ denotes the restriction of
  $\tilde g$ to the annihilator of the span of $d\rho,d v$.} at $S$,
and $\alpha$ is a one-form on $M$ of the form
$\alpha=dv/2+O(v)+O(\rho)$ near $S$.
\end{itemize}
\end{definition} 

The zero-set $S=\{v=0,\rho=0\}$ is called the \emph{light-cone at infinity} and is in fact a submanifold of $M$.

 The example of primary importance of a Lorentzian scattering space is the radial compactification of $n=1+d$-dimensional Minkowski space $\rr^{1,d}$ outlined in the introduction. Namely, writing the Minkowski metric as $dz_0^2-(d  z^2_1+ \dots + d z^2_d)$, a manifold $M$ with boundary $\p M=\{\rho=0\}$ is obtained  by making the change of coordinates $z_0=\rho^{-1}\cos\theta$, $z_i=\rho^{-1}\omega_i \sin \theta$, (valid near $\rho=0$), where $\rho=(z_0^2+z_1^2+\dots+z_d^2)^{-1/2}$ and $\omega_i$ are coordinates on the sphere $\ss^{d-2}$. Then a further change of coordinates 
\[
v=\cos 2\theta=\rho^2(z_0^2-(z_1^2+\dots+z_d^2))
\]
brings the metric into the form
\[
g= v \frac{d\rho^2}{\rho^4}-\frac{v}{4(1-v^2)}\frac{dv^2}{\rho^2}-\frac{1}{2}\left( \frac{d\rho}{\rho^2}\otimes \frac{dv}{\rho}+\frac{dv}{\rho}\otimes \frac{d\rho}{\rho^2}\right) + \frac{1-v}{2}\frac{d\omega^2}{\rho^2},
\]
which is a special case of (\ref{eq:formscmetrics}) with
$\alpha=dv/2$.

\subsection{Wave operator and $\b$-geometry}\label{ss:wb}

The main object of interest is the wave operator $\Box_g \in \Diff^2(M)$. It is convenient to introduce at once the conformally related operator
\beq\label{eq:wop}
P\defeq \rho^{-(n-2)/2} \rho^{-2} \Box_g \rho^{(n-2)/2}.
\eeq
With this definition, $P$ is a $\b$\emph{-differential operator}, that
is $P\in \Diff_\b^2(M)$ where $\Diff^k_\b(M)$ consists of differential
operators of order $k$ which are in the algebra $\cf(M)$-generated by
$\rho\p_\rho,\p_w$, using as before coordinates $(\rho,w)$ near $\p
M$. The operator $P$ is formally self-adjoint with respect to the
\emph{$\b$-density} (i.e., smooth section of the density bundle of
$\be TM$) $\rho^n |d g|$. We denote by $\bra \cdot,\cdot \ket_\ssb$ the
corresponding pairing and $L^2_\b(M)$ the Hilbert space it defines.

Let us now introduce the notions relevant for the description of the bicharacteristic flow in the $\b$-setting. To start with, the $\cf(M)$-module generated by the vector fields $\rho\p_\rho$, $\p_{w}$ can be viewed as the space of smooth sections of a bundle $\be TM$, called the $\b$-tangent bundle. The dual bundle $\be T^* M$ is called the $\b$-conormal bundle and locally near $\p M$ its sections are the $\cf(M)$-span of $\rho^{-1}d\rho$, $d w$. Since $\b$-vector fields (i.e., sections of $\be T M$) can also be considered as sections of $ T M$, there is a canonical embedding $\cf(\be TM)\to\cf(TM)$ and a corresponding dual map on covectors. Now for a submanifold $S\subset M$, the \emph{$\b$-conormal bundle} $\be N^*S$ is defined as the image in $\be T^* M$ of covectors in $T^*M$ that annihilate the image of $TS$ in $TM$. 

Specifically, in the setting of Lorentzian scattering spaces, the $\b$-conormal bundle of $S=\{ \rho=0, v=0 \}$ is easily seen to be generated by $dv$. Indeed, the vectors in $TS$ are annihilated by $dv$ and $d\rho$, and their image in $\be T^* M$ is respectively $dv$, $\rho (\rho^{-1} d\rho)$ with the latter vanishing above $\{\rho=0\}$.    

The bundles $\be T^*M\setminus\zero$, ${}^\b N^* S\setminus\zero$ have their `spherical' versions $\be S^*M$ and $\bconormal$, defined as the quotients
\[
\be S^*M\defeq (\be T^*M\setminus\zero) / \rr_+, \quad \bconormal\defeq ({}^\b N^* S\setminus\zero) / \rr_+.
\] 
by the fiberwise $\rr_+$-action of dilations, where $\zero$ is the
zero section.

Let now $p\in\cf(T^*M\setminus\zero)$ be the principal symbol of $P$ (in this paragraph the specific form of $P$ is irrelevant, only the fact that it belongs to $\Diff^m_\b(M)$ and that $p$ is real). By homogeneity, the Hamiltonian vector field of $p$ on $T^*M\setminus \zero$ extends to a vector field on $\be T^*M\setminus \zero$, which is tangent to the boundary. Specifically, it is given by (and could be defined by) the local expression
\[
H_p = (\p_\varsigma p)(\rho\p_\rho)-(\rho\p_\rho p)\p_\varsigma + \textstyle \sum_i \left( (\p_{\zeta_i}p) \p_{w_i}-(\p_{w_i} p) \p_{\zeta_i}\right),
\]
in $\b$-covariables $(\varsigma,\zeta)$ in which sections of $\be T^*M$ read $\varsigma (\rho^{-1} d\rho)+\sum_i \zeta_i dw_i$.

In order to keep track of the behavior of $H_p$ along the orbits of the $\rr_+$ action it is actually convenient to view $\be S^*M$ as the boundary of the so-called radial compactification $\be\overline{T}^*M$ of $\be T^*M$. Without giving the details of the construction (cf. \cite[Ch. 1.8]{melrose3}), the relevant feature here is that it comes with a function $\tilde\rho\in\cf(\be T^*M\setminus\zero)$, homogeneous of degree $-1$, that serves as a boundary defining function. Since $p$ is homogeneous of degree $m$, $\tilde\rho^m p$ can be restricted to fiber infinity and thus identified with a smooth function on $\be S^*M$. Now, the \emph{characteristic set} $\Sigma$ (of $P$) is the zero-set of the rescaled principal symbol ${\tilde \rho}^m p\in\cf(\be S^*M)$. The \emph{bicharacteristic flow} of $P$ is defined in the present setup as the flow $\Phi_t$ of the rescaled Hamilton vector field $\H_p\defeq{\tilde \rho}^{m-1} H_p$ in $\Sigma$. Accordingly, the (reparametrized) projections of the integral curves of $H_p$ by the quotient map in $\be T^*M \setminus \zero\to \be S^*M$ are called bicharacteristics\footnote{These are also called \emph{null} bicharacteristics in the literature.}.

\subsection{Non-trapping assumptions}

In contrast to standard real principal type estimates that are entirely local and are therefore not invalidated by the presence of trapping, the estimates that we use here to obtain the Fredholm property of $P$ on appropriate function spaces are \emph{global}, i.e.~depend on what happens at infinite times, therefore issues related with trapping are very likely to produce difficulties. To eliminate these we make use of the non-trapping geometrical setup considered in \cite{BVW,GHV,kerrds} (of which radially compactified Minkowski space is an example again):

\begin{assumption}\label{hyp1} We assume that $g$ is \emph{non-trapping} in the following sense.
\begin{enumerate}
\item $S=\{v=0,\rho=0\}$ is the disjoint sum of two components $S=S_+\cup S_-$ and moreover:
\item $\{v>0\} \subset \p M$ splits into disjoint components $X_\pm$ with $S_\pm =\p X_\pm$
\item all maximally extended bicharacteristics flow from $\bconormal_+$ to $\bconormal_-$ or vice-versa.
\end{enumerate}
\end{assumption}

The Lorentzian scattering space $(M,g)$ is then called an asymptotically Minkowski spacetime and the submanifold $S_+$ is the \emph{lightcone at future null infinity} and $S_-$ the \emph{lightcone at past null infinity}. 

\def\svgwidth{7cm}
%\sskip
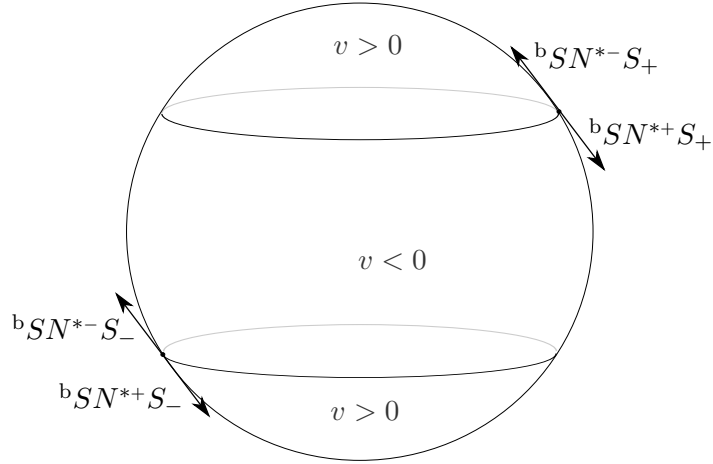
\begin{figure}[h]\label{fig4}
\begingroup%
  \makeatletter%
  \providecommand\color[2][]{%
    \errmessage{(Inkscape) Color is used for the text in Inkscape, but the package 'color.sty' is not loaded}%
    \renewcommand\color[2][]{}%
  }%
  \providecommand\transparent[1]{%
    \errmessage{(Inkscape) Transparency is used (non-zero) for the text in Inkscape, but the package 'transparent.sty' is not loaded}%
    \renewcommand\transparent[1]{}%
  }%
  \providecommand\rotatebox[2]{#2}%
  \ifx\svgwidth\undefined%
    \setlength{\unitlength}{336.55242521bp}%
    \ifx\svgscale\undefined%
      \relax%
    \else%
      \setlength{\unitlength}{\unitlength * \real{\svgscale}}%
    \fi%
  \else%
    \setlength{\unitlength}{\svgwidth}%
  \fi%
  \global\let\svgwidth\undefined%
  \global\let\svgscale\undefined%
  \makeatother%
  \begin{picture}(1,0.98787576)%
    \put(0,0){\includegraphics[width=\unitlength]{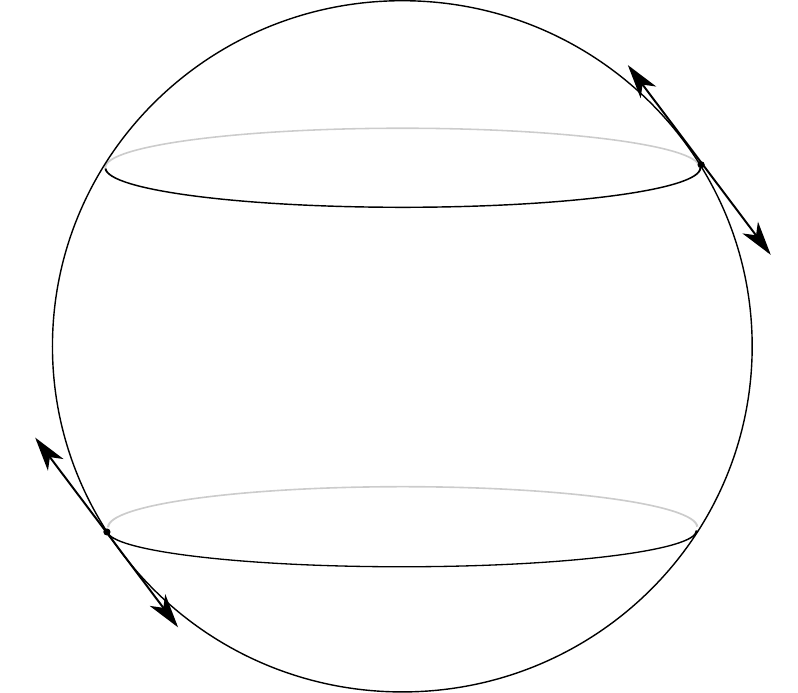}}%
    \put(0.45476981,0.7713377){\color[rgb]{0.2,0.2,0.2}\makebox(0,0)[lb]{\smash{$v>0$}}}%
    \put(0.4494318,0.07651267){\color[rgb]{0.2,0.2,0.2}\makebox(0,0)[lb]{\smash{$v>0$}}}%
    \put(0.5009072,0.36336202){\color[rgb]{0.2,0.2,0.2}\makebox(0,0)[lb]{\smash{$v<0$}}}% 
		\put(-0.1500779,0.23048785){\color[rgb]{0,0,0}\makebox(0,0)[lb]{\smash{$\sources_-$}}}%
    \put(-0.06137542,0.09999153){\color[rgb]{0,0,0}\makebox(0,0)[lb]{\smash{$\sinks_-$}}}%
    \put(0.83211029,0.74476723){\color[rgb]{0,0,0}\makebox(0,0)[lb]{\smash{$\sources_+$}}}%
    \put(0.934498,0.60094132){\color[rgb]{0,0,0}\makebox(0,0)[lb]{\smash{$\sinks_+$}}}%
  \end{picture}%
\endgroup 
\caption{An asymptotically Minkowski spacetime $M$. The radial sets are located above $S=S_+\cup S_-$ and split into sources and sinks $\sos_\pm$.}
\end{figure}

The characteristic set $\Sigma\subset {}^\b S^* M$ of $P$ splits into two connected components $\Sigma^\pm$. The radial sets (i.e., where the bicharacteristic flow degenerates) are located above  $S=S_+\cup S_-$. Each radial set $\bconormal_\pm$ splits into two components $\sos_\pm$ (corresponding to the splitting $\Sigma=\Sigma^+\cup \Sigma^-$), which act as sources (-) or sinks (+) for the bicharacteristic flow, meaning specifically that
\[
\H_p\tilde\rho=\tilde\rho\beta_0,
\] 
where $\pm\beta_0>0$ for sources, resp. sinks \cite{BVW} (see Figure 3).

We introduce the short-hand notation $\cR\defeq\textstyle\bigcup_{\pm}\sos$ for the whole radial set.

Let us remark that in this setup, a time orientation of $(M,g)$ can be fixed as follows: one specifies the future lightcone to be the one from which forward bicharacteristics (in the sense of the $\H_p$-flow) tend to $S_+$. Moreover, it was shown in \cite{semilinear} that if $\rho$ can be chosen in such way that $\rho^{-1}d\rho$ is timelike near $X_+\cup X_-$ (with respect to $\rho^2 g$) then the interior of $M$, $M^\inti$, is globally hyperbolic, we will however not use this assumption unless specified otherwise.

\subsection{$\b$-regularity and propagation of singularities}\label{ss:breg}

Recall that the algebra of $\b$-differential operators $\Diff_\b(M)$ is generated by vector fields tangent to the boundary (and the identity), thus setting
\[
H_\b^{k,0}(M) = \{ u\in  \cC^{-\infty}(M) : \ Au \in L^2_\b(M) \ \forall A\in\Diff_\b^k(M)\},
\]
for $k\in\nn$ gives a space of distributions (the \emph{$\b$-Sobolev space of order $k$}) that have usual Sobolev regularity of order $k$ in $M^\inti$, the interior of $M$, and are moreover regular of order $k$ at the boundary in the sense of conormality. In the above expression $\Diff_\b^k(M)$ can be replaced by $\b$-pseudodifferential operators of order $k$, $\Psi_\b^k(M)$ --- here we will not give the precise definition (instead we refer the reader to \cite{melrose,corners,propagation}), though formally one can simply think of those as operators of the form $A=a(\rho,w;\rho\p_\rho,\p_w)$, with $a$ a symbol in the usual sense. By analogy this allows one to define $\b$-Sobolev spaces of arbitrary order $m\in\rr$, and at the same time we introduce weighted ones:
\[
\bea
H_\b^{m,0}(M)& = \{ u\in \cC^{-\infty}(M) : \ A u \in L^2_\b(M) \ \forall A\in\Psi_\b^m(M)\},\\
H_\b^{m,l}(M) &= \rho^l H_\b^{m,0}(M),
\eea
\]
so that $m$ corresponds to usual Sobolev regularity in $M^\inti$ and conormal regularity at the boundary, whereas $l$ corresponds to decay near the boundary (and this agrees with the definition sketched in the introduction). The dual of $H_\b^{m,l}(M)$ can be identified with $H_\b^{-m,-l}(M)$ using the $L^2_\b(M)$ pairing $\bra \cdot,\cdot\ket_\ssb$. We have correspondingly spaces of distributions of arbitrarily low and arbitrarily high $\b$-Sobolev regularity (thus, the latter consists of distributions conormal to the boundary)
\[
H_\b^{-\infty,l}(M)\defeq \textstyle\bigcup_{m\in\rr} H^{m,l}_\b(M), \ \ H_\b^{\infty,l}(M)\defeq \textstyle\bigcap_{m\in\rr} H^{m,l}_\b(M),
\]
endowed with their canonical Fr\'echet topologies, one of which is the
dual of the other if $l$ is replaced by $-l$.

There is a notion of $\b$-Sobolev wave front set of a distribution $u\in H^{-\infty,l}_\b(M)$, denoted $\wf_\b^{m,l}(u)\subset\be S^* M$, which consists of the points in phase space in which $u$ is not in $H_\b^{m,l}(M)$. Concretely, the definition says that for $\alpha\in \be S^*M$, $\alpha\notin \wf_\b^{m,l}(u)$ if there exists $A\in\Psi_\b^{0,0}(M)$ elliptic at $\alpha$ and such that $Au\in H_\b^{m,l}(M)$, where ellipticity refers to invertibility of the principal symbol, cf. \cite{melrose,propagation,corners}. Note that locally in the interior of $M$, $\b$-Sobolev regularity and standard Sobolev regularity are just the same, so the $\b$-Sobolev wave front set coincides with the standard wave front set there. We refer to \cite[Sec. 2 \& 3]{corners} for a more detailed discussion.

The definitions of $\Psi_\b^{m,l}(M)$, $H_\b^{m,l}(M)$ and $\wf_\b^{m,l}(u)$ can be extended to allow for \emph{varying} Sobolev orders $m\in\cf(\be S^*M)$, cf. for instance \cite[App. A]{BVW}. This is particularly convenient for the formulation of propagation of singularities theorems near radial sets. We will use in particular the following result from \cite{kerrds}, cf. also the discussion in \cite{GHV}.

\begin{theorem}\label{thm:propag}Let $(M,g)$ be a Lorentzian scattering space. Let $P$ be the rescaled wave operator \eqref{eq:wop}, let us denote by $\cR_i$ any of the components of the radial sets, and let $u\in H_\b^{-\infty,l}(M)$. 
\begin{enumerate}
\item If $m<\12-l$ and $m$ is nonincreasing along the bicharacteristic flow in the direction approaching $\cR_i$, then
\[
\wf_\b^{m,l}(u)\cap\cR_i=\emptyset \ \ {\mbox{if}} \ \ \wf_\b^{m-1,l}(Pu)\cap\cR_i=\emptyset  
\]
and provided that $(U\setminus \cR_i)\cap \wf_\b^{m,l}(u)=\emptyset$
for some neighborhood $U\subset \Sigma\cap \be S^* M$ of $\cR_i$.
\item If $m_0>\12-l$, $m\geq m_0$ and $m$ is nonincreasing along the bicharacteristic flow in the direction going out from $\cR_i$ then
\[
\wf_\b^{m,l}(u)\cap\cR_i=\emptyset  \ \ {\mbox{if}} \ \ \big(\wf_\b^{m_0,l}(u)\cup\wf_\b^{m-1,l}(Pu)\big)\cap\cR_i=\emptyset.  
\] 
\end{enumerate}
\end{theorem}

Thus, there is a threshold value $m=\12-l$, and in the `below-threshold' case $m<\12-l$ one has a propagation of singularities statement similar to real principal type estimates, while in the `above-threshold' case one gets arbitrarily high regularity at the radial set provided $Pu$ is regular enough.

\section{Propagators}\init

\subsection{Inverses of the wave operator}\label{ss:invop} Theorem \ref{thm:propag} is deduced from (and is in fact equivalent to) a priori estimates involving $H_\b^{m,l}$ norms of $u$ and $H_\b^{m-1,l}$ norms of $Pu$ (plus a weaker norm of $u$ in $H_\b^{m',l}$, $m'<m$), microlocalized using $\b$-pseudodifferential operators accordingly with the stated direction of propagation. These estimates give a global statement if for each component $\Sigma^j$ of the characteristic set ($j\in\{+,-\}$) one takes $m$ to be above-threshold at one radial set within $\Sigma^j$ and below-threshold at the other \cite{HV,GHV}, one gets namely
\beq\label{eq:propagesti}
\|u \|_{H_\b^{m,l}(M)}\leq C \big(\|Pu\|_{H_\b^{m-1,l}(M)}+\|u\|_{H_\b^{m',l}(M)}\big).
\eeq
Thus, in other words, (\ref{eq:propagesti}) is obtained by `propagating estimates from one radial set to another', i.e., from where $m$ is above the threshold to where $m$ is below the threshold. 
Defining then
\beq
\cY^{m,l}\defeq H_\b^{m,l}(M), \quad \cX^{m,l}\defeq \left\{ u \in  H_\b^{m,l}(M) : \ Pu\in H_\b^{m-1,l}(M) \right\},
\eeq
by analogy to some elliptic problems \cite{propagation} one would like to conclude  a statement about $P$ being Fredholm as a map $\cX^{m,l}\to\cY^{m-1,l}$ (using a standard argument from functional analysis, see \cite[Proof of Thm.~26.1.7]{hoermander}). The problematic point (as explained in more detail in \cite{GHV}) is however that $H_\b^{m,l}$  is not compactly included in $H_\b^{m',l}$ (as opposed  for instance to $H_\b^{m,l}\hookrightarrow H_\b^{m',l'}$ for $m'<m$, $l'<l$) and therefore the corresponding remainder term is not negligible. Improved estimates (with better control on the decay of remainder terms) can be however derived by a careful analysis of the \emph{Mellin transformed normal operator} of $P$, defined as follows. 

Recall that any $P\in\Diff^k_\b(M)$ is locally given by 
\[
P=\sum_{i+|\alpha|\leq k} a_{i,\alpha}(\rho,w)(\rho\p_\rho)^i \p_w^\alpha.
\] 
Its Mellin transformed normal operator family is then
\[
\widehat N(P)(\sigma) \defeq \sum_{i+|\alpha|\leq k} a_{i,\alpha}(0,x)\sigma^i \p_x^\alpha.
\] 
A direct computation shows that in our specific case of interest, $\widehat N(P)(\sigma)\in\Diff^2(\pM)$ takes the form
\beq
\widehat N(P)(\sigma) = 4\left( (v+O(v^2)) \p_v^2+(\i\sigma+1+O(v))\p_v\right)+O(1)\p_y^2+O(1)\p_y + O(v)\p_v\p_y  
\eeq 
near $\{v=0\}$ modulo terms $O(\sigma^2)$, cf. \cite{BVW} for more explicit expressions. The crucial property is that $\widehat N(P)(\sigma)$ is  hyperbolic on $\{v< 0\}$ (and elliptic elsewhere, which is the easiest part) and its characteristic set splits into two connected components $\hat\Sigma^{\pm}$ with bicharacteristics starting and ending at radial sets. Fredholm estimates combined with a semiclassical analysis with small parameter $|\sigma|^{-1}$ are then used in \cite{GHV} to prove that $\widehat N(P)(\sigma)^{-1}$ exists as a meromorphic family and the structure of its poles determines the Fredholm (or invertibility) property of $P:\cX^{m,l}\to\cY^{m-1,l}$. In particular the following assumption is made use of.

\begin{assumption}\label{hyp2}The weight $l$ is assumed to satisfy $l\neq -\Im\,\sigma_i$ for any resonance\footnote{This is synonym for $\sigma_i$ being a pole of the meromorphic family $\widehat{N}(P)(\sigma)^{-1}$.} $\sigma_i\in \cc$ of the Mellin transformed normal operator family $\widehat{N}(P)(\sigma)$ of $P$.
\end{assumption}

Concerning the possible choices of the order defining function $m$, different choices of directions along which $m$ is increasing give different (generalized, see below) inverses of $P$. Specifically, for each of the two sinks $\sinks_\pm$, we can choose whether estimates are propagated \emph{from} it or \emph{to} it. Following the convention in \cite{positive}, let us label this choice by a set of indices $I\subset\{ + , - \}$ indicating the sinks \emph{from} which we propagate, i.e.~where high regularity is imposed (and thus also the components of the characteristic set $\Sigma=\Sigma^+\cup \Sigma^-$ along which $m$ is increasing). Then the complement $I^{\rm c}$ indicates the sinks \emph{to} which we propagate. 
We denote correspondingly $\cR_I^-$ the components of the radial set from which the estimates are propagated, and $\cR_I^+$ the remaining others. Note that by definition $\cR_{I^{\rm c}}^\mp =  \cR_{I}^\pm$. 

With these definitions at hand, the main result of \cite{GHV} states that $P:\cX^{m,l}\to\cY^{m-1,l}$ is Fredholm for any $m$ such that 
\beq\label{eq:whereregularity}
\pm m> 1/2 -l \mbox{ \ near \ } \cR_I^\mp,
\eeq
with $m$ monotone along the bicharacteristic flow as long as $l$
satisfies Hypothesis~\ref{hyp2}.
Moreover, it is shown that $P:\cX^{m,l}\to\cY^{m-1,l}$ is invertible
if $|l|$ is small and $(M,g)$ is a perturbation of the radial compactification of Minkowski space in the sense of Lorentzian scattering metrics $\cf(M;\scmetrics)$, within the closed subset of Lorentzian scattering spaces (cf. Definition \ref{def:sc}). 

We will use the shorthand notation $\cX_I$, $\cY_I$ for the spaces $\cX^{m,l},\cY^{m-1,l}$ with any choice of orders and weights $m,l$ as in (\ref{eq:whereregularity}). We will also write occasionally $P_I$ for $P$ understood as an operator $\cX_I\to\cY_I$. 

A consequence of the Fredholm property is that one can define a \emph{generalized inverse} of $P_I:\cX_I\to\cY_I$ as follows. First, one makes a choice of complementary spaces $\cW_I$, $\cZ_I$, to respectively $\Ker P_I$, $\Ran P_I$ in $\cX_I$, $\cY_I$, with $\cW_I$ of finite codimension and $\cZ_I$ of finite dimension. We define $P_I^{-1}$ to be the unique extension of the inverse of $P:\cW_I\to \Ran P_I$ to $\cY_I\to\cX_I$ such that 
\[
\Ker P_I^{-1}=\cZ_I, \quad \Ran P_I^{-1}=\cW_I.
\] 
In what follows we will choose a complementary space $\cZ_I$ consisting of $\cfd(M)$ functions, which is always possible since $\Ran P_I$ is of finite codimension and $\cfd(M)$ is dense in $\cY_I$. The property $\cZ_I\subset \cfd(M)$ then ensures  that $P P^{-1}_I$ equals $\one$ on $\cY_I$ modulo smoothing terms. To make sure that $P^{-1}_I$ is also a left parametrix\footnote{By say, left parametrix, we mean that $P_I^{-1}P$ equals $\one$ modulo terms that have smooth kernel in $M^\inti$.}, one needs the following additional property.

\begin{assumption}\label{hyp3}Assume that $\Ker P_I\subset H_\b^{\infty,l}(M)$.
\end{assumption}  

We will refer to Hypothesis \ref{hyp3} simply as \emph{smoothness of the kernel}, we will actually see in Proposition \ref{prop:faf} that it is in fact automatically satisfied in the Feynman and anti-Feynman case (the argument we use therein does however not apply to the advanced and retarded case).

The (generalized) inverses $P_I^{-1}$ corresponding to the four possible choices of $I$ are named as follows:
\begin{enumerate}
\item $I=\emptyset$ (i.e., $\cR_I^-= \sources$) --- Feynman propagator, 
\item $I=\{+,-\}$ (i.e., $\cR_I^-= \sinks$) --- anti-Feynman propagator,
\item $I=\{-\}$ (i.e., $\cR_I^-= \bconormal_-$) --- retarded (or forward) propagator,
\item $I=\{+\}$ (i.e., $\cR_I^-= \bconormal_+$) --- advanced (or backward) propagator.
\end{enumerate}
The terminology for $I=\{-\}$, resp. $I=\{+\}$ is motivated by the fact that due to its mapping properties, the corresponding inverse $P_I^{-1}$ solves the forward, resp. backward problem in the interior $M^\inti$ of $M$, and thus equals the advanced, resp. retarded propagator defined in the usual way as in the introduction (modulo smoothing terms if $P_I^{-1}$ is just a parametrix). The name \emph{Feynman propagator} for $P_{\emptyset}^{-1}$ can be justified by relating it to a Feynman parametrix in the sense of Duistermaat and H\"ormander \cite{DH}, as pointed out in \cite{GHV,positive} (and analogously for the anti-Feynman one). Here we make this precise by proving that the Schwartz kernel of $P_{\emptyset}^{-1}$ (considered as a distribution on $M^\inti\times M^\inti$) has wave front set of precisely the same form as the Feynman parametrix' of Duistermaat and H\"ormander, and therefore the two operators coincide modulo smoothing terms (at least provided $(M^\inti,g)$ is globally hyperbolic so that the assumptions in \cite{DH} are satisfied).

Such statement is closely related to the propagation of singularities along the bicharacteristic flow $\Phi_t$. In the present setting it can be formulated as follows. If $I\subset\{+,-\}$, $m,l$ are chosen consistently with $I$, $m_0>\12-l$ is a fixed constant  and $u\in\cX^{m,l}$ then  
\beq\label{eq:bpropagation}
\bea
\big(\wf_\b^{m_0,l}(u)\cap \Sigma \big)\setminus\cR_I^+ \subset \wf^{m_0-1,l}_\b(Pu)&\cup\textstyle\bigcup_{j\in I}\big(\cup_{t\geq 0}\Phi_t(\wf_\b^{m_0-1,l}(Pu)\cap\Sigma^j)\big)\\
&\cup\textstyle\bigcup_{j\in I^{\rm c}}\big(\cup_{t\leq 0}\Phi_t(\wf_\b^{m_0-1,l}(Pu)\cap\Sigma^j)\big)
\eea
\eeq
provided that $\wf^{m_0-1,l}_\b(Pu)\cap\cR_I^-=\emptyset$. The latter condition is trivially satisfied if for instance $\supp\,Pu\subset\subset M^\inti$, then in the interior of $M$ (\ref{eq:bpropagation}) reduces to 
\beq\label{eq:propagation}
\bea
\wf^{m_0}(u)\cap\Sigma\subset \wf^{m_0-1}(Pu)&\cup\textstyle\bigcup_{j\in I}\big(\cup_{t\geq 0}\Phi_t(\wf^{m_0-1}(Pu)\cap\Sigma^j)\big)\\
&\cup\textstyle\bigcup_{j\in I^{\rm c}}\big(\cup_{t\leq 0}\Phi_t(\wf^{m_0-1}(Pu)\cap\Sigma^j)\big),
\eea
\eeq
in terms of the standard Sobolev wave front set $\wf^{m_0}(u)\subset S^* M^\inti$ (since the restriction of $\wf^{m_0-1,l}_\b$ to $M^\inti$ is precisely $\wf^{m_0}$). Therefore, disregarding singularities lying on $\diag_{T^*M^\inti}$ (the diagonal in $T^*M^\inti\times T^*M^\inti$), one expects that the primed wave front set of the Schwartz kernel of $P_I^{-1}$, denoted $\wf'(P_I^{-1})$, is contained in
\[
\kC_I\defeq \textstyle\bigcup_{j\in I}\big(\cup_{t\geq 0}\Phi_t(\diag_{T^*M^\inti})\cap\pi^{-1}\Sigma^j\big)\cup\textstyle\bigcup_{j\in I^{\rm c}}\big(\cup_{t\leq 0}\Phi_t(\diag_{T^*M^\inti})\cap\pi^{-1}\Sigma^j\big),
\]
where $\Phi_t$ operates on the left factor and $\pi:\Sigma\times\Sigma\to\Sigma$ is the projection to the left factor\footnote{Here one can equivalently take the projection to the right factor.}. 

In other words $\kC_I$ consists of pairs of points $((y,\eta),(x,\xi))$ such that  $(y,\eta)$, $(x,\xi)\in\Sigma$ are connected by a bicharacteristic and such that on the component $\Sigma^j$, $(y,\eta)$ comes after $(x,\xi)$ respective to the Hamilton flow if $j\in I$, and $(x,\xi)$ comes after $(y,\eta)$ otherwise. 

\begin{proposition} \label{prop:wfPI} Assume Hypotheses \ref{hyp1}, \ref{hyp2} and \ref{hyp3} and global hyperbolicity of $(M^\inti,g)$. Then:
\begin{enumerate}
\item\label{prop:wfPIit1} $\wf'(P_I^{-1})=(\diag_{T^*M^\inti})\cup\kC_I$ for $I\subset\{+,-\}$;
\item\label{prop:wfPIit2} $\wf'(P_\emptyset^{-1}-P_{\papm}^{-1})= \cup_{t\in\rr}\Phi_t(\diag_{T^*M^\inti})\cap\pi^{-1}\Sigma^\pm$.
\end{enumerate}
\end{proposition}
\proof In the case of retarded/advanced propagators, statement (\ref{prop:wfPIit1}) follows from \cite{DH}, so we only have to show (\ref{prop:wfPIit1}) in the (anti-)Feynman case. We start by proving (\ref{prop:wfPIit2}).

Let $\delta_x$ be the Dirac delta distribution supported at some point
$x\in M^\inti$. For any $I$ we can choose the order defining function
$m$ in $\cX_I=\cX^{m,l}$ in such way that $\delta_x\in \cY_I$. Even
more, we can arrange that $\delta_x$ is at the same time in
$\cY_\emptyset$ and in $\cY_{+}$. Then $P_I^{-1}\delta_x\in\cX_I$ for $I=\emptyset$ and $I=\{+\}$. Consequently, the distribution $(P_\emptyset^{-1}-P_{\pap}^{-1})\delta_x$ has above-threshold regularity microlocally in $\Sigma^-$ near $S_+$. Since it also solves the wave equation (modulo smooth terms), this implies by propagation of singularities
\beq
\wf( (P_\emptyset^{-1}-P_{\pap}^{-1})\delta_x)\subset \Sigma^+.
\eeq
In fact, by propagation of singularities estimates
(which are \emph{uniform} estimates), this holds in the sense of the uniform wave front set for the
family
\beq
\{(P_\emptyset^{-1}-P_{\pap}^{-1})\delta_x:\ x\in K\},
\eeq
$K$
compact in $M^\inti$. By this we mean that for $A\in\Psi^{0}(M)$ of
compactly supported Schwartz kernel and with $\WF'(A)\cap\Sigma^+=\emptyset$, the set
\beq\label{eq:tempwfdif}
\{A(P_\emptyset^{-1}-P_{\pap}^{-1})\delta_x:\ x\in K\}\ \text{is
  bounded in}\ \cf. 
\eeq
On the level of the Schwartz kernel
$(P_\emptyset^{-1}-P_{\pap}^{-1})(y,x)=((P_\emptyset^{-1}-P_{\pap}^{-1})\delta_x)(y)$,
which holds in a distributional sense, (\ref{eq:tempwfdif}) yields
\beq\label{eq:tempwfdif2}
\wf' (P_\emptyset^{-1}-P_{\pap}^{-1}) \subset (\Sigma^+\cup\, \zero) \times T^*M^\inti,
\eeq
as can be seen e.g.\ by using the explicit Fourier transform
characterization of the wave front set, using appropriate
pseudodifferential operators in \eqref{eq:tempwfdif}.
We now use \cite[Thm.~1]{positive}, which states (for parametrices,
which our inverses are)  that $\i^{-1}(P_\emptyset^{-1}-P_{\papm}^{-1})$ differs from a positive operator by a smooth term. Disregarding this smooth error, one can write a Cauchy-Schwarz inequality for $|\bra f,(P_\emptyset^{-1}-P_{\pap}^{-1})g\ket_\ssb|$ in terms of $|\bra f,(P_\emptyset^{-1}-P_{\pap}^{-1})f\ket_\ssb|$, $|\bra g,(P_\emptyset^{-1}-P_{\pap}^{-1} )g\ket_\ssb|$ for any test functions $f,g$. This allows us to get estimates for the wave front set in $\zero\times (T^*M^\inti \setminus \zero)$ from estimates in $(T^*M^\inti \setminus \zero)\times (T^*M^\inti \setminus \zero)$, and also to get a symmetrized form of the wave front set\footnote{It is worth mentioning that this sort of argument was already used for instance in \cite{radzikowski2,FW,SV}.}, in particular (\ref{eq:tempwfdif2}) gives
\beq\label{eq:disjointwf1}
\wf' (P_\emptyset^{-1}-P_{\pap}^{-1}) \subset (\Sigma^+\cup\,\zero) \times (\Sigma^+\cup\,\zero).
\eeq
The analogous argument gives correspondingly
\beq\label{eq:disjointwf2}
\wf' (P_\emptyset^{-1}-P_{\pam}^{-1}) \subset (\Sigma^-\cup\,\zero) \times (\Sigma^-\cup\,\zero).
\eeq
Observe that the two wave front sets (\ref{eq:disjointwf1}), (\ref{eq:disjointwf2}) are disjoint. In view of the identity
\[
(P_\emptyset^{-1}-P_{\pap}^{-1})-(P_\emptyset^{-1}-P_{\pam}^{-1})=P_{\pam}^{-1}-P_{\pap}^{-1}
\]
this implies that $\wf' (P_\emptyset^{-1}-P_{\papm}^{-1})$ equals $(\Sigma^\pm\cup\,\zero) \times( \Sigma^\pm\cup\,\zero)\cap\wf'(P_{\pam}^{-1}-P_{\pap}^{-1})$. On the other hand, using the exact form of $\wf'(P_{\papm}^{-1})\setminus \diag_{T^*M^\inti}=\kC_{{\papm}}$ one obtains $\wf'(P_{\pam}^{-1}-P_{\pap}^{-1})=\kC_{\pap}\cup\kC_{\pam}$, thus
\beq
\label{eq:disjointwf3}
\bea
\wf' (P_\emptyset^{-1}-P_{\papm}^{-1}) &= (\Sigma^\pm \times \Sigma^\pm)\cap (\kC_{{\pap}}\cup\kC_{{\pam}})\\
&= \cup_{t\in\rr}\Phi_t(\diag_{T^*M^\inti})\cap\pi^{-1}\Sigma^\pm.
\eea
\eeq
The exact form of $\wf' (P_\emptyset^{-1})$ is concluded from (\ref{eq:disjointwf3}) and $\wf'(P_{\papm}^{-1})=\diag_{T^*M^\inti}\cup\kC_{{\papm}}$ by means of the two identities $P_\emptyset^{-1}=(P_\emptyset^{-1}-P_{\papm}^{-1})+ P_{\papm}^{-1}$.\qeds

Concerning the $\b$-wave front set, it would require more work to make precise statements about the Schwartz kernel of $P_I^{-1}$ (in the sense of manifolds with boundaries), we still have however at our disposal information on the $\b$-wave front set of $P_I^{-1}f$ given the $\b$-wave front set of $f$. For our purposes it is sufficient to observe that the generalized inverse $P_I^{-1}$ (defined using some $m,l$ chosen consistently with $I$) adds singularities only at the radial set. Specifically, by propagation of singularities (\ref{eq:bpropagation})
\beq\label{eq:htp}
\wf^{m_0,l}_\b(P_I^{-1}f)\subset \cR_I^+, \quad f\in\Ran P_I\cap H_\b^{m_0-1,l}(M)
\eeq
for $m_0>\12-l$, so in particular if $f\in H_\b^{\infty,l}(M)$ then $\wf_\b^{\infty,l}(P^{-1}_I f)\subset \cR^+_I$. 

\section{Symplectic spaces of smooth solutions}\init

\subsection{Solutions smooth away from $\cR$}\label{ss:ssaf}

A particularly useful way to construct solutions of $Pu=0$ is to take $u=(P_I^{-1}-P_{I^\c}^{-1})f$ for $f\in \Ran P_I\cap\Ran P_{I^\c}$, where the operators are considered on spaces with orders $m,l$, resp. $m^{\rm c},l$ corresponding to $I$, resp. $I^{\rm c}$ (i.e., $m,l$ are such that \eqref{eq:whereregularity} holds and similarly for $m^{\rm c},l$). Then for $m_0>\12-l$ and $f\in\Ran P_I\cap \Ran P_{I^\c}\cap H_\b^{m_0-1,l}(M)$, by \eqref{eq:htp} we have $\wf^{m_0,l}_\b(u)\subset\cR$. In particular, this applies if $m_0<\min_{\be S^*M}\max\{m,m^{\rm c}\}$ and $f\in\Ran P_I\cap \Ran P_{I^\c}$, provided that $\min_{\be S^*M}\max\{m,m^{\rm c}\}>\12-l$.

In what follows it will be convenient to take $m$ to be constant, $m_*>\12-l$, outside a compact subset of a small neighborhood $U_+$ of the outgoing radial set $\cR_I^+$, and similarly for $m^{\rm c}$, with $U_+\cap U_-=\emptyset$ for the respective neighborhoods. Then $\min_{\be S^*M}\max\{m,m^{\rm c}\}=m_*$, and the conclusion of the previous paragraph applies even with $m_0=m_*$.

We will see that the so-obtained space of solutions can be equivalently defined as
\beq\label{eq:defsol}
\SolI(P)\defeq \big\{ u\in \cW_I+\cW_{I^\c}: \ Pu=0, \ \wfb^{m_0,l}(u)\subset \cR\big\},
\eeq
where we recall that $\cW_I$ is a complement of $\Ker P_I$. Note that by definition $\SolI(P)={\Sol}_{I^\c}(P)$. If $P_I$ is invertible then the condition $u\in \cW_I+\cW_{I^\c}$ in (\ref{eq:defsol}) reduces to  $u\in \cX_I+\cX_{I^\c}$. In the case when $P_I$ is merely a Fredholm operator, the main reason to use $\cW_I$ in the definition is the validity of the following lemma.

\begin{lemma}\label{lem:microvanish}Assume Hypothesis \ref{hyp3}. If $u\in\SolI(P)$ is microlocally in $H^{m',l}_\b(M)$ near $\cR_I^-$ for $m'>\12-l$ then $u=0$.  
\end{lemma}
\proof By assumption $u\in\cX_I$ and $Pu=0$, hence $u\in\Ker P_I$ by definition of $P_I:\cX_I\to\cY_I$. Using Hypothesis \ref{hyp3} this implies $u\in\cX_{I^\c}$, and repeating the previous argument one gets $u\in\Ker P_{I^\c}$. This contradicts that $u\in \cW_I +\cW_{I^\c}$ unless $u=0$.\qeds

We will use Lemma \ref{lem:microvanish} repeatedly. For instance, let $Q_I\in \Psi_\b^{0,0}$ be microlocally the identity near $\cR_I^-$ and microlocally vanishing near the remaining components $\cR_I^+$ of the radial set. For any $u\in\SolI(P)$,
\[
u= Q_I u + (\one - Q_I)u= Q_I u + P_I^{-1}P(\one-Q_I)u + (\one-P_I^{-1}P)(\one-Q_I)u.
\]
Since $(\one-Q_I)u$ belongs to $\cX_I$, the term $(\one-P_I^{-1}P)(\one-Q_I)u$ is in the null space of $P$, so in fact we have
\[
u= Q_I u - P_I^{-1} P Q_Iu \mod \Ker P_I,
\]
and hence
\beq\label{eq:tempQI}
u= Q_I u - P_I^{-1} P Q_Iu \mod \cX_I\cap\cX_{I^\c}
\eeq
by Hypothesis \ref{hyp3}. Rewriting this in the form $u=Q_I u- P_I^{-1} [ P, Q_I]u$ (modulo irrelevant terms) we conclude that $-P_I^{-1}[P,Q_I]u$ agrees with $u$ microlocally at $\cR_I^+$, and so does $P_{I^{\rm c}}^{-1}[P,Q_I]u - P_I^{-1}[P,Q_I]u$. The latter is in $\SolI(P)$ (because $[P,Q_I]u=PQ_I u=-P(\one-Q_I)u\in\Ran P_I\cap \Ran P_{I^\c}$), therefore by Lemma \ref{lem:microvanish} (using $\cR_I^+=\cR_{I^\c}^-)$ we obtain
\beq\label{eq:GInverse}
 (P_{I^{\rm c}}^{-1}- P_I^{-1})[P,Q_I]=\one \ \mbox{ \ on \ } \SolI(P).
\eeq
For the sake of compactness of notation we define $G_I\defeq P_I^{-1} - P_{I^{\rm c}}^{-1}$, in terms of which the above identity reads
\beq\label{eq:invGI}
-G_I [P,Q_I] = \one \ \ \mbox{ \ on \ } \SolI(P).
\eeq

\begin{proposition}\label{prop:exact}Assume Hypotheses \ref{hyp1}, \ref{hyp2} and \ref{hyp3}. Then the map $G_I$ induces a bijection
\beq\label{eq:arrows0}
\frac{\Ran P_I\cap \Ran P_{I^\c}}{P(\cX_I\cap \cX_{I^\c})}\xrightarrow{\makebox[2em]{$[G_I]$}}\SolI(P)
\eeq
\end{proposition}
\proof
We first need to check that $G_I$ induces a well-defined map on the quotient, i.e.~$G_I (\Ran P_I\cap \Ran P_{I^\c}) \subset \SolI(P)$ (which we already know) and $G_I P(\cX_I \cap \cX_{I^\c})=0$. The latter follows from the identity
 \beq\label{eq:eqrang}
P(\cW_I\cap \cW_{I^\c})=P(\cX_I\cap \cX_{I^\c}),
\eeq
(this is true because the spaces $\cW_I\cap \cW_{I^\c}$ and $\cX_I\cap \cX_{I^\c}$ differ only by elements of $\Ker P_I$ and $\Ker P_{I^\c}$) and the fact that $P_I^{-1} P=\one$ on $\cW_I$.

Surjectivity of $[G_I]$ means 
\[ 
G_I (\Ran P_I\cap \Ran P_{I^\c}) \supset \SolI(P).
\]
but this follows readily from (\ref{eq:invGI}), taking into account that $[P,Q_I]$ is smoothing near the radial set. Injectivity of $[G_I]$ means that the kernel of $G_I$ acting on $\Ran P_I\cap \Ran P_{I^\c}$ equals $P (\cW_I\cap \cW_{I^\c})$. Indeed if $u\in \Ran P_I\cap \Ran P_{I^\c}$ and $G_I u =0$ then setting $w= P_I^{-1} u$ we have $u=Pw$, with $w\in\cW_I$. On the other hand $w=P_{I^\c}^{-1} u$ hence it is also in $\cW_{I^\c}$.\qeds

To simplify the discussion further it is convenient to eliminate the dependence of the spaces $\cX_I$, $\SolI(P)$ and $\Ran P_I$ on the specific choice of Sobolev orders $m,m^\c$ by taking the intersection over all possible orders (satisfying the extra assumption stated after Lemma \ref{lem:microvanish}). With this redefinition, $\wf^{\infty,l}_\b(u)\subset\cR$ for all $u\in\SolI(P)$. Furthermore, Proposition \ref{prop:exact} remains valid and in the special case when $P_I$ and $P_{I^\c}$ are invertible (this is true for instance when $(M^\inti,g)$ is globally hyperbolic) one gets instead of (\ref{eq:arrows0}) the more handy  statement that there is a bijection
\beq\label{eq:arrows2}
\frac{H_{\b}^{\infty,l}(M)}{P H_{\b}^{\infty,l}(M)}\xrightarrow{\makebox[2em]{$[G_I]$}}\SolI(P).
\eeq
The case $I=\{-\}$ in (\ref{eq:arrows2}) is the analogue of the well-known characterization of smooth space-compact\footnote{By space-compactness one means that the restriction to a Cauchy surface has compact support.} solutions of the wave equation on globally hyperbolic spacetimes as the range of the difference of the advanced and retarded propagator acting on test functions (see e.g.~\cite[Thm.~3.4.7]{BGP}). 

In what follows we will consider pairings between elements of spaces such as $\cX_I$, $\cX_{I^\c}$ and for that purpose we fix $l=0$ for the weight respective to decay. As shown in \cite{positive}, the formal adjoint of $P_I^{-1}$ is $P_{I^{\rm c}}^{-1}$, possibly up to some obstructions caused by the lack of invertibility of $P:\cX_I\to\cY_I$ in the case when it is merely Fredholm. In addition to that, there is a positivity statement in the Feynman case, more precisely:

\begin{theorem}[\cite{positive}]\label{lem:skewadjoint} Assume Hypotheses \ref{hyp1}, \ref{hyp2}. As a sesquilinear form on $\Ran P_I\cap\Ran P_{I^\c}$, $G_I=P_I^{-1}-P_{I^{\rm c}}^{-1}$ is formally skew-adjoint. Moreover if $I=\emptyset$ then $\i^{-1}\bra \cdot, G_I\cdot\ket_\ssb$ is positive on $\Ran P_I\cap\Ran P_{I^\c}$. 
\end{theorem}

The relevance of Proposition \ref{prop:exact} and Theorem \ref{lem:skewadjoint} in QFT stems from the conclusion that $\bra\overline{\cdot}, G_I\cdot\ket_\ssb$ induces a well-defined symplectic form (in particular non-degenerate, thanks to the injectivity statement of Proposition \ref{prop:exact}) on the quotient space 
\[
\kV_I\defeq\Ran P_I\cap \Ran P_{I^\c}/P(\cW_I\cap \cW_{I^\c}),
\]
which can be then transported to $\SolI(P)$ using the isomorphism in (\ref{eq:arrows0}). In the case $I=\{-\}$ the resulting structure is interpreted as the canonical symplectic space of the classical field theory and is the first ingredient in the construction of non-interacting quantum fields. The next step is to specify a pair of \emph{two-point functions} on $\kV_I$, defined in the very broad context below. 

\begin{definition}\label{def:2pf}Let $\kV$ be a complex vector space equipped with a (complex) symplectic form, and let $G$ be the associated anti-hermitian form. One calls a pair of sesquilinear forms $\Lambda^\pm$ on $\kV$ \emph{bosonic (resp. fermionic) two-point functions} if $\Lambda^+-\Lambda^-=\i^{-1} G$ (resp. $\Lambda^++\Lambda^-=\i^{-1} G$) and $\Lambda^\pm\geq 0$ on $\kV$. 
\end{definition}

Note that in the fermionic case one needs to have necessarily $\i^{-1} G\geq 0$. Once $\Lambda^\pm$ are given, the standard apparatus of quasi-free states and algebraic QFT can be used to construct quantum fields, see Appendix \ref{secapp1} or \cite{derger,haag,KM}; here we will rather focus on the two-point functions themselves. 

The main case of interest is the symplectic space $\kV_I$ with $I=\{-\}$ or equivalently $I=\{+\}$ and bosonic two-point functions $\Lambda_I^\pm$ on it. The physical reason one is interested in the case $I=\{\pm\}$ is that the Schwartz kernel $G_I(x,y)=\pm(P^{-1}_-(x,y)-P^{-1}_+(x,y))$ vanishes for space-like related $x,y\in M^\inti$ and in consequence the relation $\Lambda^+_I-\Lambda^-_I=\i^{-1} G_I$ translates to the property that quantum fields commute in causally disjoint regions.

In contrast, two-point functions on $\kV_I$ in the cases $I=\emptyset$, $I=\{+,-\}$ have not been considered before to the best of our knowledge. We argue that since $\i^{-1} G_I$ is positive in the Feynman case, from the purely mathematical point of view it is natural to consider then \emph{fermionic} two-point functions $\Lambda^\pm_I$ (despite their lack of obvious physical interpretation in the present context). In later sections we will indeed construct fermionic two-point functions (in particular satisfying $\Lambda^+_I+\Lambda_I^-=\i^{-1} G_I$ for $I=\emptyset$) for which however the quantity $\Lambda^+_I-\Lambda^-_I$ equals $\i(P^{-1}_+(x,y)-P^{-1}_-(x,y))$ merely modulo terms smooth in $M^\inti$ (in the special case of Minkowski space one finds $\i(P^{-1}_+(x,y)-P^{-1}_-(x,y))$ exactly though, i.e.~the smooth remainders are absent).

In our setup, rather than with abstract sesquilinear forms on $\kV_I$ it is much more convenient to work with operators $\Lambda^\pm_I$ that map continuously, say, $H^{m',0}_\b\to H^{-m',0}_\b$ for large $m'$, these then define a pair of (hermitian) sesquilinear forms $\bra\cdot,\Lambda^\pm_I\cdot\ket_\ssb$ on $\kV_I$ if $\Lambda^\pm_I$ is formally self-adjoint on $\Ran P_I\cap \Ran P_{I^\c}$ with respect to $\bra\cdot,\cdot\ket_\ssb$ and 
\beq\label{eq:beeq}
\bra \phi, \Lambda^\pm_I P \psi\ket_\ssb = 0 \quad \forall\,\phi\in\Ran P_I\cap \Ran P_{I^\c}, \ \psi\in\cW_I\cap \cW_{I^\c}.
\eeq
The sesquilinear forms $\bra\cdot,\Lambda^\pm_I \cdot\ket_\ssb$ are two-point functions on $\kV_I$ if they satisfy
\beq\label{eq:beeq2}
(-1)^{I(+)}\Lambda^+_I + (-1)^{I(-)}\Lambda^-_I=\i G_I, \quad \bra\cdot,\Lambda^\pm_I \cdot\ket_\ssb \geq 0 \ \mbox{ \ on \ } \Ran P_I\cap \Ran P_{I^\c}
\eeq 
where we employed the notation
\[
(-1)^{I(\pm)}\defeq\begin{cases} 1 \mbox{\ if \ } \pm\in I, \\
-1 \mbox{\ otherwise},\end{cases}
\]
so that one gets bosonic two-point functions in the retarded/advanced case and fermionic ones in the Feynman/anti-Feynman case.

In QFT on curved spacetime one is primarily concerned about the subclass of Hadamard two-point functions, which in the present setup can be defined as follows (conforming to the discussion above, two-point functions will be considered to be operators instead of sesquilinear forms).  

\begin{definition}\label{def:hadamard}We say that $\Lambda^\pm_I: H^{m',0}_\b(M)\to H^{-m',0}_\b(M)$ are \emph{Hadamard two-point functions} for $P$ if they satisfy (\ref{eq:beeq}), (\ref{eq:beeq2}) and if moreover 
\beq\label{eq:LXpm0}
\wf'(\Lambda^{\pm}_I)=\textstyle\bigcup_{t\in\rr}\Phi_t(\diag_{T^*M^\inti})\cap\pi^{-1}\Sigma^\pm
\eeq
over $M^\inti\times M^\inti$.
\end{definition}

\begin{remark}\label{rem:hadamard}In practice (in the setup of the assumptions from Proposition \ref{prop:wfPI}), if we are given a pair of operators $\Lambda^\pm_I$ satisfying (\ref{eq:beeq}) and (\ref{eq:beeq2}), then to ensure the Hadamard condition (\ref{eq:LXpm0}) it is sufficient to have $\wf'(\Lambda^\pm_I)\subset (\Sigma^\pm \cup\, \zero)\times T^*M^\inti$, as can be shown by the same arguments as in the proof of Proposition \ref{prop:wfPI}.
\end{remark}

The wave front set condition (\ref{eq:LXpm0}) will be called the \emph{Hadamard condition}, in agreement with the terminology used on globally hyperbolic spacetimes, cf. \cite{radzikowski,sanders,SV} for the various equivalent formulations. From the point of view of applications in QFT (renormalization in particular, see \cite{HW,KM,dang} and references therein), one of the key properties of Hadamard two-point functions is that any two differ by an operator whose kernel is smooth in $M^\inti\times M^\inti$. This statement (known on globally hyperbolic spacetimes as Radzikowski's theorem \cite{radzikowski}) is easily shown using the identity
\[
(-1)^{I(+)}(\Lambda^+_I-\tilde\Lambda^+_I)+(-1)^{I(-)}(\Lambda^-_I-\tilde\Lambda^-_I)=\i G_I -\i G_I=0
\] 
for any two pairs of Hadamard two-point functions $\Lambda^\pm_I$, $\tilde\Lambda^\pm_I$. Indeed, the terms in parentheses have disjoint primed wave front sets in the interior of $M$, so in fact $\Lambda^+_I-\tilde\Lambda^+_I$ and $\Lambda^-_I-\tilde\Lambda^-_I$ have smooth kernel in $M^\inti$.

\subsection{Time-slice property}

Let us consider again the identity 
\beq\label{eq:invGI2}
G_I [P,Q_I] = \one \ \ \mbox{ \ on \ } \SolI(P),
\eeq
which we proved to be true for any pseudo-differential operator $Q_I\in\Psi_\b^{0,0}(M)$ that is microlocally the identity near $\cR_I^-$ and microlocally vanishes near $\cR_I^+$. In the cases $I=\{+\}$, $I=\{-\}$, $Q_I$ can actually be chosen to be a multiplication operator and one can ensure that $[P,Q_I]$ vanishes in a neighborhood of $S=S_+\cup S_-$, so this way one can characterize $\SolI(P)$ as the range of $G_I$ acting on functions supported away from $S$.    

\begin{proposition}\label{prop:timeslice}Assume Hypotheses \ref{hyp1}, \ref{hyp2}. Suppose $I=\{+\}$ or $I=\{-\}$ and let $Q_I\in \cf(M)$ be equal $0$ near $S_-$ and $1$ near $S_+$. Then for any $u\in \Ran P_I\cap \Ran P_{I^\c}$ there exists $\tilde u \in\Ran P_I\cap \Ran P_{I^\c}$ s.t. $[u]=[\tilde u]$ in $\Ran P_I\cap \Ran P_{I^\c}/P(\cX_I\cap \cX_{I^\c})$ and
\beq
\supp (\tilde u) \subset \supp(Q_I)\cap \supp (\one-Q_I).
\eeq
\end{proposition}
\proof It suffices to set $\tilde u = [P,Q_I] G_I u$, then it is clear that this has the requested support properties. Furthermore
$G_I(\tilde u - u)=0$ by \eqref{eq:invGI2}, thus $\tilde u - u \in P (\cX_I\cap \cX_{I^\c})$ by the injectivity statement of Proposition \ref{prop:exact}.\qeds

In the case when $M^\inti$ is globally hyperbolic this statement implies that for any $[u]\in H^{\infty,0}_\b(M)/P H^{\infty,0}_\b(M)$ one can find a representative $\tilde u$ supported in an arbitrary neighborhood of a Cauchy surface. This fact (with $\cf_\c(M^\inti)$ in place of $H^{\infty,0}_\b(M)$) is known as the \emph{time-slice property}, a particularly useful consequence is that this allows one to construct two-point functions by specifying their restriction to a small neighborhood of a Cauchy surface.  

\section{Parametrization of solutions on the lightcone at infinity}\label{section5}\init

\subsection{Mellin transform}\label{ss:mellin} In what follows we collect some elementary facts on the Mellin transform that will be needed later on.

 Recall that for $u\in\cC_\c^\infty((0,\infty))$ the Mellin transform is defined by the integral
\[
(\cM_\rho u)(\sigma)\defeq \int_0^\infty \rho^{-\i\sigma-1}u(\rho) d\rho. 
\]
It extends to a unitary operator $\rho^l L^2_\b(\rr_+)\to L^2(\{\Im \sigma=-l\})$ whose inverse can be expressed using the integral formula
\beq\label{eq:imt}
u(\rho)=(2\pi)^{-1}\int_{\{\Im\sigma=-l\}}\rho^{\i\sigma} (\cM_\rho u)(\sigma) d\sigma,
\eeq
and it intertwines the generator of dilations $\rho D_\rho$ with multiplication by $\sigma$, i.e.~$\rho D_\rho= \cM_\rho^{-1} \sigma \cM_\rho$. 
 
Let us denote by $\kS_{-l}(\cc)$ the space of all complex functions $u$, holomorphic in $\Im\,\sigma>-l$ and rapidly decreasing (Schwartz) as $\sigma\to\infty$ in strips, i.e., we require 
\[
\forall N,k,M\in\nn, \ \bra\sigma\ket^N \p_\sigma^k  u|_{\{\sigma: \ \Im\,\sigma\in(-l,M)  \}}\in L^\infty.  
\] 
If $E$ is a Fr\'echet space we denote by $\kS_{-l}(\cc;E)$ the corresponding space of $E$-valued functions. 

If the Mellin transform of $u$ belongs to $\kS_{-l}(\cc)$ then by \eqref{eq:imt} $\rho^{-l}u$ is bounded near $\rho=0$, and by a simple reduction to this case we get the following estimate.

\begin{lemma}\label{lem:mellinl}If $\cM_\rho u\in \kS_{-l}(\cc)$ then $\rho^{-l}  (\log \rho)^k  (\rho\p_\rho)^j u(\rho)$ is bounded near $\rho=0$ for any $j,k\in\nn$.   
\end{lemma}

\subsection{Asymptotic data of solutions}\label{ssec:asymptoticdata}

Let now $l\geq0$ be any order satisfying Hypothesis \ref{hyp2}. For a brief moment let us consider the space of all solutions with wave front set only in the radial set, i.e.
\beq\label{eq:sold}
\Sol(P)\defeq\{ u\in H^{-\infty,l}_\b(M): \ Pu=0, \ \wf^{\infty,l}_\b(u)\subset \cR \}.
\eeq
This is simply the space $\SolI(P)$ considered in Subsect. \ref{ss:ssaf} plus possible elements of $\Ker P_I$ and $\Ker P_{I^\c}$. These solutions enjoy the following properties:
\begin{enumerate} 
\item by below-threshold propagation of singularities they belong to $H_\b^{m,l}(M)$ for all $m<\12-l$;
\item as proved in \cite{BVW} they are `\b-Lagrangian' distributions\footnote{Note that components of $\bconormal$ are not even Legendre in $\be S^*M$ since the symplectic structure degenerates at $\p M$ in the $\b$-normal directions, so $\bconormal$ has dimension $n-2$ if $n$ is the dimension of $M$: both the boundary defining function $\rho$ and its $\b$-dual variable $\sigma$ vanish on $\bconormal$.} associated to $\cR$ in the sense that 
\[
A_1 A_2\dots A_k \Sol(P)\subset H_\b^{m,l}(M), \quad \forall k\in\nn, A_j\in\kM(M), 
\]
where $\kM(M)\subset\Psi^1_\b(M)$ is the space of $\b$-pseudodifferential operators whose principal symbols vanish on the radial set $\cR$. More explicitly, $\kM(M)$ can be characterized as the $\Psi^0_\b(M)$-module generated by $\rho\p_\rho$, $\rho\p_v$, $v\p_y$, $\p_y$ and $\one$.
\end{enumerate}

Let $\eta_\pm\in \cf(M)$ be  smooth cutoff functions of a neighborhood of $S_\pm$ in $M$. For the moment we restrict our attention to $S_+$, keeping in mind that the discussion for $S_-$ is analogous.   

For a solution $u\in\Sol(P)$, cutting it off with $\eta_+$ and taking the Mellin transform\footnote{Near the boundary $M$ admits a product decomposition of the form $[0,\epsilon)_\rho\times \p M$, we can then take $\eta_+$ supported in, say, $\rho<\epsilon/2$, which makes the Mellin transform of $\eta_+ u$ well defined.} in $\rho$ one obtains a family of functions $\cM(\eta_+ u)(\sigma)$ that is holomorphic in $\Im\,\sigma>-l$ with boundary value at $\Im\,\sigma=-l$ lying in the $H^m$-based Lagrangian space
\[
\{ f\in H^{m}(\p M): \ A_1 A_2\dots A_k f \in H^{m}(\p M), \ A_j\in\kM(\pM) \},
\]
and such that $\cM(\eta_+ u)(\sigma)$ rapidly decreases as $\sigma\to\infty$ (where $\kM(\pM)$ is generated by $v\p_y$, $\p_y$). Furthermore,  as shown in \cite{BVW}, $\cM(\eta_+ u)(\sigma)$ is necessarily a classical conormal distribution  in the sense that it is given by the sum of two oscillatory integrals of the form
\[
\int\e^{\i v\gamma}|\gamma|^{\i\sigma-1}\tilde a^\pm (\sigma,v,y,\gamma) d\gamma
\]
modulo $\kS_{-l}(\cc;\cf(\pM))$, with
$\tilde a^\pm$ (Schwartz function of $\sigma$ with values in)
classical symbols\footnote{Here we use $L^\infty$-based symbols, so a
  symbol $a$ of order $0$ satisfies $|D_y^\alpha D_v^k D_\gamma^N
  a|\leq C_{\alpha k N}\bra\gamma\ket^{-N}$ for all $\alpha$, $k$,
  $N$.} of order $0$ in $\gamma$. Here $\tilde a^\pm$ are supported in
$\pm\gamma>0$, corresponding to the half of $\bconormal_+$ considered
($\sinks_+$ versus $\sources_+$).
Thus, inverting the Mellin transform, and absorbing a factor of $2\pi$
into a newly defined $\tilde a^\pm$, $\eta_+ u$ itself is of the form
\[
J(\tilde a^\pm)=\int_{\Im\sigma=-l}\int\rho^{\i\sigma}\e^{\i v\gamma}|\gamma|^{\i\sigma-1}\tilde a^\pm (\sigma,v,y,\gamma)\, d\gamma\,d\sigma,
\]
modulo elements of $H_\b^{\infty,l}$. We call such distributions
weight $l$
{\em b-conormal distributions} of symbolic order $0$ associated to the half of $\bconormal_+$ considered
($\sinks_+$ versus $\sources_+$).
Note that if $\tilde a^\pm$ vanishes to order $k$ at $v=0$ then
integration by parts in $\gamma$ allows one to conclude that $J(\tilde
a^\pm)=J(\tilde b^\pm)$ where $\tilde b^\pm$ now take values of
classical conormal symbols of order $-k$. Then, by an asymptotic
summation argument (which is just the $\sigma$-dependent version of
the standard argument for conormal distributions, here conormal to $v=0$, see e.g.~\cite[Prop.~2.3]{melrosenotes} or \cite[Eq. (3.35)]{vasyminicourse}) one sees that the $v$ dependence of $\tilde a^\pm$ can be essentially
completely eliminated in that one can write the integrand as
$\chi_0(v)$ times a $v$ independent symbol, with $\chi_0\equiv 1$ near $0$
and of compact support, again modulo $\kS_{-l}(\cc;\cf(\pM))$. In particular, the leading term of the
asymptotic expansion of $\tilde a^\pm$ as $\gamma\to\pm\infty$ is
recovered by simply taking the Fourier transform of the Mellin
transform of $\eta_+ u$ and letting $\gamma\to\pm\infty$. Furthermore, analogous statements apply if
$\tilde a^\pm$ is a classical symbol of order $s$. In particular, the
isomorphism properties of the Fourier and Mellin transforms show that
when $\tilde a^\pm$ is a classical symbol of order $s$, $J(\tilde
a_\pm)$ is in $H_\b^{m,l}(M)$ if $m<\12-l-s=-\12-(l+s-1)$, with
$l+s-1$ being the symbolic order of the symbol $|\gamma|^{\i\sigma-1}\tilde a_\pm$.

In terms of $u\in\Sol(P)$, this means that for $v$ and $\rho$ near $0$, $\eta_+ u$ is the sum of two integrals of the form 
\beq\label{eq:froma}
\int \rho^{\i\sigma} \e^{\i v \gamma} |\gamma|^{\i\sigma-1} a^{\pm}(\sigma,y)\chi^{\pm}(\gamma) d\gamma d\sigma 
\eeq 
with $a^\pm\in \kS_{-l}(\cc;\cf(\pM))$, modulo terms that belong to $H^{m',l}_\b(M)$ for some
$m'>\12-l$ (indeed, any $m'<\frac{3}{2}-l$) and for this reason will turn out to be irrelevant for the analysis that follows. Above, $\chi^\pm$ are smooth functions with support in $\pm [0,\infty)_\gamma$.

In the reverse direction, taking the inverse Mellin and Fourier transform yields two maps
\beq\label{eq:rhoo}
\Sol(P)\ni u \mapsto a^+(\sigma,y)\in \tilde\cI^{l}_{+}, \ \ \Sol(P)\ni u \mapsto a^-(\sigma,y)\in \tilde\cI^{l}_{+},
\eeq
where we have introduced the notation
\[
\begin{aligned}
\tilde\cI^{l}_{\pm}\defeq \big\{ a \in\, &\cf(\overline{\cc_{-l}}\times S_\pm): \ \overline{\partial}a=0, \\ &\forall M,N,k\in\nn, \ B\in\Diff(S_\pm), \ \bra\sigma\ket^N \p_\sigma^k B a \traa{\{\sigma: \ \Im\,\sigma\in(-l,M)  \}}\in L^\infty \big\}
\end{aligned}
\]
for the principal symbols of conormal distributions considered here. Above, $\cc_{-l}= \{ \sigma\in\cc : \Im\,\sigma>-l\}$ and the Cauchy--Riemann operator $\overline{\p}$ acts in the first variable (i.e., $\sigma$) in the domain where $l$ is such that no resonances of the Mellin transformed inverse of $P$ have imaginary part in $[-l,l]$.

Now, we make a choice of components $\cR_I^-$ in the radial set from which the estimates are propagated, labelled as usual by $I\subset\{+,- \}$ and set
\[
\tilde\cI_I\defeq \tilde\cI^{l}_\pm\oplus \tilde\cI^{l}_\pm,
\]
where the signs are chosen in such way that the number of pluses (resp. minuses) reflects the number of components of $\cR_I^+$ in $S_+$ (resp. $S_-$).  Accordingly, we have a map (denoted $\varrho_I$) that assigns to a solution its pair of data on $\cR_I^+$
\beq\label{eq:mapasy}
\Sol(P)\ni u \mapsto \varrho_I u = ( a,a') \in \tilde\cI_I.
\eeq

We will show that the map $\varrho_I:\SolI(P)\to \tilde\cI_I$ is in fact bijective, possibly after removing a finite-dimensional subspace from $\tilde\cI_I$. 

Injectivity is a consequence of Lemma \ref{lem:microvanish} (note that the hypotheses of this lemma are the reason why we consider here the restricted solution space $\SolI(P)$ instead of $\Sol(P)$), so we focus on surjectivity. Let $\tcP^0_I$ be the map defined for $(a,a')\in \tilde\cI_I$, by applying formula (\ref{eq:froma}) to $a$ and $a'$ (with the signs chosen consistently with $I$), multiplying the resulting distributions by $\eta_+$ or $\eta_-$ (consistently with $I$), and then adding them up. Then $w=\tcP^0_I(a,a')$ belongs to $H_\b^{m,l}(M)$ for $m<\12-l$ and its wave front set is in $\cR$. Moreover, $w$ is regular under $\kM$. The especially non-obvious part of this statement is regularity with respect to $\rho D_v$, which uses the holomorphicity: $\rho D_v$ applied to (\ref{eq:froma}) yields indeed
\begin{equation}\begin{aligned}\label{eq:rDv-contour-shift}
& \int_{\Im\,\sigma=-l}\rho^{\i(\sigma-\i)}\e^{\i v \gamma}|\gamma|^{\i(\sigma-\i)-1}a^{\pm}(\sigma,y)\chi^\pm(\gamma) d\gamma d\sigma\\
& = \int_{\Im\,\sigma=-l+1}\rho^{\i(\sigma-\i)}\e^{\i v \gamma}|\gamma|^{\i(\sigma-\i)-1}a^{\pm}(\sigma,y)\chi^\pm(\gamma) d\gamma d\sigma\\
& = \int_{\Im\,\sigma=-l}\rho^{\i\sigma}\e^{\i v \gamma}|\gamma|^{\i\sigma-1}a^{\pm}(\sigma+\i,y)\chi^\pm(\gamma) d\gamma d\sigma.
\end{aligned}\end{equation}
One also gets that $Pw\in H_\b^{m,l}$ (two orders better than a priori
expected, this follows from $P$ being equal to $-4D_v(vD_v + \rho
D_\rho)$ modulo $\kM^2$).
We can improve this further:

\begin{lemma}\label{lem:morereg}
Suppose $l\in\rr$.
There is a continuous linear map $\tcP_I:\tilde\cI_I\to H_\b^{m,l}(M)$,
for all $m<\12-l$,
such that $P\circ\tcP_I:\tilde\cI_I\to H_\b^{\infty,l}(M)$ and
\[
\tcP_I-\tcP_I^0:\tilde\cI_I\to H_\b^{m+1,l}(M)
\]
for all $m<\12-l$.
\end{lemma}

The proof of Lemma \ref{lem:morereg} is given in Appendix \ref{app2}.

We now define the \emph{Poisson operator}
\beq\label{eq:poisson}
\cP_I\defeq(P_I^{-1}-P_{I^{\rm c}}^{-1}) P \tilde\cP_I.
\eeq
Let us analyze its mapping properties. First, $P \tilde\cP_I$ maps $\tilde\cI_I$ to $\Ran P_{ I}$ directly from the definition as $\tilde\cP_I$ maps into $\cX_I$ by virtue of Lemma \ref{lem:morereg}. Furthermore $P \tilde\cP_I$ maps also to $\cY^{\infty,l}=\bigcap_m \cY^{m,l}$, which is a subset of $\cY_{I^\c}$. Since $P:\cX_I\to\cY_I$ is Fredholm, the kernel of $P \tilde\cP_I$ is finite dimensional and has thus a complement $\cK_I\subset\tilde\cI_I$. On $\cK_I$, $P \tilde\cP_I$ is injective, so  the pre-image of $\cZ_{I^\c}$ (where we recall that $\cZ_{I^\c}$ is a complement of $\cY_{I^\c}$) is finite dimensional. Taking the pre-image of $\Ran P_{I^\c}$ and adding to it elements of $\Ker P \tilde\cP_I$ we obtain a subspace of $\tilde\cI_I$: 
\[
\cI_I\defeq (P \tilde\cP_I)^{-1}\Ran P_{I^\c} + \Ker P \tilde\cP_I,
\] 
which has a finite dimensional complement and such that $P \tilde\cP_I \cI_I\subset \Ran P_{I^\c}$. Thus, the Poisson operator \eqref{eq:poisson} maps
\[
\cP_I : \cI_I \to \Sol_I(P).
\]
We will prove that $\varrho_I$ maps $\Sol_I(P)\to\cI_I$ and that it does so bijectively, with inverse $\cP_I$. We will need two  auxiliary lemmas, the proof of which is deferred to Appendix \ref{app2}.

\begin{lemma}\label{lem:morereg2} The operator $\tcP_I \circ \varrho_I$ acts on $\Sol(P)$ as a pseudodifferential operator that is microlocally the identity near $\cR^+_I$ and microlocally vanishes near $\cR^-_I$, modulo terms that map to $H_\b^{m',l}(M)$ for some $m'>\12-l$.
\end{lemma}

\begin{lemma}\label{lem:morereg3} The map $(a,a')\mapsto[w]=[\tcP_I(a,a')]$ is injective, with the equivalence class considered modulo $H_\b^{m+1,l}(M)$, $-\12+l<m<\12+l$.
\end{lemma}

Now, since in the sense stated in the above lemma, $\tcP_I \varrho_I$ is microlocally the identity near $\cR^+_I$ and microlocally vanishes near $\cR^-_I$, arguing as in the paragraph below \eqref{eq:tempQI} we conclude that $P \tcP_I \varrho_I$ maps $\SolI(P)$ to $\Ran P_I\cap \Ran P_{I^\c}$. This in turn implies that $\varrho_I$ maps to $\cI_I$. On the other hand using (\ref{eq:GInverse}) we get
\beq\label{eq:GInverse2}
 -(P_{I^{\rm c}}^{-1}- P_I^{-1})P\tilde \cP_I \varrho_I=\one \ \mbox{ \ on \ } \SolI(P),
\eeq
that is $\cP_I \varrho_I=\one$ on $\SolI(P)$. Thus, to deduce surjectivity of $\varrho_I$ we need to show that $\cP_I$ is injective. To that end, observe that $\cP_I(a,a')=\tcP_I(a,a')$ at $\cR^+_I$ modulo
$H^{m+1,l}_\b$ terms with $\mathfrak{M}(M)$ regularity. Thus, the injectivity of $\cP_I$ follows from the injectivity of $[\tcP_I]$ stated in  Lemma \ref{lem:morereg3}.

We have thus proved:

\begin{proposition}\label{prop:bijec} Assume Hypotheses \ref{hyp1}, \ref{hyp2} and \ref{hyp3}. Then the map  $\SolI(P)\ni u \mapsto \varrho_I u\in\cI_I$  defined in (\ref{eq:mapasy}) is bijective with inverse $\cP_I$.
\end{proposition}

We now consider the pairing formula for smooth approximate solutions, i.e.~for $u$ satisfying 
\beq\label{eq:appros}
u\in H_\b^{-\infty,0}(M), \ \ Pu\in H_\b^{\infty,0}(M), \ \ \wf^{\infty,0}_\b(u)\subset\cR;
\eeq 
the computations below are closely related to \cite{positive}.
To this end we will need a family of operators $\cJ_r$ belonging to $\Psi_\b^{-N}$ for $r\in(0,1]$ (and $N$ sufficiently large), uniformly bounded in $\Psi_\b^0$ for $r\in(0,1]$ and tending to $\one$ as $r\to0$ in $\Psi^\epsilon_\b$ for any $\epsilon>0$, so that $[P,\cJ_r]\to 0$ in $\Psi_\b^{1+\epsilon}$. Let us take concretely $\cJ_r$ to have principal symbol $j_r=(1+r|\gamma|)^{-N}$ near the radial sets. Then, in terms of the pairing $\bra\cdot,\cdot\ket_\ssb$ defined in Subsect.~\ref{ss:wb},
\beq\label{eq:pairings}
\bea
\i^{-1}(\bra P u_1,u_2\ket_\ssb -\bra u_1, P u_2\ket_\ssb )&=\i^{-1} \lim_{r\to 0}(\bra \cJ_r P u_1,u_2\ket_\ssb - \bra\cJ_r u_1, Pu_2\ket_\ssb ) \\
& = \lim_{r\to 0}\bra \i [\cJ_r,P] u_1,u_2 \ket_\ssb,
\eea
\eeq 																							
for any $u_1,u_2$ satisfying \eqref{eq:appros} and the principal symbol of $\i[\cJ_r,P]$ is
\[
-H_p j_r = (\sgn\gamma)N r(1+r |\gamma|)^{-1}j_r H_p\gamma.
\]
Moreover, $H_p|\gamma|=(\sgn\gamma) H_p\gamma$ is positive at sinks, negative at sources. Concretely, in our case, as $p$ is given by $-4\gamma(v\gamma+\sigma)$ modulo terms that vanish quadratically at the radial set $\cR$, $H_p\gamma$ is given by $4\gamma^2$ modulo terms vanishing at $\cR$. Hence, $-H_p j_r$ equals $4\gamma^2(\sgn\gamma)N r(1+r|\gamma|)^{-1}j_r$ modulo such terms, thus the sinks correspond to $\gamma>0$, whereas the sources to $\gamma<0$. 

Now, $u_1$ and $u_2$ have module regularity of the same type as already discussed for $\Sol(P)$, so the result of the computation of (\ref{eq:pairings}) is unaffected if $P$ is changed by terms in $\kM^2$ (provided they preserve the formal self-adjointness). Moreover, $u_i$ can be replaced by distributions $\tilde u_i$ with $u_i-\tilde u_i\in H_\b^{m+1,l}$, $P\tilde u_i \in H_\b^{m,l}$ with wave front set in the radial sets. So in particular, for each $i$ we may replace $u=u_i$ by $\tilde\cP_\emptyset(a^+_-,a^-_-)+\tilde\cP_{\{+,-\}}(a^+_+,a^-_+)$, where $a^\pm_\pm$ are the $\b$-conormal principal symbols discussed before, with the superscript denoting the component of the characteristic set and the subscript the component of the radial set: $\cR_\emptyset^-$ versus $\cR_\emptyset^+$.    

Therefore, as the Mellin transform and Fourier transform are isometries up to constant factors,  we can reexpress (\ref{eq:pairings}) as
\begin{align*}
= & \lim_{r\to 0} \Const \sum_{\pm}\int   4\gamma^2 N r (1+r|\gamma|)^{-1} j_r |\gamma|^{\i\sigma-1}|\gamma|^{-\i\sigma-1} \\
 & \phantom{\lim_{r\to 0} \Const \sum_{\pm}\int}\times \Big(\chi^+(\gamma)^2 \sum_{\pm} \overline{a^{\pm}_{1,+}}{a^{\pm}_{2,+}} -\chi^-(\gamma)^2\sum_\pm  \overline{a^{\pm}_{1,-}}{a^{\pm}_{2,-}}\Big)|dh(y)|d\gamma d\sigma\\
= &\lim_{r\to 0} \Const \sum_{\pm} \left(  \int  4 Nr (1+r|\gamma|)^{-1} j_r \chi^+(\gamma)^2 d\gamma \right)\left( \int \overline{a^{\pm}_{1,+}}{a^{\pm}_{2,+}}  |dh(y)| d\sigma \right)\\
 & \phantom{\lim_{r\to 0} \Const \sum_{\pm} } - \left(  \int 4 Nr (1+r|\gamma|)^{-1} j_r \chi^-(\gamma)^2 d\gamma \right)\bigg( \int \overline{a^{\pm}_{1,-}}{a^{\pm}_{2,-}}  |dh(y)| d\sigma \bigg)
\end{align*}
where $h$ is the metric on $S_\pm$ and the integral in $\sigma$ is over $\Im\sigma=0$. Integrating by parts and then applying the dominated convergence theorem gives
\begin{align*}
= & \lim_{r\to 0} \Const \sum_{\pm} \left( \int -4\frac{d}{d\gamma}(j_r)\chi^+(\gamma)^2 d\gamma \right) \left( \int \overline{a^{\pm}_{1,+}}{a^{\pm}_{2,+}} |d h(y)| d\sigma\right) \\
& \phantom{\lim_{r\to 0} \Const \sum_{\pm}}- \left( \int -4 \frac{d}{d\gamma}(j_r)\chi^-(\gamma)^2 d\gamma \right) \left( \int \overline{a^{\pm}_{1,-}}{a^{\pm}_{2,-}} |d h(y)| d\sigma\right) \\
= & \lim_{r\to 0} \Const \sum_{\pm} \left( \int -4 j_r\frac{d}{d\gamma}\chi^+(\gamma)^2 d\gamma \right) \left( \int \overline{a^{\pm}_{1,+}}{a^{\pm}_{2,+}} |d h(y)| d\sigma\right) \\
& \phantom{\lim_{r\to 0} \Const \sum_{\pm}} - \left( \int -4 j_r\frac{d}{d\gamma}\chi^-(\gamma)^2 d\gamma \right) \left( \int \overline{a^{\pm}_{1,-}}{a^{\pm}_{2,-}} |d h(y)| d\sigma\right) \\
= & \,  \FourConst \sum_{\pm}  \left( \int \overline{a^{\pm}_{1,+}}{a^{\pm}_{2,+}} |d h(y)| d\sigma - \int \overline{a^{\pm}_{1,-}}{a^{\pm}_{2,-}} |d h(y)| d\sigma \right).
\end{align*}																																																											
This means that for $u_1=\tcP_I(a_1^+,a_1^-)$, and $u_2\in\Sol(P)$  with asymptotic data $\varrho_I u=(a_2^+,a_2^-)$ we have
\beq\label{eq:argg}
\bra P \tcP_I (a_1^+,a_1^-),u_2\ket_\ssb = \FourConst\i \sum_\pm (-1)^{I(\pm)}\int \overline{a_1^\pm} {a_2^\pm} |d h(y)| d\sigma,
\eeq
where we have used the notation introduced before
\[
(-1)^{I(\pm)}=\begin{cases} 1 \mbox{\ if \ } \pm\in I, \\
-1 \mbox{\ otherwise}.\end{cases}
\]
If instead $(a_2^+,a_2^-)$ are the asymptotics of $u_2$ at $\cR_I^+=\cR_{I^{\rm c}}^-$ then
\[
\bra P \tcP_{I^{\rm c}} (a_1^+,a_1^-),u_2\ket_\ssb = -\FourConst\i  \sum_\pm (-1)^{I(\pm)}\int \overline{a_1^\pm} {a_2^\pm} |d h(y)| d\sigma.
\]
This gives in the former case
\beq\label{eq:rhoequals}
\varrho_I u_2 = \FourConst\i \begin{pmatrix}{(-1)^{I(+)}}&{0}\\{0}&{(-1)^{I(-)}}\end{pmatrix} (P \tcP_I)^* u_2
\eeq
and so if $u_2$ belongs to the restricted solution space $\SolI(P)$,
\[
\bea
u_2&=\FourConst\i  \cP_I \begin{pmatrix}{(-1)^{I(+)}}&{0}\\{0}&{(-1)^{I(-)}}\end{pmatrix} (P \tcP_I)^* u_2\\
 &= \FourConst\i   (P_{I}^{-1}- P_{I^{\rm c}}^{-1}) P \tilde\cP_I \begin{pmatrix}{(-1)^{I(+)}}&{0}\\{0}&{(-1)^{I(-)}}\end{pmatrix}  (P \tilde\cP_I)^* u_2.
\eea
\]
In particular,
\[
 (P_{I}^{-1}- P_{I^{\rm c}}^{-1})=  \FourConst\i  (P_{I}^{-1}- P_{I^{\rm c}}^{-1}) P \tilde\cP_I \begin{pmatrix}{(-1)^{I(+)}}&{0}\\{0}&{(-1)^{I(-)}} \end{pmatrix}(P \tilde\cP_I)^* (P_{I}^{-1}- P_{I^{\rm c}}^{-1}),
\]
hence using (\ref{eq:rhoequals}) again,
\[ 
(P_{I}^{-1}- P_{I^{\rm c}}^{-1})=  \i(\FourConst)^{-1} (P_{I}^{-1}- P_{I^{\rm c}}^{-1}) \varrho_I^* \begin{pmatrix}{(-1)^{I(+)}}&{0}\\{0}&{(-1)^{I(-)}}\end{pmatrix} \varrho_I (P_{I}^{-1}- P_{I^{\rm c}}^{-1})
\]
Denoting now 
\beq\label{eq:eeeq}
q_I\defeq (\FourConst)^{-1} \begin{pmatrix}{(-1)^{I(+)}}&{0}\\{0}&{(-1)^{I(-)}}\end{pmatrix},
\eeq
and recalling that $G_I= P_I^{-1}-P_{I^{\rm c}}^{-1}$, this can be rewritten as $\i G_I = -G_I \varrho_I^* q_I \varrho_I G_I$. In the sense of sesquilinear forms on $\Ran P_I\cap\Ran P_{I^\c}$, $\i G_I$ is formally self-adjoint so this gives
\beq\label{eq:gqrho}
\i G_I = G_I^* \varrho_I^* q_I \varrho_I G_I.
\eeq
In summary:

\begin{theorem}\label{prop:iso} Assume Hypotheses \ref{hyp1}, \ref{hyp2} and \ref{hyp3}. Let $I\subset\{ +, -\}$ and suppose $l=0$ is not a resonance in the sense of Hypothesis \ref{hyp2}. There are isomorphisms of symplectic spaces
\beq\label{eq:arrows}
\frac{\Ran P_I\cap \Ran P_{I^\c}}{P(\cX_I\cap \cX_{I^\c})}\xrightarrow{\makebox[2em]{$[G_I]$}}\SolI(P)\xrightarrow{\makebox[2em]{$\varrho_I$}}\cI_I,
\eeq
where the symplectic form on the first one is given by $\bra\overline{\cdot},G_I\cdot\ket_\ssb$ and on the last one by \eqref{eq:eeeq}.
\end{theorem}

As an aside, observe that if we get back to equation \eqref{eq:argg} specifically in the Feynman or anti-Feynman case, then the pairing is definite and we obtain that for any approximate solution $u$ with asymptotic data $\varrho_I u=(a^+,a^-)$,  the quantity $\bra P \tcP_I (a^+,a^-),u\ket_\ssb$ vanishes if and only if $(a^+,a^-)=0$. In particular, if $u\in\Ker P_I$ (so that $u$ is regular at $\cR_I^-$) then
\[
\bra P \tcP_I (a^+,a^-),u\ket_\ssb=\bra \tcP_I (a^+,a^-),Pu\ket_\ssb=0
\]
so $(a^+,a^-)=0$. This implies $u$ has above-threshold regularity at $\cR_I^+$; it is also regular at $\cR_I^-$ so in fact by above-threshold propagation estimates (i.e., (2) of Theorem \ref{thm:propag}) we get:

\begin{proposition}\label{prop:faf}In the Feynman ($I=\emptyset$) and anti-Feynman case ($I=\{+,-\}$), Hypothesis \ref{hyp3} is satisfied for $l=0$, i.e.~$\Ker P_I\subset H_\b^{\infty,0}(M)$.  
\end{proposition}

\subsection{Hadamard two-point functions}\label{ss:hada}

The second arrow in (\ref{eq:arrows}) means that the symplectic space $\kV_I$ is isomorphic to $\cI_I$ equipped with the symplectic form $\i^{-1} q_I$, which is more tractable in applications.

Let us denote
\[
\pi^+= (\FourConst)^{-1}\mat{\one}{0}{0}{0}, \quad \pi^-= (\FourConst)^{-1}\mat{0}{0}{0}{\one},
\]
and for $I\subset\{+,-\}$ consider the pair of operators
\beq\label{eq:lpm}
\Lambda^\pm_I\defeq G^*_I \varrho^*_I \pi^\pm \varrho_I G_I  : H^{\infty,0}_\b(M)\to H^{-\infty,0}_\b(M). 
\eeq
They satisfy $P\Lambda^\pm_I=\Lambda^\pm_I P=0$, $(-1)^{I(+)}\Lambda^+_I+ (-1)^{I(-)}\Lambda^-_I=\i G_I$ and $\Lambda^\pm_I\geq 0$ when identified with sesquilinear forms on $\Ran P_I\cap \Ran P_{I^\c}$ via the product $\bra \cdot, \cdot\ket_\ssb$. We will prove that they also satisfy the wave front set condition required from Hadamard two-point functions.

\begin{theorem}Assume Hypotheses \ref{hyp1}, \ref{hyp2} and \ref{hyp3}. \label{prop:Had} The pair of operators $\Lambda^\pm_I$ defined in (\ref{eq:lpm}) satisfy 
\[
\wf'(\Lambda^\pm_I)\subset (\Sigma^\pm\cup\,\zero)\times (\Sigma^\pm\cup\,\zero),
\]
which implies the Hadamard condition if $(M^\inti,g)$ is globally hyperbolic. Thus in that case, if $I=\{\pm\}$ then $\Lambda^\pm_I$ are Hadamard two-point functions for $P$ (cf.~Definition \ref{def:hadamard}).
\end{theorem}
\proof We assume for simplicity that all the operators $P_I$ are invertible, otherwise one simply needs to use projections to the finite-dimensional spaces $\Ker P_I$ and $\cZ_I$ to legitimize the arguments that follow. We consider the case $I=\{+\}$, the remaining ones being analogous, and we skip the subscript $I$ for brevity of notation. 

First observe that for any $v\in \cX_+\cap\cX_-$, the distribution $f=\tilde\cP\pi^+\varrho G v$ has above-threshold regularity at $\sinks_-$, $\sources_-$ (due to the definition of $\tilde\cP$) and also at $\sources_+$ (due to the presence of $\pi^+$). Now $\Lambda^+ v = (\one - P_+^{-1} P) f$ differs from $f$ by a term regular at $\bconormal_+$, thus $\Lambda^+ v$ is regular near $\sources_+$. It also solves the wave equation, so by propagation of singularities $\wf(\Lambda^+ v)\subset \Sigma^+$ in $M^\inti$.   

Applying this to $v=\delta_x$, this means on the level of the Schwartz kernel that $\wf'(\Lambda^+)\subset (\Sigma^+ \cup \,\zero)\times T^*M^\inti$, and in the same way one  gets $\wf'(\Lambda^-)\subset (\Sigma^- \cup \,\zero)\times T^*M^\inti$. Proceeding as in the proof of Proposition \ref{prop:wfPI}  we obtain the assertion.\qeds 

As already outlined in the introduction, the two-point functions $\Lambda^\pm_+$ and $\Lambda^\pm_-$ constructed from asymptotic data $\varrho_+$ and $\varrho_-$ can be thought as analogues of two-point functions constructed in other setups \cite{Mo1,Mo2,characteristic,GW3} for the conformal wave equation and for the massive Klein-Gordon equation (rather than for the wave equation considered here). A common feature of all these constructions is that the two-point functions are distinguished once the asymptotic structure of the spacetime is given, in particular they do not depend on the precise choice of coordinates and boundary defining function.

\subsection{Blow-up of $S$}\label{ss:blowup}

In the setting of Definition~\ref{def:sc}, a convenient way to specify the asymptotic data of a solution of the wave equation is based on the radiation field blow-up proposed by Baskin, Vasy and Wunsch in \cite{BVW} in the context of asymptotic expansions for the Friedlander radiation fields (much in the spirit of Friedlander's work \cite{friedlander}). In what follows we briefly discuss how this can be used in our situation to provide a more geometrical description of the data $\varrho_I u$ (for a restricted class of solutions), starting with the following example. Namely, on Minkowski space $\rr^{1+d}$ with coordinates $(t,x)$, a convenient choice of new coordinates is $s=t-|x|$, $y=x/|x|$, $\rho=(t^2+|x|^2+1)^{-1/2}$. These make sense locally near the \emph{front face} $\ff=\{\rho=0\}$, and asymptotic properties of solutions can be described in terms of their restriction to $\ff$, multiplied first by a $\rho^{-(n-2)/2}$ factor to make this restriction well-defined. The step that consists of multiplying a solution $u$ by $\rho^{-(n-2)/2}$ can be interpreted as replacing the original metric by a conformally related one, which extends smoothly to $\{\rho=0\}$, and then considering $u$ as a solution for the conformally related wave operator.

In the general setting of Lorentzian scattering spaces, recalling that
$\rho$ is a boundary defining function of $\p M$ and $(v,y)$ are
coordinates on $\pM$ with $S=\{\rho=0,v=0\}$, the analogue of this
construction consists of introducing  coordinates $(s,y)$ with
$s=v/\rho$, valid near a boundary hypersurface `$\ff$' (the
\emph{front face}) of a new manifold that replaces $M$, constructed as
the sum of $M\setminus S$ and the \emph{inward-pointing spherical
  normal bundle} of $S$. More precisely, one replaces $M$ with a
manifold with corners $[M;S]$ (the \emph{blow-up of $M$ along $S$},
cf. \cite{melrose}), equipped in particular with a smooth map
$[M;S]\to M$ called the \emph{blow-down} map which is
a diffeomorphism between the interior of
the two spaces. It is possible to canonically define $[M;S]$ in such
way that `polar coordinates' $R=(v^2+\rho^2)^{1/2}$,
$\vartheta=(\rho\cdot v)/R$ are smooth, and smooth functions on $M$
lift to smooth ones on $[M;S]$ by the blow-down map. The boundary
surface of interest $\ff$ is simply defined as the lift (i.e.\ inverse
image) of $S$ to $[M;S]$ (see Figure 4), and near its interior, $(\rho,s,y)$ constitute a well-defined system of coordinates indeed.

\def\svgwidth{10cm}
%\sskip 
\begin{figure}[h]\label{fig3}
\begingroup%
  \makeatletter%
  \providecommand\color[2][]{%
    \errmessage{(Inkscape) Color is used for the text in Inkscape, but the package 'color.sty' is not loaded}%
    \renewcommand\color[2][]{}%
  }%
  \providecommand\transparent[1]{%
    \errmessage{(Inkscape) Transparency is used (non-zero) for the text in Inkscape, but the package 'transparent.sty' is not loaded}%
    \renewcommand\transparent[1]{}%
  }%
  \providecommand\rotatebox[2]{#2}%
  \ifx\svgwidth\undefined%
    \setlength{\unitlength}{627.70153261bp}%
    \ifx\svgscale\undefined%
      \relax%
    \else%
      \setlength{\unitlength}{\unitlength * \real{\svgscale}}%
    \fi%
  \else%
    \setlength{\unitlength}{\svgwidth}%
  \fi%
  \global\let\svgwidth\undefined%
  \global\let\svgscale\undefined%
  \makeatother%
  \begin{picture}(1,0.24799698)%
    \put(0,0){\includegraphics[width=\unitlength]{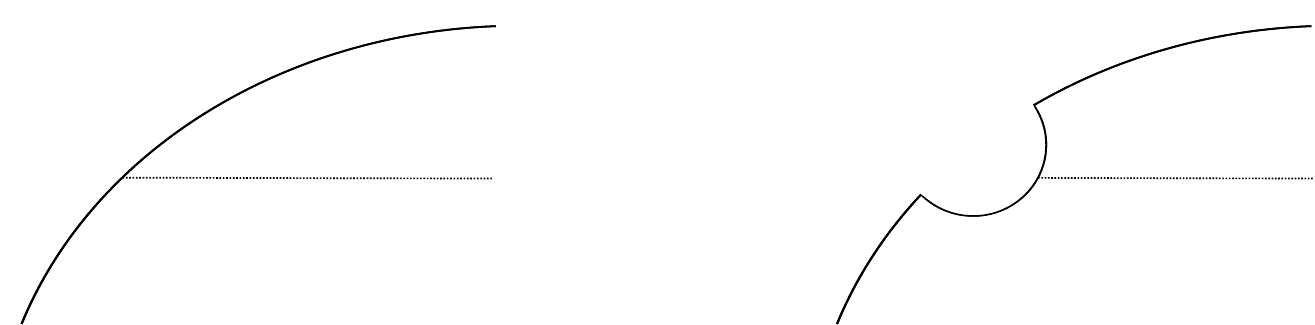}}%
    %\put(0.28341959,0.12408488){\color[rgb]{0,0,0}\makebox(0,0)[lb]{\smash{$S^+$}}}%
    \put(0.25912964,0.23111089){\color[rgb]{0.2,0.2,0.2}\makebox(0,0)[lb]{\smash{\scalebox{0.8}{$X_+$}}}}%
    %\put(0.90441132,0.11893479){\color[rgb]{0,0,0}\makebox(0,0)[lb]{\smash{$S^+$}}}%
    \put(0.87754651,0.2298231){\color[rgb]{0.2,0.2,0.2}\makebox(0,0)[lb]{\smash{\scalebox{0.8}{$X_+$}}}}%
    \put(-0.00010314,0.06206516){\color[rgb]{0.2,0.2,0.2}\makebox(0,0)[lb]{\smash{\scalebox{0.8}{$X_-$}}}}%
    \put(0.60573158,0.05309934){\color[rgb]{0,0,0}\makebox(0,0)[lb]{\smash{\scalebox{0.8}{$X_-$}}}}%
    \put(0.76071262,0.11970261){\color[rgb]{0,0,0}\makebox(0,0)[lb]{\smash{$\ff$}}}%
    \put(0.48406757,0.10229508){\color[rgb]{0,0,0}\makebox(0,0)[lb]{\smash{$\longrightarrow$}}}%
  \end{picture}%
\endgroup 
\caption{The radiation field blow-up of $M$ along $S=S_+\cup S_-$. The
blow-down map goes in the reverse of the direction of the arrow.}
\end{figure}
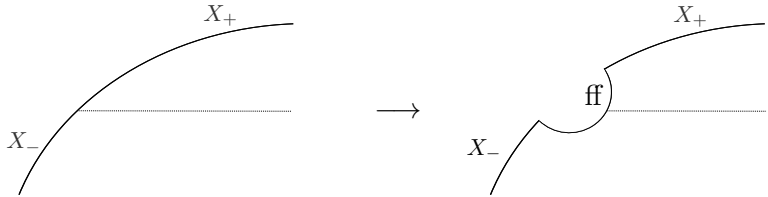

Although the metric $g$ (lifted using the blow-down map) is ill-behaved as $\rho$ tends to $0$, rescaling it by a conformal factor $\rho^2$ yields a Lorentzian metric $\rho^2 g$ which is smooth down to $\rho=0$. Note that if $u(\rho,v,y)$ solves $Pu=f$, then $u(\rho,\rho s, y)$ is a solution of the inhomogeneous Klein-Gordon equation conformally related to $\Box_g$. 

It can be argued that the restriction of $u$ to the front face is well-defined for $u\in\Sol_I(P)$ at least if $l>0$. Indeed, let $\tilde\cP_0$ be the analogue of the map $\tilde\cP_I$ acting on full symbols rather than on principal symbols (see also \eqref{eq:analo} in Appendix \ref{app2}) in the blown-up setting. In the case $l>0$, $u$ can be (locally) expressed as $\tilde\cP_0 a$ modulo some decaying terms, and since $\tilde\cP_0$ maps to distributions which are conormal to the front face (in particular we get decay in the $L^2_\b$ sense due to the assumption $l>0$), the restriction to $\ff$ makes sense. 

Now, recall that in our discussion of the asymptotic data $\varrho_I$, the starting point was the expression 
\beq\label{eq:froma2}
\int \rho^{\i\sigma} \e^{\i v \gamma} |\gamma|^{\i\sigma-1} a^{\pm}(\sigma,y)\chi^{\pm}(\gamma) d\gamma d\sigma
\eeq
for elements of $\Sol(P)$, valid (near $S$) modulo terms in $H^{m',l}_\b(M)$ for some $m'>\12-l$. Performing the $\sigma$ integral first, one obtains (up to non-zero constant factors)
\[
\int \e^{\i v \gamma}(\cM^{-1}a^\pm)(\rho |\gamma|,y)\chi^\pm(\gamma)|\gamma|^{-1} d\gamma,
\]
where $\cM^{-1}$ is the inverse Mellin transform in $\sigma$. Replacing $\gamma$ by $\nu=\rho\gamma$, one has
\[
\int \e^{\i \nu (v/\rho)}(\cM^{-1}a^\pm)( |\nu|,y)\chi^\pm(\rho^{-1}\gamma)|\nu|^{-1} d\nu.
\]
As $\rho\to 0$ this becomes
\[
\int_{\pm[0,\infty)}  \e^{\i \nu (v/\rho)}(\cM^{-1}a^\pm)(|\nu|,y)|\nu|^{-1} d\nu,
\]
which is the inverse Fourier transform in $\nu$ of $(\cM^{-1}a^\pm)(|\nu|,y)\bbbone^\pm(\nu)|\nu|^{-1}$ ($\bbbone^\pm$ being the characteristic function of $\pm[0,\infty)$) evaluated in the radiation face coordinate $s=v/\rho$:
\beq\label{eq:invftofm}
\cF^{-1}\left((\cM^{-1}a^\pm)(|\,.\,|,y)\bbbone^\pm(\,.\,)|\,.\,|^{-1} \right)(v/\rho).
\eeq
Note that the inverse Fourier transform above is well-defined because the product of $(\cM^{-1}a^\pm)(|\,.\,|,y)$ and $\bbbone^\pm(\,.\,)|\,.\,|^{-1}$ is in $L^1$ by Lemma \ref{lem:mellinl}. As the inverse Fourier transform of a distribution conormal to the origin, (\ref{eq:invftofm}) is a symbol, although it is difficult to make an exact statement for the exact class of symbols it is in since the superlogarithmic decay at the origin does not translate directly into nice estimates. 

After performing the blow-up, we can view (\ref{eq:invftofm}) as the restriction of a solution to the front face $\ff$. Thus in the reverse direction, one takes $u\traa{\ff}$, one Fourier transforms it, then restricts to the positive or negative half-lines and then Mellin transforms the result to obtain the principal symbol of the solution in the respective half of $\bconormal_\pm=\sinks_\pm\cup\sources_\pm$. This means that for any $u$ with well-defined restriction $u\traa{\ff}$, $\varrho_I u$ can be expressed as 
\beq\label{eq:mapasy2}
\Sol(P)\ni u \mapsto \varrho_I u \defeq \left( \cM(\cF(\eta_\pm  u\traa{\ff})|\gamma|\,\bbbone^\pm), \cM(\cF(\eta_\pm  u\traa{\ff})|\gamma|\,\bbbone^\pm)\right) \in \tilde\cI_I,
\eeq
where the signs are chosen relatively to $I$, i.e.~for each component the subscript indicates $S_+$ versus $S_-$ and the sign in the superscript indicates $\sinks$ versus $\sources$, and as before, $\eta_\pm$ are smooth cutoff functions of a neighborhood of $S_\pm$ in $M$.

We remark here that specifying $u\traa{\ff}$ is analogous to setting (part of) a characteristic Cauchy problem in the sense that the conormal of $\ff$ lies in the characteristic set of $\Box_{\rho^2 g}$, this bears thus some resemblance to the construction used in \cite{Mo1,Mo2,characteristic} in the case of the conformal wave equation.

\section{Asymptotically de Sitter spacetimes}\init

\subsection{Geometrical setup}

The proof of the Fredholm property of the rescaled wave operator $P$ on asymptotically Minkowski spacetimes in \cite{BVW,semilinear,GHV} is based on a careful analysis of the Mellin transformed normal operator family $\widehat{N}(P)(\sigma)$, which is a holomorphic family of differential operators on the compact manifold $\p M$. Recall also that we used results from \cite{BVW} on module regularity of solutions of $P$, these in turn are based on the Mellin transformed version of the operator $P$. The relevant property is that for fixed $\sigma$ one has an elliptic operator in the two connected components of the region $v>0$ and a hyperbolic one in $v<0$. Furthermore, in the respective regions they can be related to the Laplacian on an asymptotically hyperbolic space and to the wave operator on an asymptotically de Sitter space by conjugation with powers of the boundary-defining functions of $S_\pm$, with $S=S_+\cup S_-$ playing the role of the asymptotically de Sitter conformal boundary. In this section we will be interested in the reverse construction, which extends a given  asymptotically de Sitter space $X_0$ (conformally compactified, with conformal boundary $S=S_+\cup S_-$) to a compact manifold $X$, and relates the Klein-Gordon operator  on the asymptotically de Sitter region to a differential operator $\hat P_X$ defined on the whole `extended' manifold $X$. The main merit of this construction is that $\hat P_X$ acts on a manifold without boundary and more importantly it fits into the framework of \cite{kerrds,semilinear}, with bicharacteristics beginning and ending at the radial sets located above $S_+$ and $S_-$.

These various relations are explained in more detail in \cite{resolvent,poisson}. Here as an illustration we start with the special case of actual $n=1+d$-dimensional Minkowski space $\rr^{1+d}$ with metric $g_{\rr^{1,d}}=d z_{0}^2-(d z_1^2+\dots+ d z_{d}^2)$. Its radial compactification is a compact manifold $M$ with boundary $\p M=\ss^{d}$, and with $\rho=(z_0^2+\dots +z_d^2)^{-1/2}$ the boundary defining function, Mellin transforming the rescaled wave operator $P=\rho^{-(d-1)/2}\rho^{-2} \Box_g \rho^{(d-1)/2}$ yields a ($\sigma$-dependent) differential operator $\hat P_{\p M}$ on the boundary $\p M$
\[
\hat P_{\p M}(\sigma) \defeq \cM_\rho \rho^{-(d-1)/2}\rho^{-2} \Box_g \rho^{(d-1)/2} \cM_\rho^{-1} \in\Diff^2(\p M).  
\]
Now the crucial observation is that the region in the boundary $\ss^d$ corresponding to $z_1^2+\dots + z_d^2 > z_{0}^2$ in the interior can be identified with the de Sitter hyperboloid $z_{0}^2-(z_1^2+\dots + z_d^2) =-1$. The latter is a manifold that we denote $X_0$ and which is equipped with the de Sitter metric $g_{X_0}$, related to the Minkowski metric by 
\[
g_{\rr^{1,d}}=-dr_{X_0}^2 + r_{X_0 }^2 g_{X_0}=\frac{1}{\rho^2}\bigg( -x^2_{X_0}\left(-\frac{d\rho}{\rho}+\frac{dx_{X_0}}{x_{X_0}}\right)^2+x_{X_0}^2 g_{X_0}\bigg),
\]
where $r_{X_0}=(z_1^2+\dots + z_d^2-z_0^2)^{1/2}$ is the space-like Lorentzian distance function and
\[
x_{X_0}=\left(\frac{z_1^2+\dots + z_d^2 - z_0^2}{z_1^2+\dots + z_d^2 + z_0^2}\right)^{\12}=r_{X_0}\rho.
\] 
Here we consider the de Sitter space $X_0$ as a manifold with boundary $S=S_+\cup S_-$ (this is the so-called \emph{conformal boundary} of $X_0$) and boundary-defining function $x_{X_0}$.

Remarkably, as shown in \cite{poisson}, $\hat P_{\p M}(\sigma)$ is related to the (Laplace-Beltrami) wave operator on $X_0$ by\footnote{Note that this differs from the formulas in \cite{poisson} by a sign in front of $\sigma$, because there the Mellin transform is taken with respect to $\rho^{-1}$ instead of $\rho$.}
\[
\hat P_{\p M}(\sigma)\traa{X_0}=x^{-\i\sigma-(d-1)/2-2}_{X_0}(\Box_{X_0} - \sigma^2 -(d-1)^2/4) x_{X_0}^{\i\sigma+(d-1)/2}.
\]
In turn, the two connected regions on the boundary $\ss^{d}$ that correspond to $|z_{0}|^2>z_1^2+\dots + z_d^2$ and respectively $\pm z_0>0$ in the interior of $M$ can be identified with the two hyperboloids
\[
z_{0}^2-(z_1^2+\dots + z_d^2) =1, \ \ \pm z_0>0.
\]
These hyperboloids are in fact two copies of hyperbolic space. Here, in the compactified setting, we consider them as two manifolds $X_\pm$ with boundary $\p X_\pm=S_\pm$, with metric $g_{X_\pm}$ satisfying
\[
g_{\rr^{1,d}}=dr_{X_\pm}^2 - r_{X_\pm}^2 g_{X_\pm}=-\frac{1}{\rho^2}\bigg( -x^2_{X_\pm}\left(-\frac{d\rho}{\rho}+\frac{dx_{X_\pm}}{x_{X_\pm}}\right)^2+x_{X_\pm}^2 g_{X_\pm}\bigg),
\]
with $r_{X_+}=r_{X_-}=(z_0^2-z_1^2+\dots + z_d^2)^{1/2}$ the time-like
Lorentzian distance function and $x_{X_\pm}=r_{X_\pm}\rho$; note that
the pull-back of the Minkowski metric to the hyperboloid is the
negative of the Riemannian metric. Similarly as in the case of $X_0$, one has an identity relating $\hat P_{\p M}$ to the Laplace-Beltrami operator on $X_\pm$:
\[
\hat P_{\p M}(\sigma)\traa{X_\pm}=x^{-\i\sigma-(d-1)/2-2}_{X_{\pm}}(-\Delta_{X_{\pm}} + \sigma^2 +(d-1)^2/4) x^{\i\sigma+(d-1)/2}_{X_{\pm}}.
\]

We now consider the more general setup of asymptotically hyperbolic and asymptotically de Sitter spacetimes (note that the latter have to be thought as a generalization of `global' de Sitter space, as opposed  for instance to the static or cosmological de Sitter patch), following \cite{resolvent,poisson}.

\begin{definition}\label{def:as}Let $X_\bullet$ be a compact $d$-dimensional manifold with boundary, equipped with a metric $g$ on $X^\inti_\bullet$, and let $x$ be a boundary defining function. One says that $(X_\bullet,g)$ is:
\begin{itemize}
\item[--] \emph{asymptotically hyperbolic} if $g=x^{-2}\hat g$, where $\hat g$ is a smooth Riemannian metric on $X_\bullet$ with $\hat g(dx,dx)\traa{x=0}=1$;
\item[--] \emph{asymptotically de Sitter} if $g=x^{-2}\hat g$, where $\hat g$ is a smooth Lorentzian metric on $X_\bullet$ of signature $(1,d-1)$, with $\hat g(dx,dx)\traa{x=0}=1$, and the boundary is the union $\p X_\bullet=S_+ \cup S_-$ of two connected components, with all null geodesics in $X^\inti_\bullet$ parametrized by $t\in\rr$ tending either to $S_+$ as $t\to\infty$ and to $S_-$ as $t\to-\infty$, or vice versa.
\end{itemize}
\end{definition}

An argument from \cite{semilinear} (discussed therein for a class of asymptotically Minkowski spacetimes) can be used to show that if $(X_0,g_{X_0})$ is asymptotically de Sitter then $(X^\inti_0,g_{X_0})$ is globally hyperbolic. Moreover, it is well-known that  $X_0$ diffeomorphic to $[-1,1]\times {S_+}$ (and to $[-1,1]\times {S_-}$). 

Furthermore, one says that an asymptotically de Sitter space $(X_0,g_{X_0})$ is \emph{even} if it admits a product decomposition $[0,\epsilon)_x\times(\p {X_0})_y$ near $\p X_0$ such that
\beq\label{eq:pd1}
g_{X_0}=\frac{dx_{X_0}^2- h(x^2_{X_0},y,dy)}{x^2_{X_0}}
\eeq 
with $h(x^2_{X_0},y,dy)$ smooth. In a similar way (but with different sign in front of $h$) one defines even asymptotically hyperbolic spaces \cite{resolvent,poisson}, cf. also the work of Guillarmou \cite{guillarmou} for the original definition. It can be shown that the product decomposition (\ref{eq:pd1}) is a general feature of asymptotically de Sitter spacetimes (this is analogous to the Riemannian case treated in \cite{GL}), so the essential property in the definition of even spaces is smoothness of $h(x^2_{X_0},y,dy)$. For us what matters the most is that this amounts to requiring that $h$ is smooth with respect to a $\cf$ structure on $X$, modified with respect to the original one in such way that  $v\defeq-x_{X_0}^2$ is a valid boundary-defining function (we call it the \emph{even $\cf$ structure} on $X_0$). 

Now, suppose we are given an even asymptotically de Sitter space $(X_0,g_{X_0})$, two even asymptotically hyperbolic spaces $(X_\pm,g_{X_\pm})$ with boundary defining functions $x_{X_\pm}$, and a compact manifold $X$ (without boundary) of the form
\[
X=X_+\cup X_0\cup X_-,
\] 
where $\p X_\pm$ is smoothly identified with the component $S_\pm$ of the boundary $\p X_0=S$ of $X_0$. Next, equipping $X$ with the even $\cf$ structure on the respective components allows one to construct an asymptotically Minkowski space $(M,g)$ with $M=\rr_\rho^+\times X$ (so that $\p M = X$) and $g$ a smooth metric of the form
\[
g=\frac{1}{\rho^2} \left(v \frac{d\rho^2}{\rho^2}-\12\left(\frac{d\rho}{\rho}\otimes dv + d v \otimes\frac{d\rho}{\rho} \right) - h(-v,y,dy)\right)
\]
with $v=-x_{X_0}^2$ on $X_0$ and $v=x_{X_\pm}^2$ on $X_\pm$. The Mellin transformed (rescaled) wave operator on $M$ defines a family of differential operators $\hat P_X(\sigma)\in\Diff^2(X)$ which is related to the Laplace-Beltrami (wave) operator on $X_\pm$ and $X_0$ by  
\beq\label{eq:defphats}
\hat P_X(\sigma)\traa{X_0^\inti}=  x_{X_0}^{\i\tilde\sigma-2} \hat P_{X_0}(\sigma) x_{X_0}^{-\i\tilde\sigma}, \quad \hat P_X(\sigma)\traa{X_{\pm}^\inti}=  x_{X_\pm}^{\i\tilde\sigma-2} \hat P_{X_{\pm}}(\sigma) x_{X_\pm}^{-\i\tilde\sigma},
\eeq
where we have set $\tilde\sigma=-\sigma+\i(d-1)/2$ and
\beq
\hat P_{X_0}(\sigma) = \Box_{X_{0}}-\sigma^2-(d-1)^2/4, \quad \hat P_{X_\pm}(\sigma) = -\Delta_{X_{\pm}}+\sigma^2+(d-1)^2/4.
\eeq

On $X_0$ and $X_\pm$ we consider the respective volume densities. On $X$ there is a unique smooth density which extends the volume form on $X_0$ and $X_\pm$, multiplied first by the conformal factor $v^{(d+1)/2}$. We denote by $\bra \cdot,\cdot\ket_{X_0}$, $\bra \cdot,\cdot\ket_{X_\pm}$, $\bra \cdot,\cdot\ket_{X}$ the pairings induced from the respective densities. 

Then, we have that $\hat P_{X}(\bar\sigma)$ is the formal adjoint of $\hat P_{X}(\sigma)$ with respect to $\bra\cdot,\cdot\ket_{X}$ (see \cite[Sec.~3.1]{resolvent}), similarly $\hat P_{X_\bullet}(\bar\sigma)$ is the formal adjoint of $\hat P_{X_\bullet}(\sigma)$ with respect to $\bra\cdot,\cdot\ket_{X_\bullet}$.

Turning our attention to inverses, by global hyperbolicity of $(X_0,g_0)$, it is well known that $\hat P_{X_0}(\sigma)$ has advanced and retarded propagators\footnote{This means here that $\hat P_{X_0,\pm}(\sigma)^{-1}$ are the inverses of $\hat P_{X_0,\pm}(\sigma)$ that solve respectively the advanced, retarded inhomogeneous problem.} $\hat P_{X_0,\pm}(\sigma)^{-1}$ for any value of $\sigma$. The two operators $\hat P_{X_\pm}(\sigma)$ possess inverses $\hat P_{X_\pm}(\sigma)^{-1}$ for sufficiently large values of $|\Im\,\sigma|$ in the sense of the resolvent of the positive operator $-\Delta_{X_{\pm}}$ (on the closure of its natural domain in $L^2$), and moreover it was shown in \cite{guillarmou,MM,resolvent} that $\hat P_{X_\pm}(\sigma)^{-1}$ continues from say $\Im\sigma\gg 0$ to $\cc$ as a meromorphic family of operators (cf. also \cite{zworski} for a recent, more concise account).

On the other hand, $\hat P_X(\sigma)$ fits into the framework of \cite{kerrds}, which allows to set up a Fredholm problem in the spaces
\[
\cX^s=\{ u \in H^s(X): \ \hat P_X(\sigma)u\in \cY^{s-1} \}, \quad \cY^{s-1}=H^{s-1}(X),
\]
with the conclusion that $\hat P_X(\sigma):\cX^s\to\cY^{s-1}$ possess in particular two inverses $\hat P_{X,\pm}(\sigma)^{-1}$ in the sense of meromorphic families of operators, where the sign $+$ corresponds to requiring above-threshold regularity $s>\12 - \Im\,\sigma$ near $N^*S_+$ and below-threshold regularity $s<\12 - \Im\,\sigma$ near $N^*S_-$, while the sign $-$ corresponds to the same conditions with $N^*S_+$ and $N^*S_-$ interchanged. In a similar vein one can define Feynman and anti-Feynman inverses (as pointed out in \cite{positive}), we have thus four inverses $\hat P_{X,I}(\sigma)^{-1}$. Focusing  our attention on retarded and advanced ones, it is proved in \cite{poisson} that just as the identities (\ref{eq:defphats}) suggest, with additional subtleties in the sign of $\sigma$ (corresponding to whether the inverse is defined by analytic continuation from $\Im\,\sigma\gg 0$ or from $\Im\,\sigma\ll 0$), it holds that
\beq\label{eq:defphats2}
\bea
\hat P_{X,\pm}(\sigma)^{-1}\traa{X_0^\inti\to X_0^\inti}&= x_{X_0}^{\i\tilde\sigma} \hat P_{X_0,\pm}(\sigma)^{-1} x_{X_0}^{-\i\tilde\sigma+2}, \\
\hat P_{X,+}(\sigma)^{-1}\traa{X_{\pm}^\inti\to X_{\pm}^\inti}&= x_{X_\pm}^{\i\tilde\sigma} \hat P_{X_{\pm}}(\sigma)^{-1} x_{X_\pm}^{-\i\tilde\sigma+2},\\
\hat P_{X,-}(\sigma)^{-1}\traa{X_{\pm}^\inti\to X_{\pm}^\inti}&= x_{X_\pm}^{\i\tilde\sigma} \hat P_{X_{\pm}}(-\sigma)^{-1} x_{X_\pm}^{-\i\tilde\sigma+2},
\eea
\eeq
away from poles of $\hat P_{X,\pm}(\sigma)^{-1}$ and $\hat P_{X_{\pm}}(\sigma)^{-1}$. Here the subscript ${}\traa{X_{\bullet}^\inti\to X_\bullet^\inti}$ means that we act with $\hat P_{X,\pm}(\sigma)^{-1}$ on $\cf(X_\bullet)$ and restrict the result to the interior of $X_\bullet$, so (\ref{eq:defphats2}) contains no direct information on how $\hat P_{X,\pm}(\sigma)^{-1}$ acts between different components of $X$.  

To derive a more precise relation, \cite{poisson} makes use of asymptotic data of solutions at the common boundaries of $X_0$ and $X_\pm$. Here we will discuss the corresponding symplectic spaces in a similar way as in Subsect. \ref{ssec:asymptoticdata}, starting first with the analogues of the space of solutions smooth away from the radial set (we focus here mainly on the spaces defined using the advanced and retarded propagator).

\subsection{Symplectic spaces of solutions} Assuming $\sigma\in\rr$, the symplectic spaces associated to $\hat P_X(\sigma)$ and the various isomorphisms between them can in fact be introduced in a very similar fashion as in the asymptotically Minkowski case. We denote by ${\Sol}(\hat P_{X}(\sigma))$ the space of solutions of $\hat P_X(\sigma)u=0$ such that $\wf(u)\subset N^* S$, and set 
\beq\label{eq:okmmj}
\hat G_X(\sigma)\defeq\hat P_{X,+}(\sigma)^{-1}-\hat P_{X,-}(\sigma)^{-1}.
\eeq
From now on the dependence on $\sigma$ will often be skipped in the notation, we stress however that we always make the implicit assumption that $\sigma$ is not a pole of the two operators $\hat P_{X,+}(\sigma)^{-1}$, $\hat P_{X,-}(\sigma)^{-1}$. Using essentially the same arguments as before (this is even in many ways simpler due to $\hat P_{X,\pm}^{-1}$ being exact inverses of $\hat P_X$) we get a bijection
\beq\label{eq:usioo1}
\frac{\cf(X)}{\hat P_{X} \cf(X)}\xrightarrow{\makebox[2em]{$[\hat G_{X}]$}}{\Sol}(\hat P_{X}).
\eeq
Furthermore, $\bra \cdot,  \hat G_X \cdot \ket_X$ induces a well-defined sesquilinear form on ${\cf(X)}/{\hat P_{X} \cf(X)}$, and since $(\hat P_{X,+}^{-1})^* =\hat P_{X,-}^{-1}$ by \cite{positive}, $\hat G_X$ is anti-hermitian.  Although the method of proof of (\ref{eq:usioo1}) is fully analogous to the case of asymptotically Minkowski spacetimes, we stress that the physical outcome is much more unusual, as it allows to build a non-interacting quantum field theory governed by a differential operator that is not everywhere hyperbolic. Note also that one can obtain an analogue of \eqref{eq:okmmj} in the `Feynman minus anti-Feynman' case.

In turn, the discussion of symplectic spaces on $X_0$ is rather standard due to global hyperbolicity of the interior. Let ${\Sol}(\hat P_{X_0})$ be the space of solutions of $\hat P_{X_0}u=0$ that are smooth in the interior $X_0^\inti$.  Setting $\hat G_{X_0}\defeq \hat P^{-1}_{X_0,+}-\hat P^{-1}_{X_0,-}$, one gets isomorphisms
\beq\label{eq:usio1}
\frac{\cf_\c(X_0^\inti)}{\hat P_{X_0} \cf_\c(X_0^\inti)}\xrightarrow{\makebox[2em]{$[\hat G_{X_0}]$}}{\Sol}(\hat P_{X_0}),
\eeq
either by using well-known results (see for instance \cite{BGP}) or by repeating the proof of the asymptotically Minkowski case. As in \eqref{eq:invGI}, the inverse of the isomorphism \eqref{eq:usio1} is the operator $[\hat P_{X_0},Q]$, where $Q\in\cf(X_0)$ equals $0$ in a neighborhood of $S_+$ and $1$ in a neighborhood of $S_-$. 

The next proposition shows that the symplectic spaces (\ref{eq:usioo1}) and (\ref{eq:usio1}) are in fact isomorphic, so the content of a QFT on $X$ is induced by a QFT in the asymptotically de Sitter region.

\begin{proposition}\label{prop:isoxxo} We have isomorphisms
\beq\label{eq:usio2}
\frac{\cf(X)}{\hat P_{X}\cf(X)}\xrightarrow{\makebox[2em]{$[\hat G_X]$}}\hat G_X \cf_\c(X_0^\inti)\xrightarrow{\makebox[2em]{${}\traa{X_0}$}}(\hat G_X \cf_\c(X_0^\inti))\traa{X_0}\xrightarrow{\makebox[2em]{$x_{X_0}^{-\i\tilde\sigma}$}}{\Sol}( \hat P_{X_0}).
\eeq
\end{proposition}
\proof By \eqref{eq:usioo1}, to prove bijectivity of the first arrow we need to show that $\Sol(\hat P_X)\subset \hat G_X \cf_\c(X_0^\inti)$ (the other inclusion is straightforward).  Let $Q\in\cf(X)$ be equal $0$ in a neighborhood of $X_+$ and $1$ in a neighborhood of $X_-$. Then as in the proof of Propositions \ref{prop:exact}, we can show that
\[
\hat G_X [\hat P_X, Q ]=\one \ \mbox{ on } \Sol(\hat P_X).
\]
Since $[\hat P_X,Q]$ is supported in the interior of $X_0$, this implies that $\Sol(\hat P_X)\subset \hat G_X \cf_\c(X_0^\inti)$. 

To prove that the second arrow is bijective, we use the expression for $\hat G_X$ resulting from (\ref{eq:defphats2}). Specifically, if $f\in\cf_\c(X_0^\inti)$ then
\beq\label{eq:temslook}
 (\hat G_X f)\traa{X_0}=x_{X_0}^{\i\tilde\sigma} \hat G_{X_0} x_{X_0}^{-\i\tilde\sigma+2}f.
\eeq
By the isomorphism (\ref{eq:usio1}) this entails that $(\hat G_X f)\traa{X_0}$ determines $f$ modulo $\hat P_X \cf_\c(X_0^\inti)$, and therefore determines $\hat G_X f$ uniquely.

Bijectivity of the third arrow follows immediately from $\hat G_{X_0} \cf_\c(X_0^\inti)={\Sol}(\hat P_{X_0})$ (this is surjectivity of the first arrow in (\ref{eq:usio1})) and (\ref{eq:temslook}).\qeds

In summary, we have an isomorphism
\beq\label{eq:theiso}
\frac{\cf(X)}{\hat P_{X}\cf(X)}\xrightarrow{\makebox[2.5em]{$[R_{X_0}]$}}\frac{\cf_\c(X_0^\inti)}{\hat P_{X_0} \cf_\c(X_0^\inti)} 
\eeq
given by $R_{X_0}=[\hat G_{X_0}]^{-1} x_{X_0}^{-\i\tilde\sigma}({}\traa{X_0^\inti}\circ\, \hat G_X)$  (where $[\hat G_{X_0}]^{-1}=[\hat P_{X_0},Q]$, with  $Q\in\cf(X_0)$ being equal $0$ in a neighborhood of $S_+$ and $1$ in a neighborhood of $S_-$).

\subsection{Hadamard states}

We now discuss how the relation between symplectic spaces on $X_0$ and $X$ translates to the level of two-point functions. We denote $\hat\Sigma$ the characteristic set of $\hat P_X$ and $\hat \Sigma^\pm$ its two connected components.

In the region $X_0$ it is quite clear what a Hadamard two-point function is, we can adopt Definition \ref{def:hadamard} quite directly indeed and say that $\Lambda_{X_0}^{\pm}:\cf_\c(X_0^\inti)\to\cf(X_0^\inti)$ are (bosonic) Hadamard two-point function for $\hat P_{X_0}$ if
\beq
\hat P_{X_0}\Lambda_{X_0}^\pm=\Lambda_{X_0}^{\pm}\hat P_{X_0}=0, \ \ \Lambda_{X_0}^{+}-\Lambda_{X_0}^{-}=\i\hat G_{X_0}, \ \ \Lambda_{X_0}^{\pm}\geq 0
\eeq 
and $\wf'(\Lambda^\pm_{X_0})= \cup_{t\in\rr}\hat\Phi_t(\diag_{T^*X_0^\inti})\cap\pi^{-1}\hat\Sigma^\pm$, where $\hat\Phi_t$ is the bicharacteristic flow of $\hat P_{X_0}$ and $\pi:\hat\Sigma\times\hat\Sigma\to\hat\Sigma$ projects to the left component.
This ensures that $\Lambda_{X_0}^{\pm}$ induce well-defined hermitian forms on $\cf_\c(X_0^\inti)/\hat P_{X_0} \cf_\c(X_0^\inti)$, and agrees with the standard definition of Hadamard two-point functions on globally hyperbolic spacetimes \cite{radzikowski}.  

A similar definition can be used on $X$, the precise form of which is dictated by the behavior of the bicharacteristic flow.

\begin{definition}\label{def:onX} We say that $\Lambda_{X}^{\pm}:\cf(X)\to\cmf(X)$ are Hadamard two-point functions for $\hat P_{X}(\sigma)$ if $\hat P_{X}\Lambda_{X}^{\pm}=\Lambda_{X}^{\pm}\hat P_{X}=0$, $\Lambda_{X}^{+}-\Lambda_{X}^{-}=\i\hat G_X$, $\Lambda_{X}^{\pm}\geq 0$ with respect to $\bra \cdot,\cdot\ket_X$,  and 
\beq\label{eq:LXpm}
\wf'(\Lambda_{X}^{\pm})\subset\big( \cup_{t\in\rr}\hat\Phi_t(\diag_{T^*X})\cap\pi^{-1}\hat\Sigma^\pm  \big)\cup (\,\zero\times N^*S)\cup( N^*S\times \zero),
\eeq
where $\hat\Phi_t$ is the bicharacteristic flow of $\hat P_{X}$ and $\pi:\hat\Sigma\times\hat\Sigma\to\hat\Sigma$ is the projection to the left component.
\end{definition}

As a consequence of Proposition \ref{prop:isoxxo}, Hadamard states on $X_0$ extend to Hadamard states on $X$ in the following sense:

\begin{theorem}\label{thm:frdtox} Let $(X_0,g_{X_0})$ be an even asymptotically de Sitter space and let $\Lambda_{X_0}^{\pm}$ be Hadamard two-point functions for $\hat P_{X_0}(\sigma)$. If $\sigma$ is not a pole of $\hat P_{X_+}(\sigma)^{-1}$ nor of $\hat P_{X_-}(\sigma)^{-1}$ then $\Lambda_{X_0}^\pm$ induce canonically two-point functions $\Lambda_{X}^{\pm}$ of a Hadamard state for $\hat P_X(\sigma)$ via the isomorphism (\ref{eq:theiso}).
\end{theorem}
\proof The isomorphism \eqref{eq:theiso} induces a pair of operators $\Lambda_{X}^{\pm}:\cf(X)\to\cf(X)$, namely
\[
\Lambda_{X}^{\pm} = R_{X_0}^*\Lambda_{X_0}^\pm R_{X_0}.
\]
It is easy to see that it satisfies $\hat P_{X}\Lambda_{X}^{\pm}=\Lambda_{X}^{\pm}\hat P_{X}=0$, $\Lambda_{X}^+-\Lambda_{X}^-=\i\hat G_X$ and $\Lambda_{X}^{\pm}\geq 0$. Furthermore, $\Lambda_{X}^{\pm}\traa{X_0^\inti\to X_0^\inti}=x_{X_0}^{\i\tilde\sigma}\Lambda_{X_0}^{\pm} x_{X_0}^{-\i\tilde\sigma+2}$, so by assumption
\[
\wf'(\Lambda_{X}^{\pm})\cap (T^*X_0^\inti \times T^*X_0^\inti) = \cup_{t\in\rr}\hat\Phi_t(\diag_{T^*X})\cap\pi^{-1}\hat\Sigma^\pm .
\]
By elliptic regularity and propagation of singularities for $\hat P$ (see \cite{kerrds}) applied componentwise, we can estimate the wave front set above $X_\pm$ modulo possible terms in $\zero\times S^*X$ and $S^*X\times\,\zero$, namely: 
\beq\label{eq:tempwfx}
\bea
\wf'(\Lambda_{X}^{\pm})\subset \big(\cup_{t\in\rr}\hat\Phi_t(\diag_{T^*X})\cap\pi^{-1}\hat\Sigma^\pm \big)&\cup (\,\zero\times S^*X_+)\cup( S^*X_+\times \zero)\\
&\cup (\,\zero\times S^*X_-)\cup( S^*X_-\times \zero).
\eea
\eeq
Furthermore,  using positivity of $\Lambda_{X}^{\pm}$, for any test functions $f,g$ we can write a Cauchy-Schwarz inequality to estimate $|\bra f,\Lambda_X^\pm g\ket_X|$ in terms of $|\bra f,\Lambda_X^\pm f\ket_X|$ and $|\bra g, \Lambda_X^\pm g\ket_X|$. Therefore we can get estimates for the wave front set in $\zero\times (T^*X \setminus \zero)$ from estimates in the diagonal of $(T^*X \setminus \zero)\times (T^* X \setminus \zero)$, and also get a symmetrized form of the wave front set. In view 
of (\ref{eq:tempwfx}) and taking into account that $\hat\Sigma\cap \{ v\geq 0\} = N^*S$, we can apply this argument outside of $(\,\zero\times N^*S)\cup( N^*S\times \zero)$. This gives
\[
\wf'(\Lambda_{X}^{\pm})\subset\big( \cup_{t\in\rr}\hat\Phi_t(\diag_{T^*X})\cap\pi^{-1}\hat\Sigma^\pm  \big)\cup (\,\zero\times N^*S)\cup( N^*S\times \zero)
\]
as claimed. \qeds

In Subsection \ref{SubsectasX} we will construct two-point functions that actually satisfy a stronger estimate on the wave front set than \eqref{eq:LXpm}, see \eqref{eq:betterwf}.

\subsection{Asymptotic data on $X_0$ and $X_\pm$}

We now turn our attention to asymptotic data for solutions of $\hat P_{X_0}$ and $\hat P_{X_\pm}$, assuming $\sigma\in\rr$. Recall that ${\Sol}(\hat P_{X_0})$ is the space of solutions of $\hat P_{X_0} u=0$ that are smooth in the interior of $X_0$. By the results of \cite{poisson,desitter}, each solution $u\in{\Sol}(\hat P_{X_0})$ can be written in the form
\[
u= \tilde a^+_{X_{0}} x_{X_{0}}^{-\i\sigma+(d-1)/2} + \tilde a^-_{X_{0}} x_{X_{0}}^{\i\sigma+(d-1)/2}, \ \ \tilde a^\pm_{X_{0}}\in\cf(X_0).
\]
In order to have a similar structure on the two asymptotically hyperbolic spaces $X_\pm$, we define ${\Sol}(\hat P_{X_\pm})$ to be the space of solutions of $\hat P_{X_\pm} u=0$ that can be written as
\[
u= \tilde a^+_{X_{\pm}} x_{X_{\pm}}^{-\i\sigma+(d-1)/2} + \tilde a^-_{X_{\pm}} x_{X_{\pm}}^{\i\sigma+(d-1)/2},  \ \ \tilde a^+_{X_{\pm}},\tilde a^-_{X_{\pm}}\in\cf(X_\pm).
\]
In the case  $u\in{\Sol}(\hat P_{X_0})$, $u$ is uniquely determined by its asymptotic data $\varrho_{X_0,+} u$ at ${S_+}$, and the same is true for the $\varrho_{X_0,-}u$ data at ${S_-}$, where 
\[ 
\varrho_{X_0,\pm} u = (\varrho_{X_0,\pm}^+u,\varrho_{X_0,\pm}^-u)\defeq ( \tilde a^+_{X_0}\traa{{S_\pm}},\tilde a^-_{X_0}\traa{{S_\pm}})\in \cf({S_\pm})\oplus \cf({S_\pm}).
\]
On the other hand, as follows from the results in \cite{MM,JSB,poisson}, in each of the cases  $u\in{\Sol}(\hat P_{X_\pm})$, there are two maps $\varrho_{X_\pm}^+$ and $\varrho_{X_\pm}^-$ defined by
\[
\varrho_{X_0,\pm}^+ u\defeq \tilde a^+_{X_\pm}\traa{\p X_\pm}, \ \ \varrho_{X_0,\pm}^- u\defeq \tilde a^-_{X_\pm}\traa{\p X_\pm}.
\]
Here, \emph{any} of the two possible data $\varrho_{X_\pm}^+ u$ or $\varrho_{X_\pm}^- u$ determines $u$ uniquely. The inverse of $\varrho_{X_0,\pm}$, resp. $\varrho_{X,\pm}^+$, $\varrho_{X,\pm}^-$ is the Poisson operator denoted $\cP_{X_0,\pm}$, resp. $\cP_{X_\pm}^+$, $\cP_{X_\pm}^-$. Note that changing the sign of $\sigma$ inverses one type of data with the other, thus, displaying the dependence on $\sigma$ explicitly, 
\[
\varrho_{X_\pm}^-(\sigma)=\varrho_{X_\pm}^+(-\sigma), \ \ \cP_{X_\pm}^-(\sigma)=\cP_{X_\pm}^+(-\sigma).
\]
More details on the construction of the various Poisson operators and the relation between them can be found in \cite{poisson} and references therein. \medskip

We now have all the necessary ingredients to state the result from \cite{poisson} that describes how $\hat P_{X,\pm}^{-1}$ acts on different components of $X$. Recall that $\tilde\sigma=-\sigma+\i(d-1)/2$, and that with the conventions in this paper the subscript `$+$' vs. `$-$' in $\hat P_{X_0,\pm}^{-1}$ refers to `advanced' vs. `retarded' (i.e.~`propagating support to the past' vs. `to the future').

\begin{theorem}[\cite{poisson}]\label{thm:preciseform} The inverse $\hat P_{X,-}(\sigma)^{-1}$ exists as a meromorphic family in $\sigma$, and its poles in $\cc\setminus \i\zz$ are precisely the union of the poles of $\hat P_{X_+}(\sigma)^{-1}$ and $\hat P_{X_+}(-\sigma)^{-1}$. Furthermore,
\[
\hat P_{X,+}(\sigma)^{-1}=\begin{pmatrix} x_{X_+}^{\i\tilde\sigma}\hat P_{X_+}(\sigma)^{-1} x_{X_+}^{-\i\tilde\sigma+2} & 0 & 0 \\ 
x_{X_0}^{\i\tilde\sigma} c_{0,+}(\sigma) x_{X_+}^{-\i\tilde\sigma+2} & x_{X_0}^{\i\tilde\sigma} \hat P_{X_0,+}^{-1}(\sigma) x_{X_0}^{-\i\tilde\sigma+2} & 0 \\
x_{X_-}^{\i\tilde\sigma} c_{-,+}(\sigma)  x_{X_+}^{-\i\tilde\sigma+2} & x_{X_-}^{\i\tilde\sigma} c_{-,0}(\sigma) x_{X_0}^{-\i\tilde\sigma+2}  & x_{X_-}^{\i\tilde\sigma} \hat P_{X_-}(-\sigma)^{-1} x_{X_-}^{-\i\tilde\sigma+2}
\end{pmatrix}
\]
where
\[
\bea
c_{0,+}(\sigma)&=\cP_{X_0,+}\imath^- \varrho_{X_+}^-\hat P_{X_+}(\sigma)^{-1},\\
c_{-,+}(\sigma)&=\cP_{X_-}^-(\imath^-)^*\varrho_{X_0,-} c_{0,+}(\sigma),\\
c_{-,0}(\sigma)&=\cP_{X_-}^-(\imath^-)^* \varrho_{X_0,+} \hat P_{X_0,-}(\sigma)^{-1},
\eea
\]
and $\imath^{\pm}:\cf(\p_\bullet X_0)\to \cf(\p_\bullet X_0)\oplus\cf(\p_\bullet X_0)$ is the left/right embedding. 
The matrix notation above means that given $f\in\cf(X)$ there is a unique distribution $u$ with $\hat P_{X,+}(\sigma)^{-1} f=u$ and such that $(u\traa{X_+}, u\traa{X_0},u\traa{X_-})$ equals the matrix of $\hat P_{X,+}(\sigma)^{-1}$ applied to $(f\traa{X_+}, f\traa{X_0},f\traa{X_-})$.

\end{theorem}

There is an analogous statement for $\hat P_{X,-}^{-1}(\sigma)$, namely, it is a meromorphic family whose poles in $\cc\setminus \i \zz$ are precisely the union of the poles of $\hat P_{X_+}(\sigma)^{-1}$ and $\hat P_{X_+}(-\sigma)^{-1}$, and
\[
\hat P_{X,-}(\sigma)^{-1}=\begin{pmatrix} x_{X_+}^{\i\tilde\sigma}\hat P_{X_+}(-\sigma)^{-1} x_{X_+}^{-\i\tilde\sigma+2} & x_{X_+}^{\i\tilde\sigma} c_{+,0}(\sigma) x_{X_0}^{-\i\tilde\sigma+2} & x_{X_+}^{\i\tilde\sigma} c_{+,-}(\sigma) x_{X_-}^{-\i\tilde\sigma+2} \\ 
0 & x_{X_0}^{\i\tilde\sigma} \hat P_{X_0,-}^{-1}(\sigma) x_{X_0}^{-\i\tilde\sigma+2} & x_{X_0}^{\i\tilde\sigma} c_{0,-}(\sigma) x_{X_-}^{-\i\tilde\sigma+2} \\
0 & 0 & x_{X_-}^{\i\tilde\sigma} \hat P_{X_-}(\sigma)^{-1} x_{X_-}^{-\i\tilde\sigma+2}
\end{pmatrix}
\]
using the same matrix notation, where
\[
\bea
c_{0,-}(\sigma)&=\cP_{X_0,-}\imath^- \varrho_{X_-}^-\hat P_{X_-}(\sigma)^{-1},\\
c_{+,-}(\sigma)&=\cP_{X_+}^-(\imath^-)^*\varrho_{X_0,+}c_{0,-}(\sigma),\\
c_{+,0}(\sigma)&=\cP_{X_+}^-(\imath^-)^* \varrho_{X_0,-} \hat P_{X_0,+}(\sigma)^{-1}.
\eea
\]
In particular, $\hat P_{X,\mp}^{-1} f$ is supported in $X_\pm$ if $f$ is supported in $X_\pm$, and $\hat P_{X,\mp}^{-1} f$  is supported in $X_\pm\cup X_0$ if $f$ is supported in $X_\pm\cup X_0$ (this weaker statement was already proved in \cite{BVW}).   

Recall also that if $\sigma\in\rr$ then $\hat P_{X,+}^*=\hat P_{X,-}$ with respect to $\bra \cdot, \cdot \ket_X$, so as an aside, we conclude immediately
\[
c_{0,-}^*=c_{-,0}, \ \ c_{+,-}^*=c_{-,+}, \ \ c_{+,0}^*=c_{-,0},
\] 
where the adjoints are taken using the respective the scalar products $\bra \cdot, \cdot\ket_{X_\bullet}$.

Theorem \ref{thm:preciseform} allows us to give a formula for the extension to $X$ of the two-point functions  by means of its asymptotic data at future infinity (and an analogous statement holds for $\varrho_{X_0,-}$ data).  

\begin{proposition}\label{prop:fraelj} Let $\Lambda_{X_0}^\pm$ be two-point functions for $\hat P_{X_0}$ of the form
\beq\label{eq:fraeljk1}
\Lambda_{X_0}^{\pm}=\hat G_{X_0}^* \varrho_{X_0,+}^* \lambda^\pm_{X_0,+} \varrho_{X_0,+} \hat G_{X_0}
\eeq
for some $\lambda^\pm_{X_0,+}:\cf(S_+)^{\oplus 2}\to\cf(S_+)^{\oplus 2}$.
Then the two-point functions for $\hat P_{X}$ induced via  (\ref{eq:theiso}) are given by $\Lambda_{X}^\pm = B^* \lambda^\pm_{X_0,+} B$, where $B$ acts on $\cfd(X_+)\oplus \cf(X_0^\inti) \oplus\cfd(X_-)$ as follows:
\[
B=(\imath^- \varrho_{X_+}^-\hat G_{X_+} x_{X_+}^{-\i\tilde\sigma+2}, \varrho_{X_0,+} \hat G_{X_0}  x_{X_0}^{-\i\tilde\sigma+2},-\cS_{X_0}\imath^- \varrho_{X_-}^-\hat G_{X_-} x_{X_-}^{-\i\tilde\sigma+2}),
\] 
where  $\cS_{X_0}\defeq\varrho_{X_0,+}\cP_{X_0,-}$ is the {scattering matrix} on the asymptotically de Sitter space $(X_0,g_{X_0})$. 
\end{proposition}
\proof Let $Q\in\cf(X_0)$ be equal $0$ in a neighborhood of $S_+$ and $1$ in a neighborhood of $S_-$. By \eqref{eq:theiso}, the two-point functions for $\hat P_{X}$ induced by $\Lambda_{X_0}^\pm$ are given by $\Lambda_X^\pm=R^*_{X_0} \Lambda_{X_0}^\pm R_{X_0}$ where
\[
\bea
R_{X_0} =  [\hat P_{X_0},Q] x_{X_0}^{-\i\tilde\sigma}({}\traa{X_0^\inti}\circ\, \hat G_X).
\eea
\] 
Using (\ref{eq:fraeljk1}) we get that $\Lambda_X^\pm=B^* \lambda^\pm_{X_0,+}  B$, where
\[
\bea
B &=\varrho_{X_0,+}\hat G_{X_0} R_{X_0}= \varrho_{X_0,+}\hat G_{X_0}[\hat P_{X_0},Q] x_{X_0}^{-\i\tilde\sigma}({}\traa{X_0^\inti}\circ\, \hat G_X) \\ &=\varrho_{X_0,+} x_{X_0}^{-\i\tilde\sigma}({}\traa{X_0^\inti}\circ\, \hat G_X).
\eea
\]
Using the formula from Theorem \ref{thm:preciseform} we get (in the notation from that theorem)
\beq\label{glkjlkj} 
B=\varrho_{X_0,+}( c_{0,+} x_{X_+}^{-\i\tilde\sigma+2}, \hat G_{X_0}  x_{X_0}^{-\i\tilde\sigma+2}, - c_{0,-} x_{X_-}^{-\i\tilde\sigma+2}).
\eeq
The first component in the above expression equals 
\[
\bea
\varrho_{X_0,+}c_{0,+} x_{X_+}^{-\i\tilde\sigma+2} & =\varrho_{X_0,+}\cP_{X_0,+}\imath^- \varrho_{X_+}^-\hat P_{X_+}(\sigma)^{-1} x_{X_+}^{-\i\tilde\sigma+2}\\
&=\imath^- \varrho_{X_+}^-\hat P_{X_+}(\sigma)^{-1} x_{X_+}^{-\i\tilde\sigma+2}=\imath^- \varrho_{X_+}^-\hat G_{X_+} x_{X_+}^{-\i\tilde\sigma+2}
\eea
\]
when applied to $\cfd(X_+)$, where in the last equality we have used that $\varrho_{X_+}^- \hat P_{X_+}(-\sigma)^{-1}$ vanishes on $\cfd(X_+)$ due to mapping properties of the resolvent. Similarly, the third component in \eqref{glkjlkj}  equals
 \[
\bea
\varrho_{X_0,+}c_{0,-} x_{X_-}^{-\i\tilde\sigma+2} & = - \varrho_{X_0,+}\cP_{X_0,-}\imath^- \varrho_{X_-}^-\hat P_{X_-}(\sigma)^{-1} x_{X_-}^{-\i\tilde\sigma+2}\\
&=-\cS_{X_0}\imath^- \varrho_{X_-}^-\hat P_{X_-}(\sigma)^{-1} x_{X_-}^{-\i\tilde\sigma+2}=-\cS_{X_0}\imath^- \varrho_{X_-}^-\hat G_{X_-} x_{X_-}^{-\i\tilde\sigma+2},
\eea
\]
which finishes the proof. \qed

\subsection{Asymptotic data on $X$}\label{SubsectasX} The existence of Hadamard two-point functions for $\hat P_{X_0}$ follows from the standard abstract argument of Fulling, Narcowich and Wald \cite{FNW}, and consequently Hadamard two-point functions for $\hat P_X$ exist. In what follows, we want to construct \emph{distinguished} Hadamard two-point functions using a variant of the method worked out in previous chapters for asymptotically Minkowski spacetimes. To that end we need to identify the asymptotic data of solutions that correspond to sources and sinks for $\hat P_X$.   

The starting point is the result from \cite{poisson} which says that if $\i\sigma\notin \zz$, any $u\in{\Sol}(\hat P_X)$, i.e.~any solution of $\hat P_X u=0$ with $\wf(u)\subset N^*S$, is of the form
\beq
\label{eq:mrb}
u=(v+\i 0 )^{-\i\sigma} \tilde a^+_X + (v-\i0)^{-\i\sigma} \tilde a_X^- + \tilde a_X, 
\eeq
for some $\tilde a^\pm_X,\tilde a_X\in \cf(X)$. Furthermore, the restriction of $\tilde a^+_X$ and $\tilde a^{-}_{X}$ to either $S_+$ or $S_-$ defines a pair of smooth functions on $X$ that determine $u$ uniquely \cite[Prop.~4.11]{poisson}. We have thus two maps $\varrho_{X,\pm}$ assigning data one at $S_+$ and the other one at $S_-$, defined on ${\Sol}(\hat P_X)$ by
\[
\varrho_{X,\pm} u = (\varrho_{X,\pm}^+u,\varrho_{X,\pm}^-u)\defeq ( \tilde a^+_X\traa{S_\pm},\tilde a^-_X\traa{S_\pm})\in \cf(S_\pm)\oplus \cf(S_\pm).
\]
The $\varrho_{X,\pm}^+u$ data corresponds to sinks for $\hat P_X$ and the $\varrho_{X,\pm}^-u$ data corresponds to sources, see \cite{kerrds,positive}, so we have a setup analogous to the asymptotically Minkowski case (yet simpler, as $\sigma$ is a fixed parameter).

We can construct an  approximate Poisson operator $\tilde\cP_{X,\pm}$ by simply setting
\beq
\tilde\cP_{X,\pm} (a^+,a^-) = (v+\i 0 )^{-\i\sigma} a^+(y) + (v-\i 0 )^{-\i\sigma} a^-(y), \ \ a^+,a^-\in \cf(S_\pm)
\eeq
Note that this is a very rough approximation, in the sense that $P\tilde\cP_{X,\pm}(a^+,a^-)$ needs not even be smooth (though more precise approximate solutions can be easily constructed as asymptotic series, cf. \cite[Lem.~6.4]{BVW}), all that matters here is that it has above-threshold regularity. In fact
\[
\cP_{X,\pm}\defeq \tilde \cP_{X,\pm}- \hat P_{X,\mp}^{-1} P\tilde \cP_{X,\pm}
\]
is the corresponding Poisson operator, i.e.~the inverse of $\varrho_{X,\pm}:{\Sol}(\hat P_X)\to \cf({S_\pm})\oplus \cf({S_\pm})$. We can now adapt the arguments of Subsect. \ref{ssec:asymptoticdata} and using an analogous commutator argument show the identity
\beq\label{lkmelkmlkm}
\i\hat G_X = \hat G_X^* \varrho^*_{X,\pm} q_{X} \varrho_{X,\pm} \hat G_{X}, \ \ \mbox{ \ where \ }   q_{X}=\mat{\alpha^+}{0}{0}{-\alpha^-}, \ \alpha^+,\alpha^-\in\rr\setminus\{0\}.
\eeq
Let us denote
\[
\pi^+_X= \alpha^+ \mat{\one}{0}{0}{0}, \quad \pi^-_X= \alpha^- \mat{0}{0}{0}{\one},
\]
In analogy to Theorem \ref{prop:Had} we obtain:

\begin{theorem} \label{thm:newhada}  Assume $\sigma$ is not a pole of $\hat P_{X_+}(\sigma)^{-1}$ nor of $\hat P_{X_-}(\sigma)^{-1}$. The pair of operators
\beq\label{eq:lpmnew}
\Lambda^\pm_{X,+}\defeq \hat G^*_X \varrho^*_{X,+} \pi^\pm_X  \varrho_{X,+} \hat G_{X}
\eeq
are Hadamard two-point functions for $\hat P_X$ and consequently, 
\[
\Lambda^\pm_{X_0,+}\defeq x_{X_0}^{-\i\tilde\sigma} ( \Lambda_{X,+}^{\pm}\traa{X_{0}^\inti\to X_{0}^\inti})x_{X_0}^{\i\tilde\sigma-2}
\]
are Hadamard two-point functions for $\hat P_{X_0}$. The same statement is true for
\beq\label{eq:lpmnew2}
\Lambda^\pm_{X,-}\defeq \hat G^*_X \varrho^*_{X,-} \pi^\pm_X  \varrho_{X,-} \hat G_{X}, \ \ \Lambda^\pm_{X_0,-}\defeq x_{X_0}^{-\i\tilde\sigma} ( \Lambda_{X,-}^{\pm}\traa{X_{0}^\inti\to X_{0}^\inti})x_{X_0}^{\i\tilde\sigma-2}.
\eeq
\end{theorem}
 
As regularity is propagated from the respective radial sets, one actually gets a more precise wave front statement in the Hadamard condition proposed in Definition \ref{def:onX}, namely
\beq\label{eq:betterwf}
\wf'(\Lambda_{X}^{\pm})\subset\big( \cup_{t\in\rr}\hat\Phi_t(\diag_{T^*X})\cap\pi^{-1}\hat\Sigma^\pm  \big)\cup (\,\zero\times N^*S^\pm)\cup( N^*S^\pm\times \zero),
\eeq
where $N^*S^\pm\defeq N^*S\cap \hat\Sigma^\pm$.\medskip

At the present stage it is worth mentioning that (beside abstract existence arguments of `generic' Hadamard two-point function on globally hyperbolic spacetimes) there is a relatively simple construction named after Bunch and Davies that gives a `maximally symmetric' Hadamard two-point function on \emph{exact} de Sitter space \cite{allen,bros1,bros2}. Furthermore, the work of Dappiaggi, Moretti and Pinamonti \cite{DMP1} provides a distinguished Hadamard two-point function for a class of cosmological spacetimes that asymptotically resemble the de Sitter cosmological chart. It is presently unknown whether our construction yields the Bunch--Davies state or extensions of the Dappiaggi--Moretti--Pinamonti state to `global' asymptotically de
Sitter spacetimes; this question will be studied in a subsequent work. 

Here the main novelty, beside working on `global' asymptotically de Sitter spacetimes, is the extension of the two-point functions across the conformal boundary.

It is possible to express the `future' and `past' two-point functions $\Lambda_{X_0,+}^\pm$, $\Lambda_{X_0,-}^\pm$ using the more conventional $\varrho_{X_0,\pm}$ data. This relies on the following result from \cite{poisson} which relates $\varrho_{X,+}$, $\varrho_{X_0,+}$ and $\varrho_{X_+}^\pm$ (an analogous result holds true at past infinity).   

\begin{proposition}[\cite{poisson}] We have
\beq\label{propopopo}
\varrho_{X,+}= \frac{1}{\e^{-\pi\sigma}-\e^{\pi\sigma}}\begin{pmatrix}\one & - \e^{\pi\sigma} \cS_{X_+}^{-1} \\ -\one &  \e^{-\pi\sigma} \cS_{X_+}^{-1} \end{pmatrix}\varrho_{X_0,+} \,x_{X_0}^{-\i\tilde\sigma}\circ{}\traa{X_0} \mbox{ on }\ \Sol(\hat P_X),
\eeq
where $\cS_{X_+}\defeq \varrho^-_{X_+}\cP^+_{X_+}$ is the \emph{scattering matrix} on $X_+$.
\end{proposition}

Using \eqref{propopopo} one obtains by a direct computation that
\[
\Lambda^\pm_{X_0,+}=\frac{\alpha^\pm}{(\e^{-\pi\sigma}-\e^{\pi\sigma})^2} \hat G_{X_0}^* \varrho_{X_0,+}^* \begin{pmatrix}\one & - \e^{\pm\pi\sigma} \cS_{X_+}^{-1} \\ - \e^{\pm\pi\sigma} \cS_{X_+} &  \e^{\pm 2\pi\sigma} \end{pmatrix} \varrho_{X_0,+} \hat G_{X_0}.
\]

\subsection{QFT in the hyperbolic caps $X_\pm$}\label{lastss} The extension across the conformal boundary performed in the previous subsections raises the question of whether the symplectic space of solutions on $X_0$ is isomorphic to a symplectic space of a similar form on one of the asymptotically hyperbolic caps $X_+$ or $X_-$. We demonstrate that  this is the case if one takes \emph{two copies} of $X_+$ (or $X_-$) instead of one.

We start by observing that despite the elliptic character of $\hat P_{X_\pm}$, the similarities between the structure of the solutions of $\hat P_{X_\pm}$ and $\hat P_{X_0}$ suggest that ${\Sol}(\hat P_{X_\pm})$ could be characterized as the range of the operator 
\[
\hat G_{X_\pm}(\sigma)\defeq(\hat P_{X_\pm}^{-1}(\sigma)-\hat P_{X_\pm}^{-1}(-\sigma))=  x_{X_\pm}^{-\i\tilde\sigma} (\hat G_{X}(\sigma)\traa{X_{\pm}^\inti\to X_{\pm}^\inti})x_{X_\pm}^{\i\tilde\sigma-2}
\]
on a suitable class of functions.  We prove that this is true if one considers $\hat G_{X_\pm}$ acting on $\cfd(X_\pm)$ --- the space of smooth functions that vanish with all derivatives at the boundary $\p X_\pm = S_\pm$. 

Note that by Stone's theorem, $\hat G_{X_\pm}(\sigma)$ is a multiple of the spectral projector of the Laplacian on $X_\pm$, so all the ingredients of the next proposition are actually standard objects from spectral theory.

\begin{proposition}\label{prop:usio3}We have bijections
\beq\label{eq:usio3}
\frac{\cfd(X_\pm)}{\hat P_{X_\pm}\cfd(X_\pm)}\xrightarrow{\makebox[2em]{$[\hat G_{X_\pm}]$}}{\Sol}(\hat P_{X_\pm}).
\eeq
Moreover, $\bra \overline{\cdot} , \hat G_{X_\pm} \cdot\ket_{X_{\pm}}$ induces a well-defined symplectic form on the quotient space $\cfd(X_\pm)/\hat P_{X_\pm}\cfd(X_\pm)$.\end{proposition}
\proof We consider the case $X_+$, the other one being analogous, and prove bijectivity of the arrow (the assertion on $\bra \overline{\cdot} , \hat G_{X_\pm} \cdot\ket_{X_{\pm}}$ follows then easily).

 The inclusion $\hat G_{X_+} \cfd(X_+)\subset{\Sol}(\hat P_{X_+})$ is proved using the identity \eqref{eq:defphats} that relates $\hat P_{X_+}$ with $\hat P_X$, and the asymptotics \eqref{eq:mrb} for solutions of $\hat P_X$. We now show the reverse inclusion (which then gives surjectivity of the first arrow). Recall that by definition, any $u\in{\Sol}(\hat P_{X_+})$ can be written as $v^+ + v^-$, where $v^\pm\in x_{X_{+}}^{\pm\i\sigma+(d-1)/2}\cf(X_+)$. Observe that $\hat P_{X_+}v^+$ equals $-\hat P_{X_+}v^-$, and on the other hand,
\[
\hat P_{X_+}v^\pm \in   x_{X_{+}}^{\pm\i\sigma+(d-1)/2+2}\cf(X_+)
\]
by \eqref{eq:defphats} (recall the notation $\hat P_{X_+}=\hat P_{X_+}(\sigma)$). Consequently, $\hat P_{X_+}v^\pm\in\cfd(X_+)$ (otherwise the asymptotic behavior of $\hat P_{X_+}v^+$ and $\hat P_{X_+}v^-$ would be different). We will now use the fact that $\hat P_{X_+}^{-1}(\pm\sigma)$ maps $\cfd(X_+)$ to  $x_{X_{+}}^{\mp\i\sigma+(d-1)/2}\cf(X_+)$, which can be seen from \eqref{eq:defphats2} and the mapping properties of $\hat P_{X,\pm}^{-1}$. This implies that $w^\pm=\hat P_{X_+}^{-1}(\mp\sigma)\hat P_{X_+}v^\pm - v^\pm\in x_{X_{+}}^{\pm\i\sigma+(d-1)/2}\cf(X_+)$, so $w^\pm$ is a solution of $\hat P_{X_+} w^\pm=0$ with data $\varrho_{X_+}^\mp w^\pm=0$ and therefore vanishes. We conclude
\[
u= v^+ + v^- =  \hat P_{X_+}^{-1}(\sigma) \hat P_{X_+} v^- + \hat P_{X_+}^{-1}(-\sigma) \hat P_{X_+} v^+=(\hat P_{X_+}^{-1}(\sigma) - \hat P_{X_+}^{-1}(-\sigma))\hat P_{X_+}v^-.
\]
This yields $u= \hat G_{X_+} f$ with $f=\hat P_{X_+} v^-\in \cfd(X_+)$ as claimed.

To prove injectivity of the arrow, observe that if $f\in\cfd(X_+)$ is in the kernel of $\hat G_{X_+}$ then $\hat P_{X_+}^{-1}(\sigma)f$ equals $\hat P_{X_+}^{-1}(-\sigma)f$, with asymptotic behavior of the two distinct types at the same time, so in fact $\hat P_{X_+}^{-1}(\sigma)f\in \cfd(X_+)$. This means that $f=\hat P_{X_+} g$ with $g=\hat P_{X_+}^{-1}(\sigma)f\in \cfd(X_+)$. 
\qeds 

In what follows we consider only the `future cap' $X_+$, but all the discussion remains valid for the `past cap' $X_-$ as well.

The next proposition shows that by taking two copies of the symplectic space $\Sol(\hat P_{X_+})$ we obtain a symplectic space that is isomorphic to $\Sol(\hat P_{X})$ and hence to $\Sol(\hat P_{X_0})$.

\begin{proposition}\label{newsdj}We have isomorphisms
\beq\label{eq:usiofddf}
\begin{array}{rl}
\displaystyle\frac{\cf(X)}{\hat P_{X}\cf(X)}\xrightarrow{\makebox[4em]{$[\hat G_{X}]$}}{\Sol}(\hat P_{X}) \mathrel{\text{\raisebox{-6pt}{\rotatebox[origin=c]{-13}{$\xrightarrow{\makebox[4em]{${}\varrho_{X,+}$}}$}}}}& \\[4mm]
& \!\!\!\!\cf(S_+)^{\oplus 2}
\\[-4mm]
\bigg( \displaystyle\frac{\cfd(X_+)}{\hat P_{X_+}\cfd(X_+)}\bigg)^{\oplus 2}
\xrightarrow{\makebox[4em]{$[ \hat G_{X_+}]^{\oplus 2}$}} \left({\Sol}(\hat P_{X_+})\right)^{\oplus 2} \mathrel{\text{\raisebox{6pt}{\rotatebox[origin=c]{13}{$\xrightarrow{\makebox[4em]{${}(\varrho_{X_+}^+)^{\oplus 2}$}}$}}}} &
\end{array},
\eeq
where the symplectic form on $\cf(X)/\hat P_{X}\cf(X)$ is induced by $\hat G_X$, the symplectic form on the other quotient space is induced by $\hat G_{X_+}\oplus - \hat G_{X_+}$, and 
$\cf(S_+)\oplus\cf(S_+)$ is equipped with the symplectic form $\i^{-1}q_X$ (see \eqref{lkmelkmlkm}).
\end{proposition}
\proof Bijectivity of all the arrows was already stated, so all we need to prove is that the same symplectic form is induced by both arrows pointing to $\cf(S_+)^{\oplus 2}$.   The key fact that we will use to that end is the identity
\[
\varrho_{X_+}^+( x_{X_+}^{-\i\tilde\sigma}\circ{}\traa{X_+})\cP_{X,+} (a^+_{X},a^-_{X})=a^+_{X} + a^-_{X}, \ \ a^\pm_{X}\in\cf(S_+),
\]
which was proved in \cite[Sec. 4]{poisson}. This entails immediately that
\[
\varrho_{X_+}^+( x_{X_+}^{-\i\tilde\sigma}\circ{}\traa{X_+})\cP_{X,+} \imath ^- =\one
\]
on $\cf(S_+)$. Thus, the inverse of $\varrho_{X_+}^+ \oplus \varrho_{X_+}^+$ is the composition of the maps:
\[
\cf(S_+)^{\oplus 2 } \xrightarrow{\makebox[3.5em]{$\imath^+\oplus \imath^+ $}} \Ran (\imath^+\oplus \imath^+) \xrightarrow{\makebox[3em]{$\cP_{X,+}^{\oplus2}$}} \begin{array}{c} \Sol(\hat P_{X}) \\ \oplus \\ \Sol(\hat P_{X})\end{array}\xrightarrow{\makebox[5em]{$( x_{X_+}^{-\i\tilde\sigma}\circ{}\traa{X_+})^{\oplus 2}$}} \begin{array}{c} \Sol(\hat P_{X_+}) \\ \oplus \\ \Sol(\hat P_{X_+})\end{array}.
\]
To prove that the symplectic form on $\cf(S_+)^{\oplus 2 }$ is $\i^{-1}q_X$, it suffices to check that the corresponding symplectic form on $\Ran (\imath^+\oplus \imath^+)$ (induced by the arrows on the right of it) is $\i^{-1}(q_X\oplus -q_X)$. But these arrows are just direct sums, so we can use the already proved isomorphisms on each of the two direct sum components independently.\qeds

Thus, a pair of  fields on $X$ (or equivalently, on $X_0$) corresponds to a pair of fields on $X_+$. We can make this more precise as follows. For the sake of uniformity let us denote $\hat P_{X_+^2}\defeq \hat P_{X_+}\oplus \hat P_{X_+}$and $\hat G_{X_+^2}\defeq \hat G_{X_+}\oplus -\hat G_{X_+}$. We will say that $\Lambda^\pm_{X_+^2}$ are two-point functions for $\hat P_{X_+^2}$ if $\Lambda^+_{X_+^2}- \Lambda^+_{X_-^2}=\i \hat G_{X_+^2}$ and $\Lambda_{X_+^2}^\pm\geq 0$. 

By Proposition  \ref{newsdj}, any pair of two-point functions $\Lambda_{X}^\pm$ for $\hat P_{X}$ (or equivalently, any pair $\Lambda_{X_0}^\pm$ of two-point functions for $\hat P_{X_0}$) induces two-point functions 
\[
\Lambda^\pm_{X_+^2} \defeq R_{X_+^2}^* \Lambda^\pm_X R_{X_+^2},
\]
where $R_{X_+^2} = [\hat G_{X}]^{-1}\cP_{X,+} (\varrho_{X_+}^+\hat G_{X_+}\oplus\varrho_{X_+}^+\hat G_{X_+}) $ is the relevant isomorphism.

\begin{proposition}\label{newprop2} Let $\Lambda_{X,+}^\pm$ be the Hadamard two-point functions for $\hat P_X$ defined in Theorem \ref{thm:newhada}. The induced two-point functions for $\hat P_{X_+^2}$ are
\beq\label{eq:kjdfn}
\Lambda_{X_+^2,+}^+=\i\begin{pmatrix}\hat G_{X_+} & 0 \\ 0 & 0 \end{pmatrix}, \ \ \Lambda_{X_+^2,+}^-=\i\begin{pmatrix} 0 & 0 \\ 0 & \hat G_{X_+} \end{pmatrix}.
\eeq
\end{proposition}
\proof Recall that  $\Lambda_{X,+}^\pm=\hat G^*_X \varrho^*_{X,+} \pi^\pm_X  \varrho_{X,+} \hat G_{X}$, and so
\beq\label{erglkegrl}
\Lambda_{X_+^2,+}^\pm= B^* \pi^\pm_X  B, 
\eeq
where
\[
B=\varrho_{X,+} \hat G_{X} [\hat G_{X}]^{-1}\cP_{X,+}  (\varrho_{X_+}^+\hat G_{X_+}\oplus\varrho_{X_+}^+\hat G_{X_+})=\varrho_{X_+}^+\hat G_{X_+}\oplus\varrho_{X_+}^+\hat G_{X_+}.
\]
Since $B$ is diagonal, we conclude from \eqref{erglkegrl} that $\Lambda_{X_+^2,+}^\pm$ are diagonal matrices, and the second/first on-diagonal component vanishes. In view of $\Lambda^+_{X_+^2}- \Lambda^+_{X_-^2}=\i \hat G_{X_+^2}$ this yields \eqref{eq:kjdfn}.\qed

 \appendix
\section{}\init\label{secapp1}
\subsection{Quasi-free states and their two-point functions}

In this appendix we briefly recall the relation between quantum fields, quantum states and two-point functions in the framework of algebraic QFT. Although this is standard material which can be found in many books and review articles, see e.g.~\cite{derger,haag,KM}, it is worth stressing that there exist several equivalent formalisms --- here we follow \cite{GW,GW2} and use the complex formalism (used to describe charged fields) as opposed to the real one (used for neutral fields). The advantage of the complex formalism is that one works with sesquilinear forms, so the positivity condition for two-point functions has a very neat formulation. On the other hand, the real formalism is particularly useful if one wants to work with $C^*$-algebras rather than mere $^*$-algebras.  

Let $\kV$ be a complex vector space $\kV$ equipped with an
anti-hermitian form $G$. It is slightly more convenient to have a
hermitian form, so we set $q\defeq\i^{-1}G$. The polynomial  CCR
$*$-algebra ${\rm CCR}^{\rm pol}(\kV,q)$ (see
e.g.~\cite[Sect. 8.3.1]{derger})  is defined as the algebra generated
by the identity $\one$ and all abstract elements of the form
$\psi(v)$, $\psi^{*}(v)$, $v\in\kV$, with $v\mapsto \psi(v)$
anti-linear, $v\mapsto \psi^*(v)$ linear, and subject to the canonical commutation relations
\beq\label{eq:ccr}
[\psi(v), \psi(w)]= [\psi^{*}(v), \psi^{*}(w)]=0,  \ \ [\psi(v), \psi^{*}(w)]=  \bar{v} q w \one, \ \ v, w\in \kV.
\eeq
A state $\omega$ is a linear functional on ${\rm CCR}^{\rm pol}(\kV,q)$ such that $\omega(a^* a)\geq 0$ for all $a$ in ${\rm CCR}^{\rm pol}(\kV,q)$ and $\omega(\one)=1$.

The \emph{bosonic two-point functions} (or complex covariances)  $\Lambda^\pm$ of a state $\omega$ on the polynomial  CCR $*$-algebra are the two hermitian forms $\Lambda^\pm$ defined by
\beq\label{eq:lambda}
\bar{v}\Lambda^+ w = \omega\big(\psi(v)\psi^*(w)\big), \quad \bar{v}\Lambda^- w = \omega\big(\psi^*(w)\psi(v)\big), \quad v,w\in \kV
\eeq
Note that both $\Lambda^\pm$ are positive and by the canonical commutation relations one has always $\Lambda^+ - \Lambda^- = q$. Crucially, there is reverse construction, namely if one has a pair of hermitian forms $\Lambda^\pm$ such that  $\Lambda^+ - \Lambda^- = q$ and $\Lambda^\pm\geq 0$ then there exists a state $\omega$ such that (\ref{eq:lambda}) holds, and this assignment is one-to-one for the class of \emph{bosonic quasi-free states}, see e.g.~\cite{araki,derger}.

Once a state $\omega$  is fixed, the \emph{GNS construction} provides: a Hilbert space $\mathfrak{H}$, unbounded operators $\hat\psi(v)$, $v\in\kV$, such that $v\mapsto\hat\psi(v)$ is anti-linear (on a common dense domain in $\mathfrak{H}$), and a vector $\Omega\in \mathfrak{H}$ in the common domain of $\hat\psi(v)$ such that 
\beq
\label{reccr}
\bar{v}\Lambda^+ w = \bra \Omega, \hat\psi(v)\hat\psi^*(w)\Omega\ket_\mathfrak{H}, \quad \bar{v}\Lambda^- w = \bra \Omega,\hat\psi^*(w)\hat\psi(v)\Omega\ket_\mathfrak{H}, \quad v,w\in \kV,
\eeq
and
\beq
\label{reccr2}
[\hat\psi(v), \hat\psi(w)]= [\hat\psi^{*}(v), \hat\psi^{*}(w)]=0,  \ \ [\hat\psi(v), \hat\psi^{*}(w)]=  \bar{v} q w \one, \ \ v, w\in \kV
\eeq
on a suitable dense domain. 
In the case when $\kV$ is a quotient space of the form $\cf_\c(M)/P \cf_\c(M)$ for some $P\in\Diff(M)$ (or a similar quotient, such as the space $H^{\infty,0}_\b(M)/P H^{\infty,0}_\b(M)$ considered in the main part of the text), then, disregarding issues due to unboundedness of $\hat\psi(v)$, $\cf_\c(M)\ni v\mapsto \hat\psi(\bar{v})$ can be interpreted as an operator-valued distribution that solves $P\hat\psi=0$. The distributions $\hat\psi$ are the (non-interacting) \emph{quantum fields} and are the main object of interest from the physical point of view. Note that although they are solutions of a differential equation, their analysis differs from usual PDE techniques, as $\hat\psi$ take values in operators on a Hilbert space $\mathfrak{H}$ that is not given a priori, but is constructed simultaneously with $\hat\psi$.

\subsection{Proof of auxiliary lemmas}\label{app2} We give below the proof of several auxiliary lemmas used in the main part of the text. We use various notations introduced in Section \ref{section5}.

Let us denote by $\kS^{-l}(\cc;S^s_{\rm cl})$ the space of holomorphic functions in $\Im\sigma>-l$, Schwartz in strips as $\sigma\to\infty$ (cf.~Subsect.~\ref{ss:mellin}), taking values in classical symbols of order $s$. Recall that to any $\tilde a\in \kS^{-l}(\cc;S^s_{\rm cl})$ we assigned the oscillatory integral   
\[
J(\tilde a)=\int \rho^{\i\sigma} \e^{\i v \gamma} |\gamma|^{\i\sigma-1} \tilde a(\sigma,v,y,\gamma) d\gamma d\sigma.
\]

\begin{lemma}\label{lem:ap1} Let $Q\in\Diff_\b^j(M)$. For any $\tilde a\in\kS^{-l}(\cc;S^s_{\rm cl})$ there is $\tilde b\in \kS^{-l}(\cc;S^{s+j}_{\rm cl})$ such that
\beq\label{eq:lemap1}
QJ(\tilde a) = J(\tilde b) \mod H_\b^{\infty,l}(M), 
\eeq
 and $\tilde b$ differs from 
\beq\label{eq:lemap2}
\sigma_{\b,j}(Q)(0,0,y,\sigma,\gamma,0) \tilde a(\sigma,v,y,\gamma)
\eeq
by a classical symbol of order $s+j-1$ (where the variables are the local coordinates $(\rho,v,y,\sigma,\gamma,\eta)$ on the $\b$-cotangent bundle). Furthermore, if $j=1$ and in addition $Q\in\kM(M)$, then $\tilde b\in \kS^{-l}(\cc;S^{s}_{\rm cl})$ (rather than merely $\tilde b\in \kS^{-l}(\cc;S^{s+1}_{\rm cl})$).
\end{lemma}
\proof 
The first statement is straightforward to see for multiplication operators by $\cf$ functions
on $\pM$, as well as for the vector fields $\rho D_\rho, D_v,
D_{y_j}$: indeed, due to the Mellin transform this amounts to a
$\sigma$-dependent version of the standard  regularity statement for
conormal distributions, conormal to $v=0$. In addition, the statement
holds for multiplication
by powers $\rho^k$ of $\rho$ which in fact increase the
domain of holomorphy, and indeed on $\Im\sigma=-l$ (and in the
corresponding upper half plane) yields a similar term but with $\tilde
b$ now of order $s-k$ by a contour shift argument similar to \eqref{eq:rDv-contour-shift}. Thus, for finite Taylor expansions of arbitrary
$\cf$ functions on $\pM$ one has the same multiplication property,
with the symbolic order improving as one increases the power of
$\rho$, so in fact the symbols arising from the full formal Taylor series can be asymptotically
summed. One also sees by rewriting multiplication by $\rho^k$ times an
element $\phi$ of $\cf(M)$ of support in $\rho<\epsilon$ as a convolution on the
Mellin transform side that $\rho^k\phi J(\tilde a)$ is in fact in
$H_\b^{m,l}$ for any $m<\12-l-s+k$.
Combining this with the asymptotic summation statement, using that
b-conormal distributions of symbolic order $s-k$ lie in
$H_\b^{m,l}$ for any $m<\12-l-s+k$, we see that (modulo
$H_\b^{\infty,l}$) multiplication by a $\cf$ function indeed gives a
distribution of the stated form.

Next, notice that 
\[
\sigma_{\b,j}(Q)(0,0,y,0,\gamma,0) \tilde a(\sigma,v,y,\gamma)
\]
differs from \eqref{eq:lemap2} by an element of $\kS^{-l}(\cc;S^{s+j-1}_{\rm cl})$, and thus
\beq\label{eq:aaaaa}
\tilde b = \sigma_{\b,j}(Q)(0,0,y,0,\gamma,0) \tilde a(\sigma,v,y,\gamma) \mod \kS^{-l}(\cc;S^{s+j-1}_{\rm cl}).
\eeq
In the special case $j=1$, if $Q\in \kM(M)$ then by definition $\sigma_{\b,1}(Q)(0,0,y,0,\gamma,0)$ vanishes. Thus, the right hand side of \eqref{eq:aaaaa} vanishes, and we obtain in this case that the principal symbol of $\tilde b$ (of order $s+1$) vanishes, hence $\tilde b$ is of order $s$. \qed

%%The statement under the stronger assumption $Q\in\kM$ is readily seen by applying the generator vector fields $\rho D_\rho$, $\rho D_v$, $vD_v$ and $D_{y_j}$ to the oscillatory integral, with the  
%argument for $\rho D_v$ having already been discussed above. \qed

\medskip

\noindent\textbf{Proof of Lemma \ref{lem:morereg}.} This is a standard construction in microlocal analysis; see the proof of \cite[Lemma~6.4]{BVW} for a similar argument, but phrased without the explicit use of oscillatory integrals.

For $\tilde a\in\kS^{-l}(\cc;S^0_{\rm cl})$, we will iteratively solve the problem of constructing $u$ of the form
\[
u=J(\tilde a_\infty) \mod H_\b^{\infty,l}(M)
\]
with $\tilde a_\infty-\tilde a$ classical of order $-1$, and with $Pu\in H_\b^{\infty,l}$. 

We take first $\tilde a_0=\tilde a$. In what follows we will use Lemma \ref{lem:ap1} repeatedly. Let us note that in the particular case $Q=\rho D_\rho+v D_v\in \kM(M)$, given $\tilde a\in\kS^{-l}(\cc;S^s_{\rm cl})$, Lemma \ref{lem:ap1} plus a simple explicit computation for the module  generators $\rho D_\rho$ and $v D_v$ yields $\tilde b$ that differs from
$-\gamma D_\gamma \tilde a(\sigma,v,y,\gamma)$, hence from $\i s\tilde a$, by a classical 
symbol of order $s-1$. In particular, if $s=0$, this says that $\tilde b$ is a classical symbol of order $-1$.

Thus, Lemma \ref{lem:ap1} allows to conclude that for $Q_2\in\kM(M)^2$, the expression
\[
PJ(\tilde a_0)=-4D_v(\rho D_\rho
+vD_v)J(\tilde a_0)+Q_2 J(\tilde a_0)
\]
is of the form $J(\tilde r_0)$ mod $H_\b^{\infty,l}$ with $\tilde r_0$ classical of order $0$.

Using Lemma \ref{lem:ap1} again, for any $\tilde a'_1$ of order $-1$, $PJ(\tilde a'_1)$ is of
the form $J(\tilde r'_1)$ modulo $H_\b^{\infty,l}$ with $\tilde r'_1$ a symbol of order $0$,
equal to $-4\i\gamma\tilde a'_1$ modulo symbols of order $-1$. Thus,
choosing $\tilde a'_1$ such that $- 4 \i\gamma\tilde a'_1=-\frac{\i}{4}\tilde r_0$, and setting $\tilde a_1=\tilde
a_0+\tilde a_1'$, we obtain 
\[
PJ(\tilde a_1)=J(\tilde r_1) \mod H_\b^{\infty,l}(M),
\]
with $\tilde r_1$ a symbol of order $-1$, which is a one order improvement over $\tilde
r_0$ corresponding to $PJ(\tilde
a_0)$. 

Similarly, we inductively construct $\tilde
a_k=\tilde a_0+\sum_{j=1}^k\tilde a'_j$ such that 
\[
PJ(\tilde a_k)=J(\tilde r_k) \mod  H_\b^{\infty,l}(M),
\] 
with $\tilde r_k$ classical of order $-k$. This can be done because for $\tilde a'_k$ classical of order
$-k$, 
\[
P\tilde J(a'_k)=J(\tilde r'_k) \mod H_\b^{\infty,l}(M),
\]
with $\tilde r'_k$ a
symbol of order $-k+1$, equal to $-4\i k\gamma\tilde a'_k$ modulo symbols of order $-k$; the point being that as $k\neq 0$, 
$-4\i k\gamma\tilde a'_k=-\tilde r_{k-1}$ (where $\tilde r_{k-1}$ corresponds
to $PJ(\tilde a_{k-1})$) can be solved for $\tilde a'_k$.
Finally, asymptotically summing $\tilde
a'_\infty\sim\sum_{j=1}^\infty\tilde a'_j$, we see that $\tilde a_\infty=\tilde
a_0+\tilde a'_\infty$ satisfies the requirements of the lemma.
\qeds

\noindent\textbf{Proof of Lemma \ref{lem:morereg2}.} Let us introduce an analogue of the map $\tilde\cP_I$ that acts on full symbols (rather than on principal symbols):
\beq\label{eq:analo}
\tilde\cP_0 a \defeq \int \rho^{\i\sigma}\e^{\i v\gamma}\eta_+(v,\rho,y) a(\sigma,v,y,\gamma) d\gamma d\sigma, 
\eeq
and correspondingly
$$
\varrho_0 u\defeq (2\pi)^{-2}\int \rho^{-\i\sigma}\e^{-\i v\gamma}\eta_+(\rho,v,y)u(\rho,v,y)\,d\rho\,dv.
$$
Now, the already discussed statement on the regularity of solutions of $Pu=0$ (see the discussion preceding \eqref{eq:froma}) implies that they are of the form $\tilde\cP_0 a$ for some symbol $a$ as above (with the appropriate holomorphy properties) modulo $\Hb^{\infty,l}$.
If they were actually of this form (and the difference in $\Hb^{\infty,l}$ is easy to deal with in any case), one would get
$$
\tilde\cP_0\varrho_0 u=\tilde\cP_0\varrho_0\tilde \cP_0 a=\tilde\cP_0(\varrho_0\tilde\cP_0 a),
$$
and hence one is done if $\varrho_0\tilde\cP_0$ is essentially the identity. Now,
\beq\label{eq:analo2}
\varrho_0\tilde\cP_0 a=\cF_v\cM_\rho \eta_+^2\cM^{-1}\cF^{-1}a,
\eeq
so the question is whether
$$
\cF_v\cM_\rho(1-\eta_+^2)\cM^{-1}\cF^{-1}a
$$
is trivial. But it indeed is, since $\cM^{-1}\cF^{-1}$ maps symbols to
distributions which are in $\Hb^{\infty,l}$ away from
$\{\rho=0,v=0\}$, thus on the support of $1-\eta_+^2$, and then
$\cF\cM$ sends these to symbols of order $-\infty$ in the required
sense.

Given this, the map $\varrho$ is simply a restriction of a rescaled version of $\varrho_0$ to $\pm\infty$ in $\gamma$; $\tilde\cP$ is (ignoring $\chi_\pm$ which just cuts everything in two) an analogous composition with extension from $\pm\infty$ (denoted by $e_\infty$), namely
$$
\varrho= r_\infty |\gamma|^{-\i\sigma+1}\varrho_0,\ \ \tilde\cP=\tilde\cP_0|\gamma|^{\i\sigma-1}e_\infty,
$$
where $r_{\infty}$ is the restriction map. Thus,
\beq
\tilde\cP\varrho=\tilde\cP_0\varrho_0+\tilde\cP_0|\gamma|^{\i\sigma-1}(e_\infty r_\infty-\one)|\gamma|^{-\i\sigma+1}\varrho_0,
\eeq
and the first term is microlocally the identity as we have seen before, while the second term maps to $\b$-conormal distributions of one lower order because $e_\infty r_\infty-\one$ maps smooth functions on the compactified line (times various irrelevant factors) to functions vanishing to first order at $\pm\infty$.\qeds

\noindent\textbf{Proof of Lemma \ref{lem:morereg2}.} Recall that we need to prove that the map $(a,a')\mapsto[w]=[\tcP_I(a,a')]$ is injective, with the equivalence class considered modulo $H_\b^{m+1,l}$, $-\12+l<m<\12+l$. This can be readily seen from the computation in \eqref{eq:analo2} which gives injectivity of the auxiliary map $\tilde \cP_0$, and hence the stated injectivity of $[\tcP_I]$ in the equivalence class modulo $H_\b^{m+1,l}$.\qeds

\medskip

\subsection*{Acknowledgments} The authors would like to thank Jan
Derezi\'nski, Jesse Gell-Redman and Christian G\'erard for useful
discussions and particularly Peter Hintz for valuable comments on the
manuscript; the are also are grateful to the referees for their helpful remarks. A.\,V. gratefully acknowledges support from the NSF under
grant number DMS-1361432. M.\,W. gratefully acknowledges the
France-Stanford Center for Interdisciplinary Studies for financial
support and the Department of Mathematics of Stanford University for
its kind hospitality. The authors are also grateful to the Erwin
Schr\"odinger Institute in Vienna for its hospitality during the
program ``Modern theory of wave equations'', 2015; their stay at ESI greatly
facilitated the completion of this paper.


\begin{thebibliography}{plain}

\bibitem{allen} B. Allen, {\em Vacuum States in de Sitter Space}, Phys. Rev. D 32, 3136 (1985).

\bibitem{araki} H. Araki, M. Shiraishi, {\em On quasi-free states of canonical commutation relations I}, Publ. RIMS Kyoto Univ. 7 (1971/72), 105-120.

\bibitem{BGP} C. B\"ar, N. Ginoux, F. Pf\"affle, {\em Wave equation on Lorentzian
manifolds and quantization}, ESI Lectures in Mathematics and Physics, EMS (2007).

\bibitem{BS} C. B\"ar, A. Strohmaier, {\em An index theorem for Lorentzian manifolds with compact spacelike Cauchy boundary}, to appear in Amer. J. Math., preprint \texttt{arXiv:1506.00959}, (2017).

\bibitem{BS2} C. B\"ar, A. Strohmaier, {\em A rigorous geometric derivation of the chiral anomaly in curved backgrounds}, Commun. Math. Phys., Volume 347 (3) (2017), 703--721.

\bibitem{BVW} D. Baskin, A. Vasy, J. Wunsch, \textsl{Asymptotics of radiation fields in asymptotically Minkowski space}, Amer. J. Math., 137 (5), (2015), 1293--1364.

\bibitem{BW} D. Baskin, F. Wang, {\em Radiation fields on Schwarzschild spacetime}, Commun. Math. Phys. 331:477--506 (2014). 

\bibitem{BGMS} M. Bertola, V. Gorini, U. Moschella and R. Schaeffer, {\em Correspondence between
Minkowski and de Sitter Quantum Field Theory}, Phys. Lett. B 462 (1999) 249--253.

\bibitem{boer} J. de Boer, S. N. Solodukhin, {\em A holographic reduction of Minkowski space-time}, Nucl. Phys. B 665 (2003), 545.

\bibitem{bros1} J. Bros, U. Moschella, {\em Two point functions and quantum fields in de
Sitter universe}, Rev. Math. Phys. 8, 327 (1996).

\bibitem{bros2} J. Bros, U. Moschella, J. P. Gazeau, {\em Quantum field theory in the de
Sitter universe}, Phys. Rev. Lett. 73, 1746 (1994).


\bibitem{BJ} M. Brum, S. E. Jor\'as, {\em Hadamard state in Schwarzschild-de Sitter spacetime}, Class. Quantum Grav. 32, no. 1 (2014).

\bibitem{dang} N.V. Dang, {\em Renormalization of quantum field theory on curved space\-times, a causal approach}, Ph.D. thesis, Paris Diderot University, (2013).

\bibitem{DD} C. Dappiaggi, N. Drago, {\em Constructing Hadamard states via an extended M{\o}ller operator}, Lett. Math. Phys. 106 (11), (2016), 1587--1615.

\bibitem{DMP1} C. Dappiaggi, V. Moretti and N. Pinamonti, {\em Distinguished quantum states in a class
of cosmological spacetimes and their Hadamard property}, J. Math. Phys. 50 (2009) 062304.

\bibitem{DMP2} C. Dappiaggi, V. Moretti and N. Pinamonti, {\em Rigorous construction and Hadamard
property of the Unruh state in Schwarzschild spacetime}, Adv. Theor. Math. Phys. 15 (2011) 355.

\bibitem{derger} J. Derezi\'nski, C. G\'erard, {\em Mathematics of Quantization and Quantum Fields}, Cambridge Monographs in Mathematical Physics, Cambridge University Press (2013).

\bibitem{DS} J.~Derezi\'nski, D.~Siemssen, \emph{Feynman propagators on static spacetimes}, to appear in Rev. Math. Phys., DOI:10.1142/S0129055X1850006X, preprint \texttt{arXiv:1609.00192}, (2018).

\bibitem{DH} J.J. Duistermaat, L. H\"{o}rmander, \textsl{Fourier integral
operators II}, Acta Math. {\bf 128}  (1972), 183--269.

\bibitem{dyatlov} S. Dyatlov, {\em Asymptotics of linear waves and resonances with applications to black holes}, Commun. Math. Phys. 335 (2015), 1445--1485. 

\bibitem{FW} C. J. Fewster, R. Verch, \textsl{A quantum weak energy inequality for Dirac fields in
curved spacetime}, Commun. Math. Phys. 225, (2002), 331.

\bibitem{friedlander} F.G. Friedlander, \textsl{Radiation fields and hyperbolic scattering theory}, Math. Proc. Cambridge Philos. Soc., 88, (1980), 483--515.

\bibitem{FNW} S.A. Fulling, F.J. Narcowich, R.M. Wald, {\em Singularity structure of the two-point function in quantum field theory in curved space-time, II}, Annals of Physics, {\bf 136} (1981), 243--272. 
                                                                                                                                                                                                                                                              
\bibitem{characteristic} C. G\'erard, M. Wrochna, \textsl{Construction of Hadamard states by characteristic Cauchy problem}, Anal. PDE, 9 (1) (2014), 111--149.

\bibitem{GW} C. G\'erard, M. Wrochna, \textsl{Construction of Hadamard states by pseudo-differential calculus}, Commun. Math. Phys. \textbf{325} (2) (2014), 713--755.

\bibitem{GW2} C. G\'erard, M. Wrochna, \textsl{Hadamard states for the linearized Yang-Mills equation on curved spacetime}, Commun. Math. Phys. 337 (1) (2015), 253--320.

\bibitem{GW3} C. G\'erard, M. Wrochna, \textsl{The massive Feynman propagator on asymptotically Minkowski spacetimes}, to appear in Amer. J. Math., preprint \texttt{arXiv:1609.00192}, (2016).

\bibitem{GW4} C. G\'erard, M. Wrochna, \textsl{Hadamard property of the \emph{in} and \emph{out} states for Klein-Gordon fields on asymptotically static spacetimes},  Ann. Henri Poincar\'e 18 (8), (2017), 2715--2756.

\bibitem{GHV} J. Gell-Redman, N. Haber, A. Vasy, \textsl{The Feynman propagator on perturbations of Minkowski space}, Commun. Math. Phys. 342 (1), (2016), 333--384.

\bibitem{GL} C. R. Graham, J. M. Lee, \textsl{Einstein metrics with prescribed conformal infinity on the ball}, Adv. Math., 87 (2), (1991), 186--225.

\bibitem{guillarmou} C. Guillarmou, \textsl{Meromorphic properties of the resolvent on asymptotically hyperbolic manifolds}, Duke Math. J., 129 (1), (2005), 1--37.

\bibitem{haag} R. Haag, {\em Local quantum physics: Fields, particles, algebras}, Texts and Monographs in Physics, Springer (1992).

\bibitem{HV} N. Haber, A. Vasy, \textsl{Propagation of singularities around a Lagrangian submanifold of radial points}, Microlocal Methods in Mathematical Physics and Global Analysis. Springer Basel, (2013), 113-116.

\bibitem{hintz} P. Hintz, \textsl{Global analysis of linear and nonlinear wave equations on cosmological spacetimes}, PhD thesis, Stanford University (2015).

\bibitem{semilinear} P. Hintz, A. Vasy, \textsl{Semilinear wave equations on asymptotically de Sitter, Kerr-de Sitter and Minkowski spacetimes}, Anal. PDE, 8 (8), (2015), 1807--1890.

\bibitem{HW} S. Hollands, R.M. Wald, {\em Quantum fields in curved spacetime}, in: General Relativity and Gravitation: A Centennial Perspective, Cambridge University Press (2015).

\bibitem{hoermander} L. H\"ormander, {\em The analysis of linear partial differential operators I-IV}, Classics
in Mathematics, Springer (2007).

\bibitem{JJM} A. Jaffe, C.~J\"akel, R.E. Martinez {\em Complex classical fields: a framework for reflection positivity}, Commun. Math. Phys. 329 (2014), 1--28.

\bibitem{JR1} A. Jaffe, G. Ritter, {\em Quantum Field Theory on curved backgrounds. I. The Euclidean functional integral}, Commun. Math. Phys. 270 (2007), 545--572.

\bibitem{JR2} A. Jaffe, G. Ritter, {\em Reflection positivity and monotonicity}, J. Math. Phys. 49 (2008), 052301, 1--10.

\bibitem{JSB} M. Joshi, A. S\'a Barreto, \textsl{Inverse scattering on asymptotically hyperbolic manifolds}, Acta Math., 184 (1), (2000), 41--86.

\bibitem{Kay} B.S. Kay, {\em The principle of locality and quantum field theory on (non globally
hyperbolic) curved spacetimes}, Rev. Math. Phys. (Special Issue), (1992), 167--195.


\bibitem{KL} B.S. Kay, U. Lupo, {\em Non-existence of isometry-invariant Hadamard states for a Kruskal
black hole in a box and for massless fields on 1+1 Minkowski spacetime with a uniformly
accelerating mirror}, Class. Quantum Grav. 33 (21), (2016), 215001. 

\bibitem{KW} B.S. Kay,  R.M. Wald,  {\em Theorems on the uniqueness and thermal properties of stationary,
nonsingular, quasifree states on spacetimes with a bifurcate Killing horizon}, Phys. Rep.
207, 49 (1991).


\bibitem{KM} I. Khavkine, V. Moretti, {\em Algebraic QFT in curved spacetime and quasifree Hadamard states: an introduction}, in: Advances in Algebraic Quantum Field Theory, Springer (2015).

\bibitem{MM} R. Mazzeo, R. Melrose, \textsl{Meromorphic extension of the resolvent on complete spaces with asymptotically constant negative curvature}, J. Funct. Anal., 75 (2), (1987), 260--310.

\bibitem{melrose} R. Melrose, \textsl{The Atiyah-Patodi-singer index theorem}, Vol. 4. Wellesley: AK Peters, (1993).

\bibitem{melrose2} R. Melrose, \textsl{Spectral and scattering theory for the Laplacian on asymptotically Euclidean spaces}, Lecture Notes in Pure and Appl. Math., vol. 161, Dekker, New York (1994), 85--130.

\bibitem{melrose3} R. Melrose, \textsl{Geometric scattering theory}, Vol. 1, Cambridge University Press (1995).

\bibitem{melrosenotes} R. Melrose, \newblock Lecture notes for `18.157: {I}ntroduction to microlocal analysis'.
\newblock Available at 
  \texttt{http://math.mit.edu/\~{}rbm/18.157-F09/18.157-F09.html}, (2009).

\bibitem{Mo1} V. Moretti, {\em  Uniqueness theorem for BMS-invariant states of scalar QFT on the null boundary of asymptotically flat spacetimes and bulk-boundary observable algebra correspondence}, Commun. Math. Phys. {\bf 268} (2006), 727-756.

\bibitem{Mo2} V. Moretti, {\em  Quantum out-states holographically induced by asymptotic flatness: invariance under space-time symmetries, energy positivity and Hadamard property}, Commun. Math. Phys. {\bf 279} (2008), 31-75.

\bibitem{MS} U. Moschella, R. Schaeffer, {\em Quantum Theory on Lobatchevski Spaces}, Class. Quant. Grav., 24:3571--3602, (2007).

\bibitem{radzikowski} M. Radzikowski, {\em Micro-local approach to the Hadamard condition in quantum field theory on curved space-time}, Commun. Math. Phys. {\bf 179} (1996), 529--553.

\bibitem{radzikowski2} M.  Radzikowski, {\em A Local to global singularity theorem for quantum field theory on curved
space-time}, Commun. Math. Phys. 180, 1 (1996).

\bibitem{rehren} K.-H. Rehren, {\em Boundaries in relativistic quantum field theory}, to appear in the proceedings of the XVIII International Congress on Mathematical Physics, Santiago de Chile, July 2015,  preprint \texttt{arXiv:1601.00826} (2016).

\bibitem{sanders} K. Sanders, {\em Equivalence of the (generalized) Hadamard and microlocal spectrum condition for (generalized) free fields in curved space-time}, Commun. Math. Phys. {\bf 295} (2010), 485--501.

\bibitem{sanders2} K. Sanders,  {\em On the Construction of Hartle-Hawking-Israel States Across a
Static Bifurcate Killing Horizon}, Lett. Math. Phys. 105, 4 (2015), 575--640.

\bibitem{SV} H. Sahlmann, R. Verch, {\em Microlocal spectrum condition and Hadamard form for vector-valued quantum fields in curved spacetime}, Rev. Math. Phys., 13(10) (2001), 1203-1246.

\bibitem{strominger} A. Strominger, {\em The dS/CFT Correspondence}, JHEP 0110, (2001), 341--346.

\bibitem{resolvent} A. Vasy, \textsl{Microlocal analysis of asymptotically hyperbolic spaces and high energy resolvent estimates}, MSRI Publications, vol. 60, Cambridge University Press, (2012).

\bibitem{forms} A. Vasy, \textsl{Analytic continuation and high energy estimates for the resolvent of the {L}aplacian on forms on
asymptotically hyperbolic spaces}, Adv. Math. 306, (2017), 1019--1045. 

\bibitem{positive} A. Vasy, \textsl{On the positivity of propagator differences}, Ann. Henri Poincar\'e 18 (3), (2017), 983--1007. 

\bibitem{kerrds} A. Vasy, \textsl{Microlocal analysis of asymptotically hyperbolic and Kerr-de Sitter spaces, (With an appendix by S. Dyatlov)}, Inventiones Math., 194, (2013), 381--513.

\bibitem{corners} A. Vasy, \textsl{Propagation of singularities for the wave equation on manifolds with corners}, Ann. of Math. (2) 168 (3), (2008), 749--812.
 
\bibitem{propagation} A. Vasy, \textsl{Propagation phenomena}, Lecture Notes, Stanford University (2014).

\bibitem{poisson} A. Vasy, \textsl{Resolvents, Poisson operators and scattering matrices on asymptotically hyperbolic
and de Sitter spaces}, J. Spect. Theory, vol. 4 (4), (2014), 643--673.

\bibitem{desitter} A. Vasy, \textsl{The wave equation on asymptotically de Sitter-like spaces}, Adv. Math., 223 (1), (2010), 49--97.

\bibitem{vasyminicourse} A.~Vasy, \textsl{A minicourse on microlocal analysis for wave propagation}, in: Asymptotic Analysis in General Relativity, London Mathematical Society Lecture Note Series 443, Cambridge University Press, (2018).

\bibitem{wald} R.M. Wald, {\em Dynamics in nonglobally hyperbolic, static space‐times}, J. Math. Phys., 21, 2802-2805 (1980).

\bibitem{zahn} J. Zahn, {\em Generalized Wentzell boundary conditions and quantum field theory}, Ann. Henri Poincar\'e, 19, (2018), 163.
 
\bibitem{zworski} M. Zworski, \textsl{Resonances for asymptotically hyperbolic manifolds: Vasy's method revisited}, J. Spect. Theor. 6, (2016), 1087--1114.


\end{thebibliography}
\end{document}